\documentclass[12pt]{article}
\usepackage{cite}
\usepackage{amsmath, amsthm, amssymb,slashed}
\usepackage{ifpdf}
\ifpdf
  \usepackage[pdftex]{graphicx}
  \usepackage{epstopdf}
\else
  \usepackage[dvips]{graphicx}
\fi
\parskip=\baselineskip
\def\Bbb{\mathbb}
\def\Tr{{\rm Tr}}
\def\16{{\bf 16}}
\def\1{{\bf 1}}
\def\2{{\bf 2}}
\def\3{{\bf 3}}
\def\4{{\bf 4}}
\def\bar{\overline}
\def\tilde{\widetilde}
\def\R{{\Bbb{R}}}\def\Z{{\Bbb{Z}}}
\def\N{{\mathcal N}}
\def\hat{\widehat}
\def\frak{\mathfrak}
\font\teneurm=eurm10 \font\seveneurm=eurm7 \font\fiveeurm=eurm5
\newfam\eurmfam
\textfont\eurmfam=\teneurm \scriptfont\eurmfam=\seveneurm
\scriptscriptfont\eurmfam=\fiveeurm

\font\teneusm=eusm10 \font\seveneusm=eusm7 \font\fiveeusm=eusm5
\newfam\eusmfam
\textfont\eusmfam=\teneusm \scriptfont\eusmfam=\seveneusm
\scriptscriptfont\eusmfam=\fiveeusm

\font\tencmmib=cmmib10 \skewchar\tencmmib='177
\font\sevencmmib=cmmib7 \skewchar\sevencmmib='177
\font\fivecmmib=cmmib5 \skewchar\fivecmmib='177
\newfam\cmmibfam
\textfont\cmmibfam=\tencmmib \scriptfont\cmmibfam=\sevencmmib
\scriptscriptfont\cmmibfam=\fivecmmib
\def\cmmib#1{{\fam\cmmibfam\relax#1}}
\numberwithin{equation}{section}

\def\d{\mathrm d}
\def\C{{\Bbb C}}
\def\Z{{\Bbb Z}}
\def\A{{\mathcal A}}
\def\bar{\overline}     
\begin{document}

\begin{titlepage}
\begin{flushright}
hep-th/yymm.nnnn
\end{flushright}
\vskip 1.5in
\begin{center}
{\bf\Large{Analytic Continuation Of Chern-Simons Theory }}\vskip
0.5cm {Edward Witten} \vskip 0.05in {\small{ \textit{School of
Natural Sciences, Institute for Advanced Study}\vskip -.4cm
{\textit{Einstein Drive, Princeton, NJ 08540 USA}}}}

\end{center}
\vskip 0.5in
\baselineskip 16pt
\date{November, 2009}

\begin{abstract}
The title of this article refers to analytic continuation of three-dimensional Chern-Simons
gauge theory away from integer values of the usual coupling parameter $k$, to explore questions such as the volume conjecture, or analytic continuation of three-dimensional quantum gravity (to the extent that it can be described by gauge theory) from Lorentzian to Euclidean signature.  Such analytic continuation can be carried out
by generalizing the usual integration cycle of the Feynman path integral.  Morse theory or Picard-Lefschetz
theory gives a natural framework for describing the appropriate integration cycles.   An important part of the analysis involves flow equations that turn out to have a surprising four-dimensional symmetry.   After developing a general framework, we describe some specific examples
(involving the trefoil and figure-eight knots in $S^3$).  We also find that the space of possible
integration cycles for Chern-Simons theory can be interpreted as the ``physical Hilbert space''
of a twisted version of $\N=4$ super Yang-Mills theory in four dimensions.
\end{abstract}
\end{titlepage}
\vfill\eject \tableofcontents
\section{Introduction}

\def\CC{{\mathcal C}}
\def\n{{\mathfrak n}}
\def\m{{\mathfrak m}}
\def\cJ{{\cmmib J}}
 This paper is devoted to the analytic continuation of
three-dimensional Chern-Simons gauge theory as a function of its
coupling parameters. One motivation for this study concerns
three-dimensional quantum gravity, which plausibly has at least
some sort of relationship to three-dimensional Chern-Simons theory
with gauge group $SL(2,\C)$. In that context, Euclidean quantum
gravity is obtained by analytic continuation to imaginary values
of the coupling parameter $s$ that is introduced in section
\ref{overview}.\footnote{And the singularity at $s=\pm i\ell$ that
we will discuss in section \ref{sing} is at least a cousin of the
``chiral point'' of three-dimensional gravity, which has been much
studied recently \cite{Strom}.} However, for imaginary $s$, the
Chern-Simons path integral, like the Euclidean quantum gravity
path integral in any dimension, does not appear to be convergent.
Our analysis in this paper will explain how to make sense of this
sort of path integral, at least in the Chern-Simons case.

Another motivation is to understand aspects of the Jones
polynomial of knots\footnote{The usual mathematical terminology is
that a knot is an embedded circle in a three-manifold (usually
 a three-sphere) while a union of disjoint
embedded circles is called a link.  Except in section \ref{tensoring}, all
statements about knots in
this paper have rather immediate analogs for links, but for
brevity we will refer to knots.}
 and its generalizations that have not
yet been explained from the point of view of three-dimensional
gauge theory.  The Jones polynomial of a knot was originally
defined as a Laurent polynomial in a complex variable $q$ -- a
rational function of $q$ with poles only at $q=0$ and $q=\infty$.
Chern-Simons gauge theory with gauge group $SU(2)$ (or another
compact Lie group) reproduces the Jones polynomial, and its
generalizations, at values of $q$ of the form $\exp(2\pi
i/(k+h))$, where $k$ is a positive integer, and $h$ (which is 2
for $SU(2)$) is the dual Coxeter number of the gauge group. The
Jones polynomial and related invariants are determined by their
values for these special values of $q$, but it would be desirable
to have a gauge theory explanation of the existence of a natural
analytic continuation. It has generally been expected that the
explanation has something to do with the relation between
Chern-Simons theories with gauge groups $SU(2)$ or $SL(2,\C)$, but
a really clear picture has not been available. A better
understanding appears to require analytic continuation of
Chern-Simons theory beyond the values of the parameters at which
the path integral can be most simply understood.

Finally, in the last decade or more, striking new results have
been obtained
\cite{Kash,MM,MMOTY,KT,Gu,Mm,GLe,GL,DK,GuM,V,Zh,SG,JH,Mura,km,DGLZ,DF,Z,kmtwo}
  about the analytic continuation of invariants
related to the Jones polynomial.  (A useful review is \cite{Mura};
partial physics-based explanations with some relevance to the
present paper are \cite{Gu,DGLZ,DF}. While it is impractical to
give complete references on Chern-Simons theory, some
contributions comparing the asymptotic or in some cases exact
behavior of the Chern-Simons partition function to geometry, but
not focusing on analytic continuation, are
\cite{FG,LJe,BG,Ro,Roz,Roza,JA,Roz1,JR,Marino,BW,BT, Beasley}.) The
colored Jones polynomial $J_n(q)$ of a knot is interpreted in
Chern-Simons gauge theory in terms of the expectation value of a
Wilson loop operator associated to the $n$-dimensional
representation of $SU(2)$. Let $q^n=\exp(2\pi i\gamma)$, and
consider the behavior of $J_n(q)$ in a scaling limit with $n$
tending to infinity for fixed $\gamma$. The behavior of this limit
for certain values of $\gamma$ can be naturally understood in
$SU(2)$ Chern-Simons theory, but many of the more recent results
certainly appear to be more directly related to the theory with
gauge group $SL(2,\C)$. For example, for $\gamma$ near 1, and at
least for some knots, the large $n$ behavior appears to be
dominated by an $SL(2,\C)$-valued flat connection that does not
take values in $SU(2)$.  Although there has not been a completely
clear picture of how and why the scaling limit of $J_n(q)$ should
be related to $SL(2,\C)$ Chern-Simons theory, the idea that there
is such a relationship has been a fruitful source of new ideas and
generalizations \cite{Gu}. Again, it has been clear that a fuller
picture will require a better understanding of the analytic
continuation of Chern-Simons theory.

In this paper, we will definitely not obtain complete answers to
all of the above questions, and among other things we do not have
a general answer concerning the volume conjecture.  (It is true,
as will become clear, if a certain integer-valued coefficient is
always nonzero, but we do not know why this would always be so.)
 But we do hope to make some things clearer.  We
show that critical points of the analytically continued
Chern-Simons functional should be used to determine -- by steepest
descent -- suitable integration cycles for the quantum path
integral.   For a given value of the coupling parameters, not all
critical points should be included; doing so would lead to
immediate contradiction with known facts about the knot
polynomials.  Instead, one should start with a real integration
cycle for real coupling and then, as one varies the coupling into
the complex plane, one varies the integration cycle so that the
integral remains convergent. In the process, one encounters subtle
Stokes phenomena that govern the appearance and disappearance of
critical points. The Stokes phenomena can be described, in
principle, by solving steepest descent equations that in the
context of Chern-Simons theory amount to four-dimensional elliptic
differential equations that generalize the instanton equation.
This gives a suitable framework for analytically continuing the
$SL(2,\C)$ Chern-Simons theory as a function of its coupling
parameter, and for better understanding its relation to the
$SU(2)$ theory.

Section \ref{overview} of the paper is devoted to a more complete
overview of some of these matters.  

In section
\ref{stokes}, we describe the analytic continuation of oscillatory
integrals in finite dimensions and their analysis by steepest
descent.  This is an application of the framework of
  \cite{Arnold,Fedorjuk,Pham,BH,BH2,Howls}, which is reviewed here for
convenience.  Various aspects of this framework have been applied
to quantum field theory from different points of view
\cite{CV,VHI,FHKS,MSS,GMN}.
 Floer theory (for reviews see for instance \cite{KM,MH}) is a prototype for the application
 of Morse theory to middle-dimensional cohomology of function spaces,
 which appears
 in our approach to Chern-Simons gauge theory.

In section \ref{backtocs},  we apply what
we have learned to the analytic continuation of
Chern-Simons gauge theory on a generic three-manifold.  A striking result
emerges: the gradient flow equations relevant to analytic continuation of
three-dimensional Chern-Simons theory actually have four-dimensional symmetry.

In  section
\ref{anajones}, we consider the special case of knot invariants in
$S^3$, where stronger conclusions are possible because, in the
absence of the knot, there are no non-trivial flat connections.
The main examples considered in section \ref{anajones} are the trefoil
knot and the figure-eight knot.  In each case, we consider the semi-classical limit of the colored Jones polynomial, keeping fixed the conjugacy class of the monodromy
of a flat connection around the knot.  (In other words, we consider the limit mentioned
above of $n\to\infty$ with fixed $\gamma$.)
The technique that we use to describe the Stokes phenomena relies on analyzing
the  singularities of the moduli space of flat connections -- that is, the special values of the
monodromy at which flat connections that are generically distinct become coincident.  
The semiclassical behavior near such a special value is interesting in itself; it can be modeled
by a finite-dimensional integral associated to the relevant singularity.  We analyze several
examples of such singularities and compare the results to what has been found in the mathematical literature in studies of the semiclassical limit of the colored Jones polynomial. 
For the trefoil knot, though many interesting questions remain, the Stokes phenomena that we analyze suffice to determine the integration
cycle of the analytically continued Chern-Simons path integral for all values of $\gamma$.
For the figure-eight knot, we primarily determine what happens close to the real $\gamma$ axis.
  In both
cases, our analysis of the Stokes phenomena suffices to explain the validity of the volume
conjecture.

Finally, in section \ref{fourfive}, we make a few remarks on a
four or five-dimensional interpretation of what we have learned, with the possible
goal of a new understanding of Khovanov homology.

\section{Overview}\label{overview}

We begin by considering Chern-Simons gauge theory with
the gauge group $G$  a complex Lie group
such as $SL(2,\C)$. Let $\A$ be a connection on a $G$-bundle $E$
over a three-manifold $M$.  Such a connection has a complex-valued
Chern-Simons invariant
\begin{equation} W(\A)=\frac{1}{4\pi}\int_M \,\Tr\,\left(\A\wedge
\d\A+\frac{2}{3}\A\wedge\A\wedge A\right),\end{equation} which we
have normalized to be gauge-invariant modulo $2\pi$.  (For
$G=SL(N,\C)$, the symbol $\Tr$ denotes the trace in the
$N$-dimensional representation.  In general, it denotes a suitably
normalized quadratic form on the Lie algebra $\mathfrak g$ of
$G$.)  Just as in the case of a compact group, $W(\A)$ is
gauge-invariant modulo $2\pi$.  The indeterminacy in $W(\A)$ is
real, and is the same as it would be if $G$ were replaced by a maximal
compact subgroup $H$, because $G$ is contractible onto $H$.\footnote{The indeterminacy in
$W(\A)$ comes from its change in going around a loop in the space of connections
modulo gauge transformations, and this change only depends on the homotopy class
of the loop.  Since $G$ is contractible to  $H$, any family of
$\frak g$-valued connections on $M$ can be deformed to a family of $\frak h$-valued connections, where $\frak g$, $\frak h$ are the Lie algebras of $G$ and $H$.  So the indeterminacy
in the Chern-Simons function for gauge group $G$ reduces to that for gauge group $H$.}

\def\bA{\overline{\A}}
Thus the  imaginary part of $W(\A)$ is actually a well-defined
real number, while its real part takes values in $\R/2\pi\Z$. To
obtain a quantum field theory, we wish to start with a classical
action $I$ that is a linear combination of ${\mathrm {Re}}\,W$ and
${\mathrm {Im}}\,W$. The quantum theory is based upon integrating
the expression $\exp(i I)$, and for this function to be
well-defined, $I$ must be defined mod $2\pi$. Because of the
indeterminacy in ${\mathrm {Re}}\,W$, its coefficient must be an
integer, while (if the only desired condition is to ensure that
$\exp(i I)$ is a well-defined complex-valued function) the
coefficient of ${\mathrm {Im}}\,W$ may be an arbitrary complex
number.  The action therefore has the general form
\begin{equation}\label{mork} I = -s\,{\mathrm {Im}}\,W + \ell\, {\mathrm
{Re}}\,W,~~s\in \C,~\ell\in \Z.\end{equation} Alternatively, we can write
\begin{equation}\label{ork}I = \frac{t \,W}{2}+\frac {\tilde t\,\,
\overline W}{2}=\frac{t}{8\pi}\int_M\Tr\,\left(\A\wedge
\d\A+\frac{2}{3}\A\wedge\A\wedge \A\right) +\frac{\tilde
t}{8\pi}\int_M\Tr\,\left(\bA\wedge \d\bA+\frac{2}{3}\bA\wedge\bA\wedge
\bA\right),\end{equation} with
\begin{equation}\label{ossy} t=\ell+is, ~\tilde t =
\ell-is.\end{equation}

\def\Ai{\mathrm{Ai}}
There is more to life, however, than  making sure that $\exp(i I)$
is a well-defined complex-valued function.  We also want to
integrate this function over the space  $\mathcal Y$ of gauge
fields modulo gauge transformations.  This makes sense
straightforwardly only if $s$ is real, for then $I$ is real and the
integrand $\exp(i I)$ of the path integral is bounded.  The path
integral
\begin{equation} \label{zork} \int_{\mathcal Y}
D\A\,D\bA\,\exp(iI) \end{equation} is then an infinite-dimensional
oscillatory integral.  In this respect, it is like any Lorentz
signature Feynman integral (though here we are doing topological
field theory rather than quantum field theory on a manifold of
Lorentz signature).   If $s$ is not real, the path integral is not convergent; our task
is to make sense of it anyway.

\subsection{Finite-Dimensional Analogs}\label{analogs}

A much simpler prototype of this problem
is a finite-dimensional oscillatory integral such as the one that defines the Airy
function:
\begin{equation}\label{morok} Z_\lambda=\int_{-\infty}^\infty
{\d x}\,\exp\left(i\lambda(
x^3/3-x)\right).\end{equation} For real $\lambda$, this integral
converges, though not absolutely so, because of rapid oscillations
at infinity; when $\lambda$ becomes complex, it must be defined by
a process of analytic continuation.

The Airy integral will serve as a useful practice case in section
\ref{stokes}.  The following slightly more elaborate example is also useful.
Introduce a complex variable $z$ and a
complex-valued polynomial
\begin{equation}\label{opor}g(z)=\sum_{j=0}^n a_j
z^j.\end{equation}
Now
consider the integral
\begin{equation}\label{omor}Z_g=\int
|\d^2z|\exp\left(g(z)-\overline{g(z)}\right).  \end{equation} This
again is a convergent oscillatory integral and a closer analog of complex
Chern-Simons theory, with $-ig(z)$ and $i\bar {g(z)}$ corresponding
to the terms $t W$ and $\tilde t \,\,\overline W$ in the action
(\ref{ork}).

We stress that no contour integral is intended in (\ref{omor}) --
a contour integral could scarcely be intended here as the
integrand is not holomorphic!  Rather, if $z=u+iv$, with real
$u,v$, we integrate separately over $u$ and $v$, the integration
measure being $|\d^2z|=2 \,\d u\,\d v$.

A finite-dimensional analog of analytically continuing
Chern-Simons theory with complex gauge group as a function of $s$
is to analytically continue the Airy function $Z_\lambda$ to
complex $\lambda$.  An even closer analog is to analytically
continue the integral $Z_g$ to let $\overline g$ be independent of
$g$.  By this, we mean the following.  Define a new polynomial
\begin{equation}\label{oppor}\tilde g(z)=\sum_{j=0}^n \tilde a_j
z^j,\end{equation} and generalize the integral $Z_g$ to
\begin{equation}\label{norkly} Z_{g,\tilde g}=\int
|\d^2z|\exp\left(g(z)-\tilde g(\bar z)\right).  \end{equation}
Then $Z_{g,\tilde g}$ coincides with the original $Z_g$ if $\tilde
g=\bar g$ (by which we mean that $\tilde a_j=\bar a_j$ for all
$j$).

The natural way to analytically continue the Airy function in
$\lambda$ is to let $x$ become a complex variable and to deform
the integration contour away from the real axis so that the
integral remains convergent as $\lambda$ varies.  Similarly, to
extend the definition of $Z_{g,\tilde g}$ to $\tilde g\not= \bar
g$, we analytically continue, treating $z$ and $\bar z$ -- or
equivalently $u$ and $v$, defined by $z=u+iv$, $\bar z=u-iv$ -- as
independent complex variables. Denoting the independent complex
variables as $z$ and $\tilde z$ (rather than $z$ and $\bar z$),
the analytically continued integral is
\begin{equation}\label{zpor}Z_{g,\tilde g}=\int_{\mathcal
C}\d z\,\d\tilde z\,\exp\left(g(z)-\tilde g(\tilde
z)\right).\end{equation} The integral is over a two-dimensional
real integration cycle $\mathcal C$, which is the real slice
$\tilde z=\bar z$ if $\tilde g=\bar g$, and in general must be
deformed as $g$ and $\tilde g$ vary so that the integral remains
convergent.

This then is the framework that we will have to imitate to
understand the analytic continuation of Chern-Simons theory with a
complex gauge group.  Just as we promoted $\bar z$ to a complex
variable $\tilde z$ that is independent of $z$, we will have to
promote $\bA$ to a new $G$-valued connection $\tilde\A$ that is
independent of $\A$.  Thus we consider the classical theory with
independent $G$-valued connections $\A$ and $\tilde\A$ (we recall
that $G$ is a complex Lie group such as $SL(2,\C)$) and action
\begin{equation}\label{morko}I(\A,\tilde\A)=\frac{t}{8\pi}\int_M\Tr\,\left(\A\wedge
\d\A+\frac{2}{3}\A\wedge\A\wedge \A\right) +\frac{\tilde
t}{8\pi}\int_M\Tr\,\left(\tilde\A\wedge
\d\tilde\A+\frac{2}{3}\tilde\A\wedge\tilde\A\wedge
\tilde\A\right).\end{equation} Then we have to find an appropriate
integration cycle ${\mathcal \CC}$ in the path integral
\begin{equation}\label{zorko} \int_{\mathcal \CC}
D\A\,D\tilde\A\,\exp\left(iI(\A,\tilde\A)\right).\end{equation}
The cycle ${\mathcal \CC}$ must be equivalent to $\tilde\A=\bar\A$
if $s$ is real, and in general must be obtained by deforming that
one as $s$ varies so that the integral remains convergent.   (We
also have to learn how to deal with gauge invariance when $\A$ and
$\tilde \A$ are treated as independent.)

\def\D{{\mathcal D}}
\def\E{{\mathcal E}}
\def\M{{\mathcal M}}
\def\CC{{\mathcal C}}
\def\W{{\mathcal W}}
\def\I{{\mathcal I}}
\def\J{{\mathcal J}}
\subsection{Role Of Morse Theory}\label{morse}

Finding a suitable integration cycle in the infinite-dimensional
path integral of Chern-Simons gauge theory may seem hopeless at
first sight.  What saves the day is that suitable integration
contours can be found by Morse theory and steepest descent (or in a different
language, via Picard-Lefschetz theory). The facts we need for understanding
oscillatory integrals have been described in
\cite{Arnold,Fedorjuk,Pham,BH,BH2,Howls}  (and the aspiration of applying this
approach to Feynman integrals is stated in \cite{Pham}, p. 321).
Section \ref{stokes} of the present paper is devoted to a
hopefully self-contained exposition of this approach, with an
emphasis on the aspects that we need in Chern-Simons theory.  The
basic idea of the theory is to associate potential integration
cycles to components of the critical point set of the extended
action $I(\A,\tilde \A)$.

\def\F{{\mathcal F}}
What makes it practical to use this formalism in Chern-Simons
theory is that the critical points are accessible.  As long as $t$
and $\tilde t$ are both non-zero, the critical point equation
derived from the analytically continued action (\ref{morko}) is
\begin{equation}\label{ozorkp}\F=\tilde \F=0,\end{equation}
where $\F=\d\A+\A\wedge\A$ and
$\tilde\F=\d\tilde\A+\tilde\A\wedge\tilde\A$ are the curvatures of
the two connections.  Hence a critical point is just a pair of
flat connections, corresponding to a pair of homomorphisms from
the fundamental group of $M$ to the complex Lie group $G$. (We
postpone to section \ref{backtocs} an important detail about the
value of the Chern-Simons invariant at a critical point.)  Hence
in Chern-Simons theory on a given three-manifold $M$, the critical
points can be effectively described; this is typically not true
for oscillatory Feynman integrals.

\def\K{{\mathcal K}}
For simplicity, we assume until section \ref{mancrit} that the
critical points are isolated points. Let $\Sigma$ be the set of
critical points and denote the critical points as $p_\sigma$,
$\sigma\in \Sigma$. Then to each $p_\sigma$, we attach an
integration cycle $\J_\sigma$ that is obtained by downward or
gradient flow from $\sigma$, taking the real part of the action as
a Morse function. (Details are explained in section \ref{stokes}.)
The $\J_\sigma$, which are known as Lefschetz thimbles, are
constructed so that the integral over $\J_\sigma$ always
converges, regardless of the choice of $\sigma$ or of the parameters in the action.
Any suitable integration cycle $\mathcal C$ for the sort of
integral that we are interested in is equivalent to a linear
combination of the $\J_\sigma$ with integer coefficients:
\begin{equation}\label{lincom} \mathcal C=\sum_\sigma \n_\sigma
\J_\sigma,~~\n_\sigma\in\Z.\end{equation} The coefficients
$\n_\sigma$ can themselves be computed via Morse theory.  To do
this,  one attaches to each critical point $p_\sigma$ another
cycle $\K_\sigma$ by upwards gradient flow starting from
$p_\sigma$. The expansion coefficients in eqn. (\ref{lincom}) can
be computed as the intersection pairing of the integration cycle
$\mathcal C$ with the $\K_\sigma$:
\begin{equation}\label{pincom} \n_\sigma=\langle
{\mathcal C},\K_\sigma\rangle.\end{equation}

One might naively think that all critical points would contribute
to the integral with $\n_\sigma=1$ (or with $\n_\sigma=\pm 1$,
since in general the $\J_\sigma$ do not have natural
orientations), but this cannot be so.  For example, in
Chern-Simons theory with real $s$, there are complex critical
points at which the action $I$ has a negative imaginary part;
their inclusion would lead to a pathologically wrong exponential
growth of the partition function for large real $s$.   In fact,
for real $s$, with $\CC$ taken to be the real cycle $\CC_\R$
defined by $\tilde\A=\bA$,  real critical points (whose
contributions to the path integral are oscillatory) all have
$\n_\sigma=1$, but critical points that would make exponentially
large contributions to the path integral have $\n_\sigma=0$. These
statements will be deduced from
 (\ref{pincom}).  However,  there is no simple formula for
$\n_\sigma$ for critical points whose contributions to the real
$s$ path integral are exponentially small.

\subsection{Stokes Phenomena}\label{anacon}

Once one expresses the integration cycle $\CC$ in terms of
critical point cycles  for real values of the Chern-Simons
coupling constant $s$ (or of the appropriate parameters in the
other examples) via a formula $\CC=\sum_\sigma \n_\sigma
\J_\sigma$, the next step is to make an analytic continuation in
$s$.  There is no problem in defining the critical points
$p_\sigma$ and the corresponding Lefschetz thimbles $\J_\sigma$
for any generic value of $s$.  In fact, in Chern-Simons theory,
the critical point equation (\ref{ozorkp}) is independent of $s$
(except at $s=\pm i\ell$, where $t$ or $\tilde t$ vanishes). The
$\J_\sigma$ do depend on $s$ (their definition by steepest descent
makes use of the action, which depends on $s$), but their
definition ensures that the path integral over $\J_\sigma$ is
always convergent, even when $s$ has an imaginary part.

One can find the right coefficients in the expansion
$\CC=\sum_\sigma \n_\sigma \J_\sigma$ by starting at real $s$,
where one knows what to do, and then continuously evolving in $s$.
Here, however, one runs into the main subtlety of the theory:
Stokes phenomena. As one varies $s$, one crosses Stokes curves at
which both the Lefschetz thimbles $\J_\sigma$ and the coefficients
$\n_\sigma$ jump, in such a way that the homology cycle
$\sum_\sigma \n_\sigma \J_\sigma$ is locally constant.  Such
jumping can occur when there are paths of steepest descent that
connect two critical points.  The flow by steepest descent has a
conserved quantity (the imaginary part of the holomorphic function
whose real part is taken as the Morse function), and a flow
between two critical points is possible only when this quantity
takes equal values at the two critical points. This is reviewed in
section \ref{stokes}.

After taking the Stokes phenomena into account, one can obtain a
recipe for how the $\n_\sigma$ evolve as $s$ varies in the complex
plane.  This then gives, finally, a way to define the analytic
continuation in $s$ of the Chern-Simons path integral for the case that the gauge group is
a complex Lie group.

\subsection{Singularities Of The Analytically Continued
Theory}\label{sing}

At this stage, it is very illuminating to ask what sort of
singularities the analytically continued theory has as a function
of $s$.   Generically, the analytic continuation of an oscillatory
integral grows exponentially as the appropriate complex parameter
goes to infinity in certain directions in the complex plane. The
appropriate parameters are $s$ for complex Chern-Simons theory,
$\lambda$ for the Airy function, and the coefficients of the
polynomials $g$ and $\tilde g$
 in our other example.

Because of this exponential growth, the analytically continued
Chern-Simons path integral has an essential singularity at
$s=\infty$. Generically, there is in addition an ordinary
monodromy around $s=\infty$; it results from the fact that, by
virtue of the Stokes phenomena, the integration contour $\CC$ does
not return to itself under continuation around $s=\infty$.

Apart from an essential singularity at infinity, will analytically
continued Chern-Simons theory have other singularities as a
function of $s$?  The answer to this question is that one should
expect singularities at $s=\pm i \ell$, where the action
(\ref{ork}) becomes holomorphic or antiholomorphic.  An
oversimplified prototype of this singularity is the Gaussian
integral
\begin{equation}\label{oversi}\int |dz\wedge d\bar z| \exp(a
z^2+b\bar z^2)=\frac{\pi}{\sqrt{-ab}},\end{equation} with
singularities at $a=0$ or $b=0$. (The integral is defined by
analytic continuation starting from an oscillatory Gaussian
integral at $b=-\bar a$.)

This example is oversimplified because the analytically continued
exponent, which is $a z^2+b \tilde z^2$, has only a unique
critical point, the one at $z=\tilde z=0$.  In Chern-Simons
theory, there are in general many critical points $p_\sigma$. In
general there are distinct critical points that have the same
action at $s=\pm i\ell$, and as a result many Stokes curves pass
through the points $s=\pm i\ell$. Hence the singularities at
$s=\pm i\ell$ will be much more complicated than in
 the Gaussian example. These singularities  will not be
analyzed in this paper, but it will not be surprising if the
analytic continuation of Chern-Simons theory in general should be
defined on the universal cover of the complex $s$-plane, with the
points at $s=\pm i\ell$ (and $\infty$) omitted. Thus the partition
function would be really a function not of $s$ but of a certain
dilogarithm of $s$.

\subsection{Analytic Continuation For Compact Gauge Groups}\label{turz}

\def\U{{\mathcal U}}
Now we will discuss the analytic continuation of Chern-Simons theory with a compact
gauge group.  We consider it first in a naive way and then in relation to the theory
with a complex gauge group.

In Chern-Simons gauge theory on a three-manifold $M$ with a
compact gauge group $H$, the gauge field is a connection $A$ on an
$H$-bundle $E\to M$.  In terms of the Chern-Simons function
$W(A)$, which we normalize as usual to be well-defined modulo
$2\pi\Z$, the action is\ $I=k W$, where $k$ is taken to be an
integer so that the integrand $\exp(i I) =\exp(i k W)$ of the path
integral is well-defined.  Then one defines the path integral as
the integral over the space\footnote{\label{thelm}  A variant in
which one takes $\mathcal U$ to be the quotient of the space of
connections by the group of based gauge transformations is also
useful.  (A based gauge transformation is one that equals 1 at a
prescribed point; the advantage of dividing only by this group is
that it acts freely, so dividing by it produces no singularities.)
In this variant, one must divide the path integral by a constant
factor, the volume of the residual group of gauge transformations.
Another important variant arises in working out the contribution
to the path integral from a given gauge orbit of flat connections
on $M$; here one may take $\mathcal  U$ to be a transversal to the
given orbit in the space of all connections on $E$.  When we speak
loosely of $\U$ as the space of connections mod gauge
transformations, we refer to any of these variants that may be
most convenient in a given context.}
 $\mathcal U$ of connections mod gauge transformations:
\begin{equation}\label{zdef}Z_H(k)=\int_\U DA\,\exp(i {k} W(A)).\end{equation}
In the presence of a knot $K$ labeled by a representation $R$, one includes a holonomy
factor and considers the integral:
\begin{equation}\label{mydef}Z_H(K;k)=\int_\U DA\,\exp(i {k} W(A))\,\Tr_RP\exp\left(-\oint_KA\right).\end{equation}

As written, the integral does not make sense unless ${k}$ is an
integer.  Indeed, as $W(A)$ is defined only modulo $2\pi$, the
expression $\exp(i{k} W(A))$ makes sense as a function on $\U$ if
and only if ${k}\in\Z$.  Can we evade this limitation? A naive way
to try to do so is to replace $\U$ by its universal cover\footnote{To be more precise,
we can take $\hat \U$ to be the smallest cover of $\U$ on which the Chern-Simons function
$W(A)$ makes sense as a real-valued function.  If $H$ is connected and simply-connected,
this is the same as the universal cover.}
$\hat\U$. $\hat\U$ is the quotient of the space of connections by
the group of topologically trivial gauge transformations. The
expression $\exp(i{k} W(A))$ is invariant under such gauge
transformations, so it makes sense as a function on $\hat\U$.
However, its integral vanishes for ${k}\not\in\Z$, since the
integrand transforms with a non-trivial phase under the deck
transformations associated to the cover $\hat\U\to \U$.

To cure this problem, we complexify the gauge field, replacing $A$
by a complex-valued connection $\A$.  $\A$ is now a connection on
an $H_\C$-bundle $E_\C\to M$.  We replace $\U$ and $\hat\U$ by
their complexifications $\U_\C$ and $\hat\U_\C$, which
respectively parametrize complex-valued connections modulo
complex-valued gauge transformations or modulo topologically
trivial complex-valued gauge transformations.    Finally, in
$\hat\U_\C$, we pick an integration cycle $\CC$ and define the
integral
\begin{equation}\label{kdef}Z_{H,\CC}({k})=\int_\CC DA\,\exp(i{k}W(\A)) .\end{equation}
(When a knot $K$ is present, we denote the corresponding integral
as $Z_{H,\CC}(K;k)$.) The analytic continuation of the path
integral measure $DA$ on $\hat\U$ is a holomorphic volume form on
$\hat\U_\C$ that we continue to denote simply as $DA$. In section
\ref{littlemore}, it will be expressed in terms of the holomorphic
volume form $D\A$ used in the complex path integral (\ref{zorko}).

We will pick $\CC$ so that the integral (\ref{kdef}) converges
even when ${k}$ becomes a complex number. It will be constructed
from Lefschetz thimbles in the usual way,
\begin{equation}\label{orm}\CC=\sum_\sigma\n_\sigma\J_\sigma.\end{equation}
Here the $p_\sigma$ correspond to flat connections $\A_\sigma$ with structure
group $H_\C$,
each with a chosen lift to $\hat\U_\C$.  (The sum in (\ref{orm})
will generally be a finite sum, but see the end of section
\ref{sc} for an exception.) The analytically continued path
integral will therefore be
\begin{equation}\label{toldyou}
Z_{H}(k)=\sum_\sigma \n_\sigma Z_{H,\sigma}(k)\end{equation}
with
\begin{equation}\label{oldyou}
Z_{H,\sigma}(k)=\int_{\J_\sigma} D\A\,\exp(i{k}W(\A))
.\end{equation}

A finite sum (\ref{orm}) is not invariant under the deck
transformations of the cover $\hat\U_\C\to\hat\U$, so it is not
possible to use those deck transformations to show that
$Z_{H,\CC}(K,k)=0$ for ${k}\notin \Z$.  (Invariance under deck
transformations would require including every lift of a given
$\A_\sigma$.) As a cycle in the appropriate homology, $\CC$ will
be locally constant in $k$ (Stokes jumping of cycles $\J_\sigma$
is compensated by jumping of coefficients $\n_\sigma$), ensuring
that $Z_{H,\CC}(K,k)$ will be holomorphic in $k$.

\subsubsection{Downward Flow Again}\label{moredown}

To get an analytic continuation of the original Chern-Simons
integral (\ref{zdef}), $\CC$ should obey one more condition: it
should be chosen so that $Z_{H,\CC}(k)=Z_H(k)$ for integer ${k}$.

In a finite dimensional problem of this type, a suitable $\CC$ can
be found as follows. (For a prototype, see the review in section
\ref{anak} of the analytic continuation of the Bessel function.)
First, lift the real cycle $\U\subset\U_\C$ to a non-closed chain
$\CC_0\subset \hat\U_\C$ as follows.   Pick a real number $\alpha$
and define a point in $\CC_0$ to be
 a real connection $A$ whose Chern-Simons function
 takes values in the closed interval $[\alpha,\alpha+2\pi]$. (We are using
 the fact that on $\hat\U_\C$, the Chern-Simons function is an ordinary  complex-valued function.)  We have to use a closed interval
here or $\CC_0$ would not be closed.   Though closed, $\CC_0$ is
not a cycle because it has a non-zero boundary $\partial \CC_0$
-- consisting of real connections with Chern-Simons function
$\alpha$ or $\alpha+2\pi$.  Of course, $\partial \CC_0$ is a union of two disjoint pieces, which we will loosely call components (we do not need to know if they are connected), corresponding to those two values of the Chern-Simons function.

We want to complete $\CC_0$ to a candidate integration cycle
$\CC\subset\hat\U_\C$.   For this, we add to $\CC_0$ an additional
piece $\CC_1$ defined as follows. We define $\CC_1$ by downward
Morse flow from the boundary $\partial\CC_0$, with respect to the
usual Morse function $h={\rm{Re}}\,(ikW(\A))$. Assume first that
downward flow from $\partial\CC_0$ leads to no critical points.
Then $\CC_1$ is closed and, if we orient correctly its two
components, then $\CC=\CC_0\cup\CC_1$ has no boundary and is the
integration cycle we want.   (In the prototype of fig.
\ref{Contour} in section \ref{anak}, the two components of $\CC_1$
correspond to the two half-lines that start at $w=0$ and $w=2\pi
i$.)

$\CC_1$ fails to be closed only if a downward flow from
$\partial\CC_0$ approaches a critical point for $t\to+\infty$.
(This critical point is then in the closure of $\CC_1$, but not in
$\CC_1$ itself.)  This possibility is severely limited because the
downward flow equations have a conserved quantity, which for
Chern-Simons theory is ${\rm{Im}}\,(ikW(\A))$, while in addition,
${\rm{Re}}(ikW(\A))$ decreases along a flow. So $\CC_1$ may fail
to be closed only if there is a critical point at which $W(\A)$
has positive imaginary part and real part $\alpha$ or
$\alpha+2\pi$. For a generic choice of $\alpha$, there is no such
critical point and the above definition works and gives a suitable
cycle $\CC$.

Now consider the integral
\begin{equation}\label{pokx}\int_\CC DA\exp(ikW(\A))=\int_{\CC_0}DA\,\exp(ikW(\A))+\int_{\CC_1}DA\,\exp(ikW(\A)).
\end{equation}
Here $\CC_1$ consists of two components (downward flow from real
connections with $W(\A)$ equal to $\alpha$ or $\alpha+2\pi$) that
differ from each other by a deck transformation that shifts $W(A)$
by $2\pi$.  For $k\in\Z$, the integrand $\exp(ikW(\A))$ is
invariant under this deck transformation, but on the other hand,
the two components of $\CC_1$ are oppositely oriented (that is,
the deck transformation maps the orientation of one to minus the
orientation of the other).  So for $k\in \Z$, we have
$\int_{\CC_1}DA\,\exp(ikW(\A))=0$ and $\int_\CC
DA\exp(ikW(\A))=\int_{\CC_0}DA\,\exp(ikW(\A))$.  But for $k\in\Z$,
the function $\exp(ikW(\A))$ is a pullback from the original space
$\U_\C$ of complex-valued connections modulo gauge
transformations, and the cycle $\CC_0$ pushes down in $\U_\C$ to
the original real integration cycle $\U$. In short, $\int_\CC
DA\exp(ikW(\A))$ is equivalent to the original Chern-Simons path
integral when $k$ is an integer.  We have accomplished our goal of expressing  the 
Chern-Simons path integral $Z_H(k)$ with a compact
gauge group  in a form that makes sense when $k$ is not an integer.

$\CC$ is an unfamiliar sort of integration cycle for a path
integral, but actually, it is straightforward in concept to use
(\ref{pincom}) to re-express $\CC$  in terms of Lefschetz thimbles.  We just count the upward flows from a given critical
point $\A_\sigma$ to $\CC$.  If  $k$ is real, the flow
equations conserve ${\rm{Im}}\,(iW(\A))$, so,  given the condition that
we have imposed on $\alpha$, $\n_\sigma$ is nonzero only for
critical points for which ${\rm{Im}}\,(iW(\A))$ takes values in
the open interval $(\alpha,\alpha+2\pi)$.

\subsubsection{Indeterminacy of $\CC$}\label{gench}

The only problem with the analytic continuation that we have obtained for the Chern-Simons
path integral is that
it is not unique.

 The choice of $\CC$ and therefore the analytic
continuation of the Chern-Simons path integral $Z_H(k)$ will in
general jump in crossing the exceptional values of $\alpha$ at
which $\CC_1$ fails to be closed. This is related to a very basic
fact.  The contribution of any given critical point $p_\sigma$ to
the path integral can be multiplied by $\exp(2\pi i m k)$, for any
integer $m$, without changing its value for $k\in\Z$. This
multiplicative factor has a simple interpretation: it results from
shifting the Chern-Simons invariant of the flat connection
$\A_\sigma$ by $2\pi m$.   So, without more input, we cannot
expect analytic continuation in $k$ to be unique.

In a situation (such as that of section \ref{anajones})  in which
the Chern-Simons path integral depends on another parameter
$\gamma$ in addition to $k$, it may not be possible to make our
almost canonical choice of $\CC$
 uniformly in $\gamma$.  The reason is that it may not be possible to choose $\alpha$
 so that it varies continuously with $\gamma$ and never coincides with the value of
 ${\rm{Im}}\,(iW(\A))$ for any flat connection $\A$.  If we choose $\CC$ as above
 for some range of $\gamma$, then upon varying $\gamma$ and letting $\CC$ evolve continuously,
we will arrive at a more general  $\CC$ and hence a more general
form of the analytic continuation.

A rather general $\CC$ would entail multiplying the contribution
to the path integral from each
 representation of the fundamental
group by a finite sum
\begin{equation}\label{finsum}\sum_{m\in\Z}a_{m}\exp(2\pi i mk),
\end{equation}
where the integers $a_m$ are almost all zero and $\sum_m a_m=1$.
In general, the $a_m$ may be different for different
representations of the fundamental group.  A function such as this will naturally
arise in an example in section \ref{asjones}.

\subsubsection{Knots}\label{spurz}

There is an important case in which there is a completely natural choice of
the analytic continuation.
This is the case or knots or links in $S^3$.  As is known from any of their standard
characterizations, the usual Wilson loop invariants for knots or links
 can be analytically continued to a Laurent  polynomial
 in $q=\exp(2\pi i/(k+h))$ and with this analytic continuation, they have no essential singularity
at $k=\infty$.  Since the function $\exp(2\pi i k)$ does have an
essential singularity at $k=\infty$, the cycle $\CC$ used in
analytically continuing the knot invariants is unique if one wants
the analytically continued function to be free of an essential
singularity at infinity.

Perhaps we should spell out a detail concerning the last
paragraph.   The Jones polynomial and its generalizations are
usually normalized to equal 1
 for the unknot. From the point of view of quantum field theory, this
 definition amounts to defining the Jones polynomial of a given knot $K$ by analytic continuation of
 a ratio $Z_H(K;k)/Z_H(K_0;k)$, where $Z_H(K;k)$ was defined in  (\ref{mydef}) and $K_0$ is the unknot.
 The ratio is certainly 1 if $K=K_0$.
 It is such ratios that analytically continue to Laurent
 polynomials in $q$.  The numerator and denominator, when taken separately, have an analytic
 continuations in $k$ with no essential singularity at $k=\infty$,
 but they are not Laurent polynomials in $q$.  For example, for $H=SU(2)$,
 the denominator is $Z_H(K_0;k)=\sqrt{2/(k+2)}\sin(\pi/(k+2))$,
 which is not a Laurent polynomial in $q$, but has no essential
 singularity at $k=\infty$.
The theory developed in the present paper applies more naturally
to a path integral than to a ratio of path integrals, and
therefore we will simply study the functions $Z_H(K;k)$ rather
than ratios of such functions.

\subsection{Comparing Compact And Complex Lie Groups}\label{relation}

\def\V{\mathcal V}
If we do not require that the generalized integral
$Z_{H,\CC}({k})$ should agree with the original Chern-Simons
integral $Z_H({k})$ when $k$ is an integer, we have much more
freedom in the choice of the cycle $\CC$.  We can take $\CC$ to be
any integer linear combination of Lefschetz thimbles
\begin{equation}\label{zonfo}\CC=\sum_{\sigma\in\Sigma}\n_\sigma
\J_\sigma,\end{equation} with coefficients $\n_\sigma$ that jump
in crossing Stokes lines to ensure that the homology cycle $\CC$
varies smoothly.

The functions $Z_{H,\CC}(k)$ of eqn. (\ref{kdef}), as $\CC$ runs
over the possible choices, form a vector space $\V$. (This space
is actually infinite-dimensional when one allows for the choices
of lifts to $\C$ of Chern-Simons invariants of critical points.)
What are these functions good for, apart from analytic
continuation of Chern-Simons theory with gauge group $H$?  They
play a role somewhat analogous to conformal blocks in
two-dimensional conformal field theory. Let us compare
Chern-Simons theory of the compact group $H$ to Chern-Simons
theory of the complex Lie group $G=H_\C$.  For reasons that will
become clear, we consider first the case of a three-manifold $M$
without knots.

As we have explained, the analytic continuation of the complex
Chern-Simons theory as a function of its coupling parameters is
made using the same machinery as for a compact Lie group.
However, the complexification of a complex Lie group $G$ is
isomorphic to the product $G\times G$, and accordingly, the
analytic continuation of Chern-Simons theory of gauge group $G$ is
described by a pair of complex connections $\A,\tilde \A$.  This
assertion was part of the derivation of  eqn. (\ref{morko}), where
we also showed that the analytically continued action is the sum
of a Chern-Simons action for $\A$ and a Chern-Simons action for
$\tilde\A$.  We can thus build an integration cycle for the
analytically continued $G$-valued theory by taking the product of
two integration cycles for  the analytically continued $H$-valued
theory.  Concretely, we can consider an integration cycle for the
analytically continued $G$-valued theory of the form
$\J_\sigma\times\J_\tau$, where $\J_\sigma$ and $\J_\tau$ are
Lefschetz thimbles for the analytically continued $H$-valued
theory.  This gives a basis for the space of appropriate cycles,
and hence the partition function $Z_G(s,\ell)$ of the $G$-valued
theory can be expanded as a bilinear form in the functions
$Z_{H,\sigma}(k)$ of eqn. (\ref{oldyou}):
\begin{equation}\label{obonfo}Z_G(s,\ell)=\sum_{\sigma,\tau}\n_{\sigma,\tau}Z_{H,\sigma}(k_1)
Z_{H,\tau}(k_2).
\end{equation}
The coefficients $\n_{\sigma,\tau}$ jump in the usual way in
crossing Stokes curves.

From the formulas (\ref{ork}) and (\ref{ossy}) for the analytically continued
action, one might expect the coupling
parameters to be $k_1=t/2=(\ell+is)/2$, $k_2=\tilde
t/2=(\ell-is)/2$.  Actually there is a shift by the dual Coxeter
number $h$ of $H$ and one has
\begin{equation}\label{ozorm}k_1=t/2-h,~~k_2=\tilde
t/2-h.\end{equation} This shift enters in comparing the
integration measures $D\A$ and $DA$ used in defining $Z_G$ and
$Z_H$, or, differently put,  it reflects the fact that in
Chern-Simons theory with a compact gauge group $H$, a one-loop
correction shifts the effective value of $k$ in many formulas to
$k+h$, while as shown in \cite{DW}, there is no such shift for a
complex Lie group.  For more on this, see section
\ref{littlemore}.

\subsubsection{Knots Again}\label{knots}

It is instructive to consider what happens to this formalism if we
specialize to knots in $S^3$, again in the theory with gauge group
$G=H_\C$.  In the presence of a knot $K$, we must include in the
integrand of the path integral an extra factor that is the
holonomy of the connection $\A$ around $K$, in some representation
$R$.  What happens depends very much on what sort of
representation we pick.  For example, one might pick an irreducible unitary
representation of $G$; these of course are infinite-dimensional.
In this case, the simplification that will be explained momentarily does not arise.

We will consider here  a
finite-dimensional and irreducible but non-unitary representation that is related to an
irreducible representation of the compact group $H$.
  The action of $H$ on an
irreducible representation $R$ can be extended to an action of $G=H_\C$ in two ways; in one
case the matrix elements of the representations are holomorphic functions on $G$ and
in the other case they are antiholomorphic.  We will consider the case of
a holomorphic representation.  Given a knot $K$ and a $G$-valued
connection $\A$, holomorphy means after analytic continuation that the trace of the holonomy around $K$ is a function only of $\A$ and not
of $\tilde \A$; as usual we denote this trace as $\Tr_RP\exp\left(-\oint_K\A\right)$.  Now consider
the  path integral:
\begin{equation}\label{onk}Z_G(K;t,\tilde t)=\int_\CC D\A\,D\tilde \A \,\exp\left(itW(\A)/2+i\tilde t W(\tilde \A)/2\right)\,
     \Tr_R P\exp\left(-\oint_K\A\right).\end{equation}

We want to express this integral in terms of Lefschetz thimbles.
So consider the equations for a critical point. (We will here
proceed somewhat informally; the details of how to include the
holonomy factor in the Morse theory framework are explained in
section \ref{placeknots}.) Since the holonomy factor only depends
on $\A$, the equation for $\tilde \A$ is unaffected by the
presence of the knot. This equation says simply that the curvature
of $\tilde \A$ vanishes, and as $S^3$ is simply-connected, this
means that up to a gauge transformation we can take $\tilde\A=0$.

\def\Q{\mathcal Q}
Consequently, there is essentially only one possible integration
cycle  for $\tilde\A$; it is the Lefschetz thimble $\tilde \Q$
associated to the critical point at $\tilde\A=0$.  Hence, the
integration cycle $\CC$ is simply the product of $\tilde \Q$ with
some integration cycle $\Q$ for $\A$.  Since the integrand of the
path integral is also a product, it follows that the path integral
is the product of an integral over $\A$ and an integral over
$\tilde \A$ (rather than a sum of such products, as in the more
typical case (\ref{obonfo})).  Moreover, the integral over
$\tilde\A$ does not depend on the knot $K$.  So the general form
of $Z_G(K;t,\tilde t\,)$ is
\begin{equation}\label{rufo}Z_G(K;t,\tilde t\,)=\tilde B(\tilde t\,) B(K;t),\end{equation}
where only the second factor depends on the knot.

What will be $B(K,t)$?  Rather like the knot invariant $Z_H(K;k)$ for the compact group $H$, $B(K,t)$
is defined by an integral of the form
\begin{equation}\label{onkey} \int_{\Q} D\A\,\exp(i (t/2) W(\A))\,\mathrm{Tr}_R
P\exp\left(-\oint_K\A\right)\end{equation}
for some integration cycle $\Q\subset\hat\U_\C$.  What integration
cycle shall we use?

In perturbation theory,
$Z_G(K;t,\tilde t\,)$ is  equal to $Z_H(K;k)$ (with $k=t/2-h$, and apart from
a factor $\tilde B$ that depends only on $\tilde t$ and not on $K$).  This relation was presented as exercise 6.32 in \cite{barnatan}.  As a perturbative statement, it simply reflects the fact that we started with a holomorphic
representation whose holonomy depends on $\A$ and not $\tilde\A$.   When we compute
the expectation value of $\Tr_R\,P\exp\left(-\oint_K\A\right)$ in perturbation theory, it does not matter
 if $\tilde \A$ is absent
(the case of a compact gauge group $H$) or present but decoupled from $\A$ (the case
of the complex gauge group $G$, after analytic continuation to decouple $\bar\A$ from
$\A$).

This relation between the $G$ and $H$ theories will hold as an
exact statement, not just an asymptotic one, if we simply take
$\Q$, the integration cycle used in constructing the theory with
complex gauge group $G$, to be the same integration cycle that was
used in analytically continuing the theory with compact gauge
group $H$. This is a satisfying result, but it is perhaps best
understood as a definition.  The original path integral
(\ref{onk}) is not convergent, even for the usual real values of
the coupling parameters $s$ and $\ell$, if regarded as an integral
over the usual real cycle $\tilde\A=\bar\A$.  The reason for this
is that the holonomy factor $\Tr_RP\exp\left(-\oint_K\A\right)$ is
exponentially growing in some directions. So the choice of $\Q$
that reduces the knot invariants for $G$ to those of $H$  can be
regarded as a definition of the knot invariants for a holomorphic
representation of $G$, rather than something we deduce from a more
primitive framework. By contrast, in the case of a unitary
representation, the holonomy can be expressed as an oscillatory
integral by adding more integration variables; this is explained
in section \ref{placeknots}.  After doing so, the integral over
the usual real cycle makes sense.

\subsection{Other Real Forms}

Let $H'$ be a noncompact real form of the complex Lie group $G=H_\C$.  (For example,
if $H=SU(2)$, $H'$ might be $SL(2,\R)$.) We sometimes describe this more briefly
by calling $H'$ a noncompact real form of $H$.   Chern-Simons theory with gauge group
$H'$ is formally defined by the familiar sort of integral
\begin{equation}\label{formally}Z_{H'}(k')=\int _{\U'}DA'\,\exp(i {k'} W(A')),\end{equation}
where now $A'$ is a connection on an $H'$ bundle $E'\to M$ and $\U'$ is the
space of such connections, up to gauge transformation.

Since the complexification of $H'$ is simply  $G=H_\C$,
the complexification of $\U'$ is the familiar space $\U_\C$ of complex-valued connections.
Consequently, $\U'$ can be viewed as an integration cycle in $\U_\C$ on which the usual
Chern-Simons integral is an (oscillatory) convergent integral.  In this respect,
 it is no different from $\U$,
the space of connections modulo gauge transformations for the
compact Lie group $H$.   Either $\U$ or $\U'$ can be regarded as a 
possible integration cycle in $\U_\C$ for the Chern-Simons path integral.   ($\U$ and $\U'$ are both fixed points of real involutions acting on $\U_\C$, as
explained more fully in section \ref{littlemore}.) As such, each can be expressed as a linear combination of
the usual Lefschetz thimbles $\J_\sigma$. 

Hence, the framework for analytic continuation of $Z_{H'}(k')$ is
the same as the framework for analytic continuation of $Z_H(k)$.
We seek to expand it as
 a linear combination with integer coefficients of the functions  $Z_{H,\sigma}(k)$   of eqn. (\ref{oldyou}):
\begin{equation}\label{otonk}Z_{H'}(k')=\sum_\sigma \n'_\sigma Z_{H,\sigma}(k).\end{equation}
(The coefficients $\n'_\sigma$ will have to jump in the usual way
in crossing Stokes lines.) However, there is a subtlety in the
relation between the parameters in $Z_{H'}(k')$ and $Z_H(k)$.
This relation is not $k'=k$   but
\begin{equation}\label{zormoz}k'+h'=k+h,\end{equation} where $h'$
is a constant\footnote{Decompose the Lie algebra of $H'$ as $\frak
h'=\mathbf{k} \oplus \frak p$, where $\mathbf{k}$ is the Lie
algebra of a maximal compact subgroup of $H'$, and $\frak p$ is
its orthocomplement.   After defining the dual Coxeter number $h$
as the trace in $\frak h'$ of the square of a suitable element of
$\frak h'$, write $h=h_++h_-$, where $h_+$ and $h_-$ come from
traces in $\mathbf{k}$ and $\frak p$, respectively.  Then
$h'=h_+-h_-$.} that has been computed in different ways
\cite{DW,GW}.  Like (\ref{ozorm}), which can actually be viewed as
a special case, (\ref{zormoz}) arises from a comparison between
the complexification of the integration measures used in defining
$Z_H$ and $Z_{H'}$.  It is further discussed in section
\ref{littlemore}.

Just as in the discussion of analytic continuation of $Z_H(k)$, we
wish to pick a linear combination of critical point cycles so that
the formula (\ref{otonk}) will agree with $Z_{H'}(k')$ for
$k'\in\Z$.    The general analysis of this is similar to what it
is for the case of compact gauge group, but again there is some
simplification for the case of a knot in $S^3$. If we label our
knot by  a finite-dimensional irreducible representation of $H'$
(as opposed to a unitary representation of $H'$, which would be
infinite-dimensional), then it is natural to  pick the same
integration cycle that we used in analytically continuing
$Z_H(K,k)$. With the identification of parameters $k'+h'=k+h$,
this leads to $Z_{H'}(K,k')=Z_H(K,k)$. That relationship is
already known in perturbation theory, which is insensitive to the
difference between $H$ and $H'$ (except for a one-loop shift in
the parameters \cite{DW}).   That fact motivates our choice of
cycle.

Our discussion in section \ref{knots} can be viewed as a special case
of this, since $G=H_\C$ is a noncompact real form of the compact
Lie group $H\times H$.  The properties of Chern-Simons theory with compact
gauge group that we have used do not depend on whether the gauge group is simple, so they
hold for $H\times H$.

\subsubsection{Comparison Of Holomorphic Volume Forms}\label{littlemore}

We will now briefly provide a context for eqn. (\ref{zormoz}).

The complex Lie group $G=H_\C$ admits two real involutions that are important
here: one leaves fixed $H$ and one leaves fixed $H'$.  Similarly, writing $\frak h_\C$,
$\frak h$, and $\frak h'$ for the Lie algebras of $H_\C$, $H$, and $H'$, the space
$\U_\C$ of $\frak h_\C$-valued connections modulo gauge transformation admits
two relevant real involutions: one leaves fixed the space $\U$ of $\frak h$-valued connections
modulo gauge transformations, and one leaves fixed the space $\U'$ of $\frak h'$-valued
connections modulo gauge transformations.

Accordingly, the complexification of either $\U$ or $\U'$ gives the same space $\U_\C$.
Now let $DA$ and $DA'$ be the path integral measures for the theories with gauge
groups $H$ and $H'$, respectively.  Either one of these can be analytically continued
to a holomorphic volume form on $\U_\C$; we denote these holomorphic forms
simply as  $DA$ and $DA'$.   Our discussion of analytic continuation has concerned the
integration of these volume forms over various real cycles in $\U_\C$ (or more precisely
the integration of their pullbacks to the universal cover $\hat\U_\C$ over cycles in $\hat\U_\C$).

 $DA $ and $DA'$ are holomorphic volume forms on the same space, so their
 ratio is an invertible holomorphic function of the complexified connection $\A$.  The relationship
 is
 \begin{equation}\label{otzox}DA=DA'\,\exp(i(h-h')W(\A)).  \end{equation}
 This relationship is computed in \cite{DW} by comparing $\eta$-invariants for the
 theories with gauge group $H$ and $H'$.  A different language is used there and in most
 of the literature.  In this language, a one-loop correction (coming from the $\eta$-invariant)
 shifts the effective Chern-Simons coupling of the $H$ theory by $h$ and that of the $H'$ theory by $h'$.  The ratio of the two statements is (\ref{otzox}).

 The relation (\ref{zormoz}) is an immediate consequence of
 (\ref{otzox}).  Similarly (\ref{ozorm}) reflects the fact that the
 ratio between the holomorphic volume forms on $\U_\C$ that we
 have called $DA$ and $D\A$ is
 \begin{equation}\label{ofrox}DA=D\A\,\exp(ih W(\A)).  \end{equation}
 Alternatively, (\ref{ozorm})
 is the special case of (\ref{zormoz}) for $G=H_\C$ regarded as a noncompact real form
 of the compact Lie group $H\times H$.

 A finite-dimensional analog of the problem of comparing such
 holomorphic volume forms would involve a complex manifold $X$
 with two different real structures. For example, $X$ could be the
 complex $x$-plane, with two real structures determined by the
 involutions $x\to \bar x$ and $x\to -\bar x$.  If we set
 $x=u+iv$,
  the fixed set of the first involution is the $u$ axis $\U$, with
  real volume form $\d u$, and the fixed point of the second is
  the $v$ axis $\U'$, with real volume form $\d v$.  On the other
  hand, on $X$ we have the natural holomorphic one-form $\d x=\d u
  + i\d v$.  The analytic continuation of $\d u$ from $\U$ to $X$
  is $\d x$, and the analytic continuation of $\d v$ is $-i\d x$.
  So the relation between the two analytically continued volume
  forms is $\d v = - i \d u$.  In the context of the
  infinite-dimensional Chern-Simons path integral, such a comparison between
  volume forms leads to an $\eta$ invariant, as explained in
  \cite{Wi, DW}, and this $\eta$ invariant can be expressed in terms of the
  Chern-Simons function.

\section{Finite-Dimensional Oscillatory Integrals, Morse Theory, And Steepest
Descent}\label{stokes}

The goal of the present section is to describe in more detail the
finite-dimensional case of the theory that was sketched in section \ref{overview}.

\def\V{{\mathcal V}}
\subsection{The Airy Function}

\begin{figure}
 \begin{center}
   \includegraphics[width=3.5in]{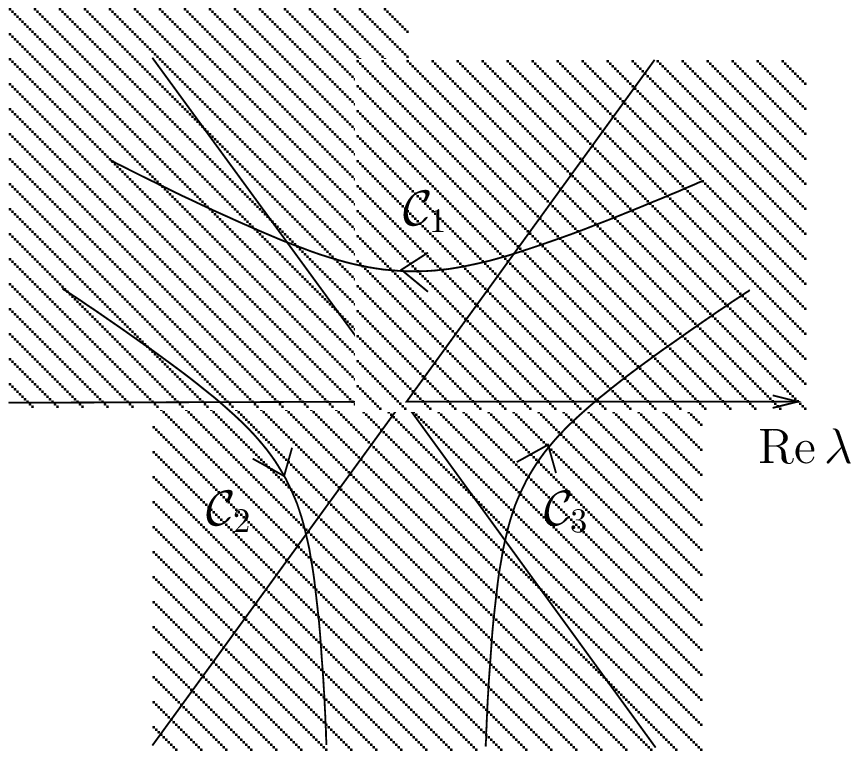}
 \end{center}
\caption{\small The complex $\lambda$ plane, with the real axis from left to right.
The hatched regions are the three ``good'' regions in which the Airy integrand vanishes at
infinity.  The contour $\CC_i$, $i=1,2,3$, connects the $i^{th}$ good region to the
$i+1^{th}$. }
 \label{Airy}
\end{figure}

\def\I{{\mathcal I}}
We begin
by examining the Airy integral, initially defined for real $\lambda$:
\begin{equation}\label{morook} Z_\lambda=\int_{-\infty}^\infty
{\d x}\exp\left(i\lambda(
x^3/3-x)\right).\end{equation}
We want to analytically continue the integral to complex $\lambda$.  This will require
deforming the integration contour away from the real $x$ axis, so we will
regard the Airy exponent
\begin{equation}\label{aexp}\I= i \lambda (x^3/3-x) \end{equation}
as a function of complex variables $x$ and $\lambda$.  For real
$\lambda$, the integration is over the real axis. As $\lambda$
varies, we will deform the integration cycle $\CC$ so as to  vary
smoothly with $\lambda$ and in such a way that the integral always
converges.

In the case of the Airy function, it is not hard to see by hand
how to do this.  We divide the $x$-plane into ``good'' regions in
which the imaginary part of $\lambda x^3$ is positive and ``bad''
regions in which it is negative.  For positive $\lambda$, this is
shown in fig. 1. The Airy integrand is exponentially small in the
good regions. The original oscillatory integral over the real $x$
axis is an integral over the boundary of two good regions.  We can
deform the contour slightly by giving $x$ a positive imaginary
part so that it will connect two good regions.  Now as we vary
$\lambda$ (keeping away from $\lambda=0$), the good regions rotate
continuously to the left or the right, and we simply rotate the
integration contour $\CC$ as $\lambda$ varies so that $\CC$ always
connects two good regions. This gives an analytic continuation of
$Z_\lambda$ as a function of $\lambda$, but it also makes it clear
that there is a monodromy around $\lambda=\infty$ (or 0).  In
fact, as $\lambda$ changes in phase by $2\pi$, the three good
regions are cyclically permuted. So $Z_\lambda$ has a monodromy of
order 3.

Because the contour $\CC$ does undergo a non-trivial monodromy as $\lambda$ varies,
it is useful to consider all possible contours in this problem.  Denote the three good
regions
as $\V_i$, $i=1,2,3$, as in the figure.
For each $i$, let $\CC_i$ be a contour connecting region $\V_i$ to
region $\V_{i+1}$.   The integrals $Z_{i,\lambda}=\int_{\CC_i}
\d x \exp(i\lambda(x^3/3-x))$ converge for all $i$ and $\lambda$, and any reasonable
contour
along which one might integrate the Airy function $\exp(i\lambda(x^3/3-x))$ (or the
product
of this function and one with slower growth at infinity) is equivalent to a linear combination of the
$\CC_i$.
However, the integrals $Z_{i,\lambda}$ are not independent.  They obey
\begin{equation}\label{polky}Z_{1,\lambda}+Z_{2,\lambda}+Z_{3,\lambda}=0. \end{equation}
This reflects the fact that the sum of contours $\CC_1+\CC_2+\CC_3$ can be deformed to
zero.

\subsubsection{Morse Theory}\label{morset}
The Airy function is simple enough that we do not really need a
more sophisticated theory. Nevertheless, we will use this simple
example as a practice case for describing a more sophisticated
approach that is useful in generalizations. To formalize the sense
in which $\CC_1+\CC_2+\CC_3=0$, it is useful to regard the $\CC_i$
as relative homology cycles.  Let $X$ denote the complex
$x$-plane, and for real $U$, let $X_U$ be the part of $X$ with
${\rm Re}(i\lambda(x^3/3-x))\leq U$. Any reasonable integration
cycle $\CC$ for the Airy integral should represent an element of
the relative homology $H_1(X,X_{-T};\Z)$ for very large $T$.  In
other words, $\CC$ is a closed cycle, and though it may not be
compact, its infinite ends should lie in $X_{-T}$.  Differently
put, ${\rm Re}(i\lambda(x^3/3-x))$ must tend to $-\infty$ at
infinity in $\CC$.

The $\CC_i$ have this property, and represent elements of the
relative homology.  Indeed, they generate this relative homology,
subject to the relation
\begin{equation}\label{koppo}\CC_1+\CC_2+\CC_3=0.\end{equation}

The fact that the $\CC_i$ generates the relative homology is
fairly obvious geometrically in this simple example.  $X$ is
contractible, but $X_{-T}$ is the union of three disjoint pieces.
A one-cycle that is non-trivial in $H_1(X,X_{-T};\Z)$ must connect
distinct pieces of $X_{-T}$.  The relation (\ref{koppo}) is also
clear geometrically.

The integration cycle  that we started with for real $\lambda$ -- the real $x$-axis, which we
will call $\CC_\R$  --  is not
quite an element of the relative homology, because at large $|x|$ it lies on the boundary of,
rather than in the interior of, the good regions of fig. (\ref{Airy}).  However, $\CC_\R$ becomes
an element of the relative homology once we displace it slightly by giving $x$ a small  imaginary
part $\epsilon$ with the same sign as $\lambda$.   This multiplies the Airy integral by
a convergence factor $\exp(-\lambda\epsilon x^2)$.  The displaced cycle
therefore represents an element of the relative homology.  And indeed since the displaced
cycle connects two of the good regions at infinity, it is equivalent to one of the $\CC_i$.

For generalizations,  we need a more powerful way to
analyze the relative homology. We consider the real part of the
Airy exponent
\begin{equation}\label{equivo} h={\mathrm {Re}}(\I)\end{equation}
as a Morse function.  A Morse function is simply a real-valued
function whose critical points are nondegenerate.  The critical
points are the points at which all first derivatives of $h$
vanish, and nondegeneracy of a critical point means that the
matrix of second derivatives is invertible at that point.  The
number of negative eigenvalues of this matrix is called the Morse
index of the critical point.

In general, because of the Cauchy-Riemann equations, the critical
points of the real part of a holomorphic function $\I$ are the
same as the critical points of $\I$.  So in this case, there are
two critical points which we call $p_\pm$, located at $x=\pm 1$.
The values of $\I$ and $h$ at $p_\pm$ are
\begin{equation}\label{tep}\I_\pm = \mp\frac{ 2i\lambda}{3}\end{equation}
and
\begin{equation}\label{beqq}h_\pm =\pm\frac{2\, {\mathrm {Im}}\lambda}{3}.\end{equation}
The critical points are nondegenerate, since
$\partial^2\I/\partial x^2\not=0$ at $x=\pm 1$. In general, in a
complex manifold of complex dimension $n$, if $h$ is a Morse
function that is the real part of a holomorphic function $\I$, its
isolated and nondegenerate critical points are all of index $n$. (The general form
of $\I$ near an isolated nondegenerate critical point is $\I=c +\sum_{i=1}^n
z_i^2$, with a constant $c$ and local complex coordinates $z_i$;
if $z_i=u_i+iv_i$ then  $h={\mathrm{Re}}\,c+\sum_{i=1}^n(u_i^2-v_i^2)$ and the
matrix of its second derivatives is diagonal with $n$ negative
eigenvalues.) So in our case, the critical points have index 1.

In general, any Morse function $h$ on a manifold $X$ can be used
to compute upper bounds on the ranks of the real homology (or
cohomology) groups of $X$: the rank of the $q$-dimensional homology of $X$ is at most
the number of critical points of $h$ of Morse index $q$.  If the differences between the
indices of distinct critical points of $h$ are never
equal\footnote{This condition ensures that in computing the
homology of $X$, there are no cancellations between critical
points of adjacent index.} to $\pm 1$, then $h$ is a perfect Morse
function and the ranks of the homology groups are equal to the upper bounds
just stated. In fact, in this case even
the integral homology can be described via the critical points.
Our Morse function is perfect, since both critical points have
index 1.

If $X$ is a compact manifold, $h$ is automatically bounded above
and below, and the critical points of a perfect Morse function
determine the ordinary homology of $X$.   In our case, $X$ is not
compact, and we are dealing with a Morse function $h$ that is
unbounded above and below.  In such a case, the critical points of
$h$ determine relative homology groups, namely the homology groups
$H_k(X,X_{-T})$, where $T$ is  a large constant.  In our example,
$h$ is a Morse function with two critical points both of index 1,
so $H_1(X,X_{-T})$ is of rank 2, and the other relative homology
groups vanish.  (Without using Morse theory, we would observe that
$H_0(X,X_{-T})=0$ since any zero-cycle is a sum of points, which
can be deformed into $X_{-T}$. And $H_2(X,X_{-T})=0$ since $X$ has
non-compact ends even after collapsing $X_{-T}$ to a point.)

For a generic perfect Morse function $h$, Morse theory also gives
a recipe for constructing a relative one-cycle $\J_\sigma$
attached to each critical point, and representing an element of
the appropriate relative homology group.  Moreover, these cycles
generate the relative homology with integer coefficients.

In general, on any manifold $X$ with real coordinates $u^i$, pick
a Riemannian metric $g_{ij}$ and consider the downward flow or gradient flow
equations
\begin{equation} \label{downflow} \frac{\d u^i}{\d t}=-g^{ij}\frac{\partial h}{\partial
u^j}.\end{equation}
These are called downward flow equations, because except for a
trivial solution that sits at a critical point for all $t$, $h$ is
always strictly decreasing along a flow:
\begin{equation}\label{ownflow}\frac{\d h}{\d t}=\sum_i
\frac{\partial h}{\partial u^i}\frac{\d u^i}{\d t}=-\sum_i\left(\frac{\partial
h}{\partial u^i}\right)^2.
\end{equation}
An important property of the flow equation is that if $u^i(t)$
equals a critical point at some $t$, then the flow equation
implies that $u^i(t)$ is constant for all $t$.  So a nonconstant
flow can only reach a critical point at $t=\pm \infty$.

Let $p$ be a nondegenerate critical point of $h$, and consider the downward flow
equations on the half-line $(-\infty,0]$ with the boundary
condition that $u^i(t)$ approaches $p$ for $t\to-\infty$. If $p$
has index $k$, the moduli space $\mathcal J_p$ of such solutions
is a $k$-dimensional manifold, since there are $k$ independent
directions of downward flow from the critical point $p$.  We think
of $\J_p$ as a submanifold of $X$, by mapping a downward flow line
$u^i(t)$ to the corresponding point $u^i(0)\in X$.  (This gives an
embedding of $\J_p$ in $X$ since, as the downward flow equation is
first order in time,  a  flow is uniquely determined by its value
at $t=0$.) An equivalent definition of $\J_p$  is that it is the
submanifold of  $X$ consisting of points
that can be reached at any $t$ by a flow  that starts at $p$ at $t=-\infty$.  Given a flow line
$u^i(t)$ that reaches a point $x\in X$ at $t=t^*$, the flow line
$u^i(t-t^*)$ arrives at $x$ at $t=0$.  A flow line defined on the
full line  $(-\infty,\infty)$ will be called a complete flow line.

Once we pick an orientation of $\J_p$, $\J_p$ will define a cycle\footnote{Our definition of $\J_p$
is slightly different from the definition usually used in Morse theory on a compact
manifold $M$.  In such a case, unless $p$ is a local minimum of $h$, there are always
complete flow lines starting at $p$ and ending at some ``lower'' critical point, and one
takes the closure of $\J_p$ to define a Morse theory cycle.   In the situation we consider,
as is explained shortly,
there generically are no complete flow lines between critical points, and it is most
convenient to define $\J_p$ as we have done.}
 in the relative homology
$H_1(X,X_{-T})$ if it is closed,
meaning that any sequence of points in $\J_p$ has a subsequence that either
converges or tends to $h=-\infty$. This fails precisely if there
is a complete flow line $l$ that starts at $p$ at $t=-\infty$
and ends at another critical point $q$ at $t=+\infty$.  In that
case, $\J_p$ is not closed,
since $l$ is contained in $\J_p$, but a sequence of points in $l$ can converge to $q$, which is
not contained  in $\J_p$.

We will determine shortly a necessary condition for a flow line to
connect two distinct critical points $p$ and $q$.  When this
condition is not satisfied,
$\mathcal J_p$ defines an element of the relative homology
$H_k(X,X_{-T})$.   $\J_p$ is known as a Lefschetz thimble.

\def\im{{\mathrm{Im}}}
\def\re{{\mathrm{Re}}}
\def\J{{\mathcal J}}
In our problem, the Morse function is the real part of a holomorphic function $\I$.
In such a case, if we pick a Kahler metric $\d s^2=|\d x|^2$, the flow equation can be
written
\begin{equation}\label{circ}\frac{\d x}{\d t}=-{\frac{\partial\overline
\I}{\partial\overline x}},~~
\frac{\d\bar x}{\d t}=-\frac{\partial \I}{\partial x}.\end{equation}
Using the chain rule and (\ref{circ}), we find
\begin{equation}\label{irc}\frac{\d\,{\mathrm{Im}}\,\I}{\d
t}=\frac{1}{2i}\frac{\d(\I-\bar\I)}{\d t}
=\frac{1}{2i}\left(\frac{\partial\I}{\partial x}\frac{\d x}{\d
t}-{\frac{\partial \bar\I}{\partial\bar  x}}{ \frac{\d \bar x}{\d
t}}\right) = 0.\end{equation} So $\im \,\I$ is conserved along a
flow.

A more conceptual explanation of the existence of this conserved quantity is as follows. 
Associated to the Kahler metric $|\d x|^2$ is the Kahler form $\omega=-i\d x\wedge \d \bar x$.
Viewing $\omega$ as a symplectic form on the $x$-plane, the gradient flow with respect to 
the Morse function $\mathrm{Re}\,\I$ is the same as the Hamiltonian flow with
Hamiltonian $\mathrm{Im}\,\I$.  Naturally, $\mathrm{Im}\,\I$ is conserved in this flow.

From (\ref{tep}), we see that in the Airy case, the values of $\im
\,\I$ at the critical points are $\im\,\I_\pm=\mp
(2/3)\re\,\lambda$.  These are equal if and only if $\lambda$ is
purely imaginary.   Hence a flow can connect the two critical
points at $x=\pm 1$ only if $\lambda$ is imaginary.   Conversely,
if $\lambda$ is imaginary, there is indeed a flow that connects the two
critical points  (the flow line is simply the part of the real
axis with $-1<x<1$). Since we  always assume that $\lambda\not=0$,
the locus of imaginary $\lambda$ is not connected but consists of
two open rays, differing by the sign of $\im \,\lambda$. We refer
to these as Stokes rays.

\subsubsection{Crossing A Stokes Ray}\label{crossing}
Away from the Stokes rays, a downward flow that starts at one critical point cannot end
at the other;
instead it always flows to $h=-\infty$.  So away from the Stokes rays, the downward flow
procedure attaches to each critical point $p_+$ or $p_-$ a relative homology one-cycle
$\J_\pm$.  Let us see what happens to the Lefschetz thimbles $\J_\pm$ in crossing a Stokes ray.

\begin{figure}
 \begin{center}
   \includegraphics[width=5.5in]{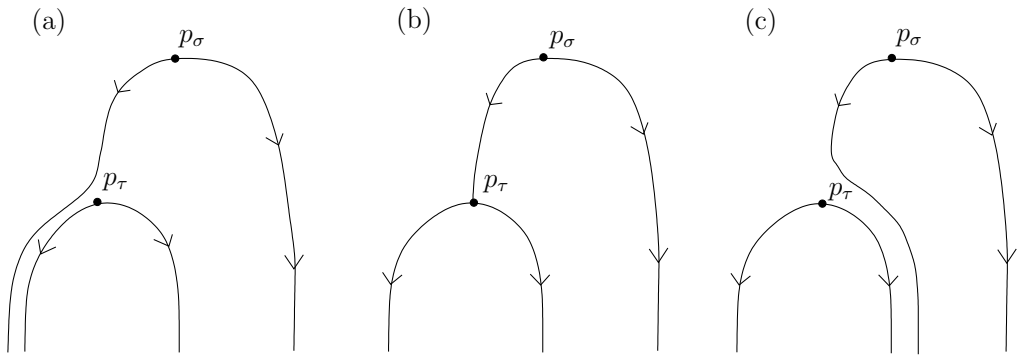}
 \end{center}
\caption{\small This figure illustrates the behavior of flow lines when one crosses
a Stokes ray.  Sketched are the downward flows from two critical points $p_\sigma$ and $p_\tau$.
The lines are sketched as flowing downward, in the direction of smaller $h$ (though it is  not
literally true that $h$ corresponds to the height in the figure, as this function has no critical points).
The behavior at a Stokes ray is depicted in (b); there is a downward flow from $p_\sigma$ to $p_\tau$.
The flows ``before'' and ``after'' crossing the Stokes ray are depicted in (a) and (c). The downward
flowing cycle $\J_\tau$ from the lower critical point is unaffected by the Stokes ray.  As for the
 cycle $\J_\sigma$ defined by flow from the upper critical point, it is ill-defined in (b) and jumps
 by $\J_\sigma\to\J_\sigma+\J_\tau$ between (a) and (c).  }
 \label{stokesray}
\end{figure}
From (\ref{tep}),  $\I_+$ and $\I_-$ are always unequal (since we
always assume that $\lambda\not=0$). On  a Stokes ray $\im
\,\I_+=\im \,\I_-$, so necessarily $\re\,\I_+\not=\re\,\I_-$, that
is $h_+\not= h_-$.  In fact, $h_+>h_-$ if $\im\,\lambda>0$, and
otherwise $h_->h_+$. As we approach a Stokes ray, let us write $q$
and $q'$ for the critical points with the larger and smaller
values of $h$, and $\J$ and $\J'$ for the corresponding
one-cycles.  Even when we reach and cross the Stokes ray, nothing
happens to $\J'$, because downward trajectories that start at $q'$
can only flow to $h=-\infty$.  But a trajectory that starts at $q$
may flow down to $q'$. (As we noted above, in the case of the Airy
function, there is such a trajectory, namely the segment of the
real axis connecting $q$ and $q'$.) When there is such a flow
line,  the definition of the cycle $\J$ attached to $q$ fails.
Since $\J$ is potentially ill-defined when $\lambda$ is on a
Stokes ray, it might jump in crossing a Stokes ray, and that is
actually what happens. How this happens is sketched  in Fig.
\ref{stokesray}.   The jump is $\J\to \J\pm \J'$, where the sign
depends on the orientations of the two cycles and the direction in
which $\lambda$ crosses the Stokes ray. The change in the pair
$\J$, $\J'$ can thus be written
\begin{equation}\label{forkey} \begin{pmatrix}\J\\
\J'\end{pmatrix}\to \begin{pmatrix} 1 & \pm 1 \\ 0 & 1\end{pmatrix}\begin{pmatrix}\J \\
\J'\end{pmatrix}.\end{equation} In section \ref{sumseries}, we
explain in another way why only $\J$ and not $\J'$ changes in
crossing the Stokes ray.

Let us compute the monodromy of the cycles $\J_+$ and $\J_-$
around $\lambda=\infty$ (or 0). In crossing the Stokes ray with
$\im\,\lambda>0$, the monodromy is strictly upper triangular:
\begin{equation}\label{poko} U_+=\begin{pmatrix} 1 & 1 \\ 0 & 1
\end{pmatrix}.\end{equation}
On the Stokes ray with $\im\,\lambda<0$, the monodromy is strictly
lower triangular:
\begin{equation}\label{noko} U_-=\begin{pmatrix} 1 & 0 \\ -1 & 1
\end{pmatrix}.\end{equation}
(If the cycles $\J_\pm$ are oriented so that the upper right
matrix element in (\ref{poko}) is $+1$, then the lower left matrix
element in (\ref{noko}) is $-1$.) The monodromy around
$\lambda=\infty$ is not simply the product $U_-U_+$, as there is
also a minus sign involving the orientation of the cycles
$\J_\pm$.  Indeed, the function $\I$ can be approximated near
$x=\pm 1$ by $\mp 2i\lambda/3+i\lambda(x\mp 1)^2$, and an
approximation to the downward flowing cycles $\J_\pm$ is given by
$x=\pm 1+ t(-i\lambda)^{-1/2}$, with real $t$.  To orient the
cycles, we need to pick a sign of $(-i\lambda)^{-1/2}$, and under
monodromy around $\lambda=\infty$, this sign is reversed.  So the
monodromy of the pair $\begin{pmatrix}\J_+\cr \J_-\end{pmatrix}$
around $\lambda=\infty$ is
\begin{equation}\label{xelf}M=-U_-U_+=\begin{pmatrix} -1 & -1 \\ 1
& 0 \end{pmatrix}.\end{equation}

In particular, the monodromy obeys $M^3=1$.  The fact that the
monodromy is of order 3 is clear without Morse theory: under
monodromy around $\lambda=\infty$, the three good regions in Fig.
\ref{Airy} undergo a cyclic permutation of order 3. What we have
gained from Morse theory is a powerful framework for
generalizations.

In this analysis, the dependence of the orientation of the
$\J_\pm$ on a sign of $\sqrt{-i\lambda}$ is inelegant, and the
resulting jumping in the sign of the $\n_\sigma$ is unthematic.
This can be avoided by multiplying the original definition of the
Airy integral by $\sqrt{-i\lambda}$. The product of this function
times the orientation of $\J_\pm$ has no jumping in sign. With
this factor included, the jumping of $\n_\sigma$ occurs only in
crossing Stokes lines, and $M$ has the opposite sign to what is
given in (\ref{xelf}) and obeys $M^3=-1$.

\subsubsection{Asymptotic Behavior Of  Integrals Over Lefschetz Thimbles}\label{asyc}

The cycles $\J_\pm$ are defined in such a way that, for any $\lambda$, the integrals
\begin{equation}\label{polik}Z_{\pm,\lambda}= \int_{\J_\pm}{\d x}\exp(\I)
\end{equation}
always converge.  Indeed, we have arranged so that $\re\,\I\to -\infty$ at infinity
along $\J_\pm$.

Not only are the integrals $Z_{\pm,\lambda}$ always convergent, but it is
straightforward to determine their asymptotic behavior as
$\lambda\to\infty$ in a fixed direction in the complex plane.  The
maximum of $h=\re\,\I$ on the cycle $\J_\pm$ is precisely at the
critical point $x=\pm 1$.  Moreover, the phase $\im\,\I$ is
stationary at $x=\pm 1$, since those are critical points of $\I$
(and in fact the phase $\im\,\I$ is entirely constant along the
cycles $\J_\pm$).  So the asymptotic behavior of $Z_{\pm,\lambda}$ is
dominated by a contribution from the critical point, giving
\begin{equation}\label{olik}Z_{\pm,\lambda}
\sim\exp(\I_\pm)(-i\lambda)^{-1/2}\sum_{t=0}^\infty b_t\lambda^{-t},
\end{equation} where $\I_\pm = \mp
2i\lambda/3$ and the coefficients $b_t$ in the asymptotic series
can be computed by expanding around the critical point.  If we
include a factor of $(-i\lambda)^{1/2}$ in the definition of the
Airy integral, as suggested at the end of section \ref{morset},
the factor of $(-i\lambda)^{-1/2}$, which comes from a Gaussian
integral, would be canceled.

\subsubsection{Sum Of Asymptotic Series}\label{sumseries}

An integral $Z_{\CC,\lambda}$ over a general cycle
$\CC=\n_+\J_++\n_-\J_-$ can be expressed asymptotically as a sum
of two asymptotic expansions:
\begin{equation}\label{zolik} Z_{\CC,\lambda}\sim \n_+
\exp(\I_+)(-i\lambda)^{-1/2}\sum_{t=0}^\infty b_t\lambda^{-t}+ \n_-
\exp(\I_-)(-i\lambda)^{-1/2}\sum_{t=0}^\infty
c_t\lambda^{-t}.\end{equation} However, generically one of these
series suffices, since in a generic direction in the complex
$\lambda$ plane, $h_+$ and $h_-$ are unequal and one of the two
series  is exponentially larger than the other. If $h_+=h_-$
(which happens for real $\lambda$), the two series compete and
must both be included, and if $\lambda$ varies so that $h_+-h_-$
changes sign, the two series exchange dominance.

On a Stokes ray, $h_+$ and $h_-$ are definitely unequal, so one
series in (\ref{zolik}) is exponentially larger than the other.
Consider for definiteness a Stokes ray with $h_+>h_-$.  The
integral in (\ref{zolik}) must vary holomorphically with
$\lambda$, so the coefficient $\n_+$ of the dominant exponential
series must be continuous.  But, since the two series are only
asymptotic series that leave exponentially small corrections
undetermined, the expansion (\ref{zolik}) in a region with
$h_+>h_-$ is  not powerful enough to ensure that $\n_-$ is
continuous.

Let us compare this expectation to (\ref{forkey}), which in the
present notation says that the jumping of $\J_+$ and $\J_-$ is
\begin{equation}\label{zorkey} \begin{pmatrix}\J_+\\
\J_-\end{pmatrix}\to \begin{pmatrix} 1 & \pm 1 \\ 0 & 1\end{pmatrix}\begin{pmatrix}\J_+\\
\J_-\end{pmatrix}.\end{equation} In order for $\n_+ \J_++ \n_-
\J_-$ to be continuous, the $\n_\pm$ transform by
\begin{equation}\label{orkey}\begin{pmatrix} \n_+ &
\n_-\end{pmatrix}\to \begin{pmatrix} \n_+ &
\n_-\end{pmatrix}\begin{pmatrix} 1 & \mp 1 \\ 0 & 1\end{pmatrix}
.\end{equation} In particular, as expected, $\n_+$ is constant and
only $\n_-$ jumps.

An equivalent way to say some of the same things is to observe that
since the integral over $\J_-$ is exponentially smaller than that
over $\J_+$, $\J_-$ cannot jump by a multiple of $\J_+$ in crossing the Stokes ray, but
$\J_+$ can jump by a multiple of $\J_-$.

\subsubsection{Expressing The Real Cycle In Terms Of Lefschetz Thimbles}

The function $Z_\lambda$ whose analytic continuation we actually
want to understand is defined for real $\lambda$ by integration
over the real cycle $\CC_\R$.  So to use the above results, we have
to express $\CC_\R$ as a linear combination of the $\J_\pm$. The
coefficients will be integers, since the $\J_\pm$ are a basis of
the relative homology with integer coefficients.  For the Airy
integral, we can find the right coefficients by inspection, but we
want to describe a more general method.

On a compact Riemann surface $X$, given a basis $\J_\pm$ of
$H_1(X)$, to express another class $\CC\in H_1(X)$ in the form
$\CC=\n_+\J_++\n_-\J_-$, we would find the coefficients $\n_+$ and
$\n_-$ via intersection pairings.   In the present case, we cannot
do precisely this, because there is no intersection pairing on the
relative homology $H_1(X,X_{-T})$. (The intersection of two
one-cycles on the two-manifold $X$ should be a zero-cycle or sum
of points; but a point vanishes in the relative homology
$H_0(X,X_{-T})$, since it can be deformed into $X_{-T}$.)

\def\K{{\mathcal K}}
There is a good substitute, however.  Indeed, Morse theory tells
us what are the duals of the downward flowing cycles from the
critical points: they are the upward flowing cycles. The upward
flowing cycle attached to a critical point $p$ is defined by
reversing the sign in the flow equation (\ref{downflow}).  That
is, one considers solutions of the upward flowing equation
\begin{equation}\frac{\d u^i}{\d t}=+g^{ij}\frac{\partial h}{\partial
u^j} \end{equation} on the half-line $(-\infty,0]$, again
requiring that $u^i(t)$ approaches $p$ for $t\to-\infty$.  The
space of possible values of $u^i(t)$  at $t=0$ now defines the
upward flowing cycle $\mathcal K_p$ attached to $p$.  It is an
$n-k$-dimensional cycle if $p$ has index $k$ (so that the Morse
function $-h$ has index $n-k$ at $p$).

The upward flowing cycles take values in a different relative
homology.  For a real number $T$, define $X^T$ as the part of the
complex $x$-plane with $h\geq T$.  The upward flowing cycle from a
critical point $p$ of the type we are considering has its infinite
ends contained in $X^T$ for large positive $T$ (as long as we keep
away from Stokes rays). So it defines an element of the relative
homology $H_k(X,X^T)$.

For $r$ and $s$ of complementary dimensions, there is a natural
intersection pairing between $H_r(X,X_{-T})$ and $H_s(X,X^T)$.
This pairing is easily evaluated for the case of interest to us --
a perfect Morse function with no flows between distinct critical
points. The absence of such flows means that if $p_\sigma$ and
$p_\tau$ are distinct critical points, the corresponding cycles
$\J_\sigma$ and $\K_\tau$ do not intersect -- such an intersection
would be a downward flow from $p_\sigma$ to $p_\tau$.  On the
other hand, for any one critical point $p_\sigma$, the downward
cycle $\J_\sigma$ and the upward cycle $K_\sigma$ have precisely
one point of intersection -- the trivial constant flow, starting
and ending at $p_\sigma$. With suitable relative orientations of
downwards and upward cycles, the trivial flow contributes $+1$ to
the diagonal intersection pairing
$\langle\J_\sigma,\K_\sigma\rangle$. So the intersection pairings
are
\begin{equation}\label{pilox}\langle \J_\sigma, \K_\tau\rangle
=\delta_{\sigma\tau}. \end{equation}

Given any element $\CC$ of the downward flowing relative homology,
we now can give a formula to determine the coefficients in the
expansion $\CC=\sum_\sigma \n_\sigma \, \J_\sigma$.  By
intersecting with $\K_\tau,$ we find simply
\begin{equation} \label{filox} \n_\tau = \langle \CC,
\K_\tau\rangle \end{equation} for all $\tau$.

\subsubsection{The Asymptotic Behavior Of The Airy Function}

We will now give some examples of this procedure.  Return to the
original Airy function
\begin{equation}\label{moroko} Z_\lambda=\int_{\CC_\R}
{\d x}\exp\left(i\lambda( x^3/3-x)\right)\end{equation}
with $\CC_\R$ being the real axis.  The critical points $p_\pm$ at
$x=\pm 1$ both lie on $\CC_\R$.  A nontrivial upward flow starting at
a critical point will end at a point with a strictly positive
value of $h=\re\,\I$; such a point will not lie on $\CC_\R$, since
$h$ vanishes identically on $\CC_\R$.  Hence the intersection of
$\CC_\R$ with either upward flowing cycle $\K_\pm$ consists precisely
of the critical point $p_\pm$.  Modulo a judicious choice of orientations,
this tells us the coefficients when we express the real cycle
$\CC_\R$ in terms of Lefschetz thimbles associated to critical points:
\begin{equation}\label{zoroko} \CC_\R=\J_++\J_-. \end{equation}

This formula expresses the integration cycle in terms of Lefschetz thimbles when $\lambda$
is real.  But when $\lambda$ is not real, we must
integrate over a more general cycle $\CC$.  As $\lambda$ varies in the complex
plane,  we vary the cycle $\CC$ smoothly so that it
always connects one good region at infinity to another.  On the
other hand, the cycles $\J_+$ and $\J_-$ have jumps described in
eqns. (\ref{poko}), (\ref{noko}) and a jump in sign involving the
orientations.  Allowing for these jumps, we get, after analytic
continuation from real $\lambda$ to a wedge-shaped region of the
complex $\lambda$ plane that does not contain Stokes rays, a
formula
\begin{equation}\label{joroko} \CC=\n_+\J_++\n_-\J_-.\end{equation}
(Instead of keeping track of Stokes rays and orientations, one can
in this particular problem let $\CC$ evolve continuously so that
the Airy integral remains convergent, and then directly use
(\ref{filox}) to determine the coefficients in any wedge-shaped
region.)  Hence in each wedge-shaped region, the Airy integral can be expressed as
\begin{equation}\label{poroko}
Z_\lambda=\n_+ Z_{+,\lambda}+\n_- Z_{-,\lambda}.\end{equation}
Together with (\ref{olik}), this determines the asymptotic
behavior of $Z_\lambda$ in each wedge-shaped region.  In
particular, $Z_\lambda$ grows exponentially at infinity if and
only if the expansion (\ref{joroko}) has a nonzero coefficient for
a critical point at which $h>0$.

Let us consider a special case in which there is {\it not}
exponential growth at infinity.  This is the Airy function with a
reversed sign for the linear term in the exponent,
\begin{equation}\label{mooroko} \tilde Z_\lambda=\int_{\CC_\R}
{\d x}\exp\left(i\lambda(
x^3/3+x)\right),\end{equation} with $\lambda$ real and  with $\CC_\R$ again equal to the real
axis. This integral is actually a special case of what one gets by
analytic continuation of the original Airy integral,
since\footnote{To be more exact, to map the original Airy function
to (\ref{mooroko}) by analytic continuation, one must increase the
argument of $\lambda$ by $2\pi(n+1/4)$, where the integer $n$ is
chosen so that the real cycle of the original Airy function maps
back to the real cycle in (\ref{mooroko}).} (\ref{moroko}) is
mapped to (\ref{mooroko}) by $\lambda\to i\lambda$, $x\to i x$.
However, we will simply treat this slightly modified example directly
according to the general recipe.

The critical points are now located at $x=\pm i$. We denote these
points as $\tilde p_\pm$.  The value of $\tilde
\I=i\lambda(x^3/3+x)$ at $\tilde p_\pm$ is now $\mp 2\lambda/3$,
so for either sign of $\lambda$, the function $\tilde
h=\re\,\tilde\I$ is positive at one critical point and negative at
the other critical point.   An upward flow from a critical point
with $\tilde h>0$ will never meet the real axis $\CC_\R$, since
$\tilde h=0$ on $\CC_\R$.  However, from the critical point with
$\tilde h<0$, there is an  upward flow to the real
axis.\footnote{\label{erf} For example, if $\lambda$ is real, the
part of the imaginary axis between $x=i$ and $x=-i$ is a flow line
that intersects the real axis.  In one direction or the other,
this is an upward flow from the critical point with negative
$\tilde h$.  If we give $\lambda$ a small imaginary part to get
away from the Stokes ray, the upward flow from that same critical
point
 still intersects the real axis (it then continues to $\tilde
h=+\infty$ rather than to the second critical point).} Write $q^+$
and $q^-$ for the critical points at which $\tilde h$ is positive
or negative, and $\tilde\J_\pm$ for the corresponding Lefschetz thimbles.
Then the expression for $\CC_\R$ in terms of Lefschetz thimbles is simply
\begin{equation}\label{yor} \CC_\R=\tilde\J_-.\end{equation}
The asymptotic behavior of $\tilde Z_\lambda$ for large real
$\lambda$ is hence
\begin{equation}\label{korny}\tilde Z_\lambda\sim
\frac{\exp\left(\tilde\I(q^-)\right)}{\lambda^{1/2}}\sum_{t=0}^\infty
a_t\lambda^{-t}.\end{equation} In particular, $\tilde Z_\lambda$
vanishes exponentially for large real $\lambda$, reflecting the
fact that there are no critical points on the real $x$-axis.

Actually, in this example, the Stokes rays are the positive and
negative real $\lambda$ axis. For real $\lambda$, there is a flow
line between the two critical points, as noted in footnote
\ref{erf}.  Hence the cycles $\tilde\J_\pm$ jump in crossing the
real $\lambda$ axis. However, the jumping only affects the cycle
$\tilde\J_+$ attached to the critical point with $\tilde h>0$.
This cycle changes by $\tilde\J_+\to\tilde\J_+\pm \tilde\J_-$. The
cycle $\tilde\J_-$ is unchanged in crossing the Stokes ray, and
hence there is no problem with the formula (\ref{yor}).

\def\S{{\mathcal S}}
\subsection{Generalization To Dimension $n$}\label{genn}

In our analysis of the Airy function, we introduced much more
machinery than was really needed.  The advantage of having done so
is that the generalization to an oscillatory integral of higher
dimension is now straightforward.

We consider a real-valued polynomial function $f(x^1,\dots,x^n)$
of $n$ real variables,  and we assume that $f$ is sufficiently
generic that (even after analytic continuation to complex values of the $x^i$) it has only finitely many critical points, which are all nondegenerate.  We want to study the
analytic continuation of the oscillatory
integral\footnote{Including the elementary prefactor
$(-i\lambda)^{n/2}$ avoids sign changes in the orientations of the Lefschetz thimbles,
as explained at the end of
section \ref{morset}.  This lets us concentrate on the more
interesting Stokes phenomena.}
\begin{equation}\label{zonko} Q_\lambda =(-i\lambda)^{n/2}\int_{\CC_\R} \d^n x
\,\exp\left(i\lambda f(x^1,\dots,x^n)\right). \end{equation} The integration
is over the cycle $\CC_\R$ defined by taking $x^1,\dots, x^n$ to be
real.  We wish to analytically continue the integral to (nonzero)
complex $\lambda$, and to determine the asymptotic behavior of the
analytically continued integral.  Depending on $f$, it may be
difficult to see how to vary the integration cycle $\CC$ as $\lambda$ is varied so that
the integral remains convergent.  So it is difficult in general to
imitate the direct approach to the Airy function by rotating $\CC$
as $\lambda$ varies.  But the approach via Morse theory can be
straightforwardly generalized.

As in the one-variable case, we regard the function $\I = i
\lambda f(x^1,\dots,x^n)$ as a function of complex variables, and
for fixed complex $\lambda$, we let $\Sigma$ be the set of its
critical points. If $p_\sigma$, $\sigma\in \Sigma$, is a critical
point, we write $\I_\sigma$ for $\I(p_\sigma)$.  Now take $h=\re
\,\I$ as a Morse function, and write $h_\sigma$ for $h(p_\sigma)$.
The critical points are all of the same index $n$, as explained in
section \ref{morset}. So $h$ is a perfect Morse function, and can
be used to describe the appropriate relative homology group in
which the possible integration cycles take values.  This group is
$H_n(X,X_{-T};\Z)$, where $X=\C^n$, and $X_{-T}$ is defined by the
condition $h\leq -T$, for large $T$.

Introducing on $\C^n$ the Kahler metric $\d s^2=\sum_i |\d
x^i|^2$, the downward flow equation (\ref{downflow}) can be
written as in (\ref{circ}):
\begin{equation}\label{lopy}\frac{\d x^i}{\d t}=-\frac{\partial
\bar\I}{\partial \bar x^i}.\end{equation} For generic $\lambda$,
there are no flows between critical points.  This follows from the
fact that $\im\,\I$ is conserved along a critical point, as shown
in eqn. (\ref{irc}). (That computation generalizes immediately to
$n$ dimensions, as long as the metric on $X$ is hermitian.  Just as in the one-dimensional
case, if $X$ is Kahler, and we use the Kahler form of $X$ as a symplectic form,
then the gradient flow with respect to the Morse function
$\mathrm{Re}\,\I$ is the same as the Hamiltonian
flow with Hamiltonian  $\mathrm{Im}\,\I$; naturally, $\mathrm{Im}\,I$ is conserved in this flow.)  If
$p_\sigma$ and $p_\tau$ are critical points with distinct values
of $\I$, then for generic $\lambda$ they have distinct values of
$\im\,\I$ and there is no flow between them. But if
$\I_\sigma=\I_\tau$, there is never a flow between $p_\sigma$ and
$p_\tau$, since $h=\re\,\I$ is always strictly decreasing along a
non-trivial flow, as explained in eqn. (\ref{ownflow}).

We define a Stokes ray to be a ray along which $\im
\,\I_\sigma=\im\,\I_\tau$ for some $\sigma,\tau\in \Sigma$.  There
are only finitely many Stokes rays (we have assumed that $\Sigma$
is a finite set, but even without this assumption, for any polynomial $f$,  $\Sigma$ has
only finitely many components, and $\I$, which is constant along
each component, has only finitely many critical values).

Just as
in the one-variable case, we attach to each critical point
$p_\sigma$ a cycle $\J_\sigma$, known as a Lefschetz thimble, by taking the union of all downward
flows that start at $p_\sigma$, or equivalently the values at
$t=0$ of downward flows $u^i(t)$ that start at $p_\sigma$ at
$t=-\infty$. As long as $\lambda$ is not on a Stokes ray, $\J_\sigma$
is closed and has its ends at $h=-\infty$. Hence, $\J_\sigma$
defines an element of the relative homology $H_n(X,X_{-T};\Z)$.
Moreover, according to Morse theory, the $\J_\sigma$ give a basis
of this relative homology.

In crossing a Stokes ray along which
$\im\,\I_\sigma=\im\,\I_\tau$, the cycles $\J_\sigma$ and
$\J_\tau$ may jump.  Such jumping involves nontrivial flows
between $p_\sigma$ and $p_\tau$. A downward flow from $p_\sigma$
to $p_\tau$  is only possible if $h_\sigma>h_\tau$. The general
form of the jumping across a Stokes ray with
$\im\,\I_\sigma=\im\,\I_\tau$, $h_\sigma> h_\tau$ is
\begin{align}\label{temo}\notag\J_\sigma & \to
\J_\sigma+\m_{\sigma\tau}\J_\tau \\
\J_\tau & \to \J_\tau.
\end{align}
Here $\m_{\sigma\tau}$ receives a contribution of $\pm 1$ from
each downward flow line from $p_\sigma$ to $p_\tau$.  (The sign of
the contribution is the sign of a certain determinant, described
at the end of section \ref{apflow}.)

Whenever the Lefschetz thimble  $\J_\sigma$ is defined, the  integral
\begin{equation}\label{ccite}Q_{\sigma,\lambda}=(-i\lambda)^{n/2}\int_{\J_\sigma}
\d^nx\,\exp(\I) \end{equation} is always convergent. Moreover, the
phase of the function $\exp(\I)$ is constant along the integration
cycle, and its modulus has a unique maximum at the critical point
$p_\sigma$.  Hence, for large $\lambda$, this integral has an
asymptotic expansion
\begin{equation}\label{cocite} Q_{\sigma,\lambda}\sim
\exp(\I_\sigma)\sum_{t=0}^\infty a_t \lambda^{-t}.\end{equation}

To find the asymptotic behavior of the analytically continued integral, we will use the expansion
\begin{equation}\label{expo}\CC=\sum_\sigma \n_\sigma \J_\sigma.
\end{equation}
For real $\lambda$, $\CC$ is simply the original real integration cycle $\CC_\R$.
 Unlike the case of the
Airy function, we do not necessarily have a convenient way to
directly determine the dependence of  $\CC$ on $\lambda$. However, once the
coefficients $\n_\sigma$ are known for real $\lambda$, we can
determine their dependence on $\lambda$ by  using the
``wall-crossing'' formulas (\ref{temo}).

To determine the coefficients for real $\lambda$, we proceed as in
the one-variable case.  For each critical point $p_\sigma$, we
define a cycle $\K_\sigma$ that consists of upward-going flow
lines that start at $p_\sigma$.  Away from Stokes rays,
$\K_\sigma$ is an element of $H_n(X,X^T)$ for very large $T$,
where $X^T\subset X$ is defined by the condition $ h\geq T$. Just
as for $n=1$, the groups $H_n(X,X_{-T})$ and $H_n(X,X^T)$ are
naturally dual, and moreover the pairings of the $\J_\sigma$ and
$\K_\tau$ are
\begin{equation}\label{thepair}\langle\J_\sigma,\K_\tau\rangle=\delta_{\sigma\tau}.\end{equation}
As a result, the coefficients in (\ref{expo}) are
\begin{equation}\label{theco}
\n_\sigma=\langle\CC,\K_\sigma\rangle.\end{equation} Thus,
$\n_\sigma$ can be computed by counting (with signs) upward flows
from $p_\sigma$ to $\CC$.

Let us analyze this formula for real $\lambda$ and the original integration cycle $\CC_\R$.
We divide the set $\Sigma$ of critical points into
three subsets: $\Sigma_\R$ consists of critical points with all
$x^i$ real, $\Sigma_-$ consists of critical points with $x^i$ not
all real and $h<0$, and $\Sigma_{\geq 0}$ consists of critical
points with $x^i$ not real and $h\geq 0$.  Note that $h=0$ along
$\CC_\R$, and in particular $h_\sigma=0$ for $\sigma\in \Sigma_\R$.
If $\sigma\in \Sigma_\R$, then the unique upward flow from
$p_\sigma$ to $\CC_\R$ is the trivial one, as $h$ would increase
along a nonconstant flow.  So in this case $\n_\sigma=1$.  If
$\sigma\in \Sigma_{\geq 0}$, there are no upward flow lines from
$p_\sigma$ to $\CC_\R$, as this would contradict the fact that $h$
increases along such lines.  Finally, if $\sigma\in \Sigma_-$,
there may or may not be ascending flow lines from $p_\sigma$ to
$\CC_\R$.

The general structure of the expression for $\CC_\R$, with real $\lambda$, is therefore
\begin{equation}\label{genexp}\CC_\R=\sum_{\sigma\in
\Sigma_\R}\J_\sigma +\sum_{\sigma\in\Sigma_-}\n_\sigma
\J_\sigma,\end{equation} with unknown integer coefficients
$\n_\sigma$. Correspondingly,  for real $\lambda$, our original integral $Q_\lambda$
is
\begin{equation}\label{enexp}Q_\lambda=\sum_{\sigma\in \Sigma_\R}
Q_{\sigma,\lambda}+\sum_{\sigma\in\Sigma_-}\n_\sigma
Q_{\sigma,\lambda}.\end{equation} Since each term on the right
hand side has the asymptotic expansion (\ref{cocite}), this
formula makes manifest the behavior for large real $\lambda$.  For
complex $\lambda$, there is a similar formula, but of course one
must use the appropriate coefficients.

The significance of the formula (\ref{enexp}) is as follows.  In
studying the function $Q_\lambda$ for large real $\lambda$, the
obvious critical points that one would expect to contribute are
the ones that lie on the real cycle $\CC_
\R$.  They indeed all
contribute with a coefficient of 1, as one would expect.  Critical
points in $\Sigma_{\geq 0}$ would make contributions to the large
$\lambda$ behavior that typically would grow exponentially and
dominate the real critical points (if $h_\sigma>0$) and that in
the exceptional case $h_\sigma=0$ would compete for large real
$\lambda$ with the real critical points (which also have $h=0$).
However, such critical points do not contribute to the integral.
Finally, there may be contributions from complex critical points
with $h$ strictly negative.  Their contributions are exponentially
small for large real $\lambda$. But there is no easy recipe to
determine which such critical points do contribute exponentially
small corrections to $Q_\lambda$, and with what weights.  Rather,
for $\sigma\in \Sigma_-$, one simply has to determine the
coefficient $\n_\sigma$ by counting (with signs) the upward flow
lines from $p_\sigma$ to $\CC_\R$.

\subsubsection{Essential Singularities}\label{essential}

Now we wish to address the following question: Under what
conditions does the analytically continued integral $Q_\lambda$
have an essential singularity at $\lambda=\infty$?

There is one special case in which there is no such singularity.
Suppose that, near infinity in the complex $\lambda$ plane, the critical
points $p_\sigma$ for which the coefficients $\n_\sigma$ are
nonzero all have $\I_\sigma=0$.  Then $Q_\lambda$ is bounded for
$\lambda\to\infty$, in any direction. In particular, it has no
essential singularity at infinity.

The hypothesis that the $\I_\sigma=0$ whenever $\n_\sigma\not=0$
is very strong, and there are not many obvious examples, other
than an oscillatory Gaussian integral with $f(x)=x^2$. However,
something similar is possible in Chern-Simons theory.

Now we want to explain a converse.  For simplicity, assume that
the values of $\I_\sigma$ at critical points are distinct. (The
purpose of this is to avoid possible cancellations.)  Let
$p_\sigma$ be a critical point with $\I_\sigma\not=0$, and suppose
that in some wedge-shaped region of the complex $\lambda$ plane,
$\n_\sigma\not=0$. We claim that in this case, the integral
$Q_\lambda$ has an essential singularity at $\lambda=\infty$. This
result is immediate if $h_\sigma>0$, for  then the critical point
$p_\sigma$ makes an exponentially large contribution to
$Q_\lambda$ in the wedge-shaped region of the $\lambda$ plane in
which we have started.
\begin{figure}
 \begin{center}
   \includegraphics[width=2.5in]{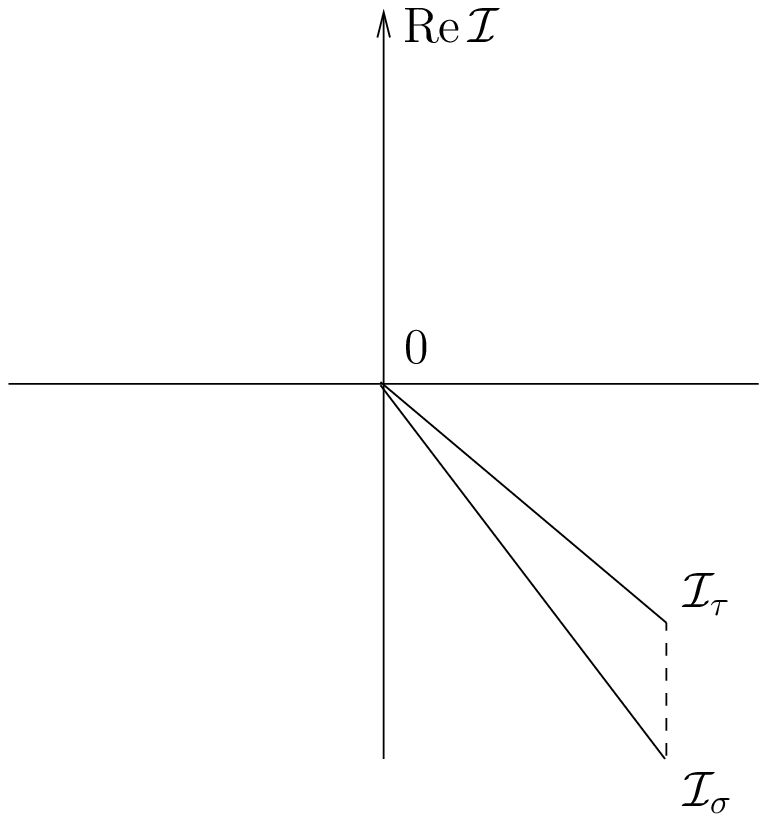}
 \end{center}
\caption{\small The complex $\I$ plane is sketched here with $h=\re\,\I$ running
vertically.
The exponents $\I_\tau$ and $\I_\sigma$ of two critical points $p_\tau$ and $p_\sigma$
have
arguments between $\pi$ and $3\pi/2$.  If
${\mathrm{Arg}}\,\I_\tau>{\mathrm{Arg}}\,\I_\sigma$,
then at a Stokes ray with $\im\,\I_\tau=\im\,\I_\sigma$, one has $h_\tau>h_\sigma$.  This
ensures that the coefficient $\n_\tau$ is unaffected in crossing the Stokes line, though
$\n_\sigma$
may jump.  The analytically continued integral will grow exponentially when $\lambda$ is
varied so
that ${\mathrm{Arg}}\,\I_\tau$ exceeds $3\pi/2$. }
 \label{Crossing}
\end{figure}

Suppose, though, that in the initially considered region of the
$\lambda$  plane, all critical points with $\n_\sigma\not=0$ have
$h_\sigma\leq 0$.  This means that for such critical points,
$\I_\sigma=r_\sigma\exp(i\phi_\sigma)$, with $r_\sigma$ positive
and $\pi/2\leq\phi_\sigma\leq 3\pi/2$.   Either the largest of the
$\phi_\sigma$ is at least $\pi$ or the smallest is no greater than
$\pi$; suppose for definiteness we are in the former situation.
(Otherwise, one makes the same argument, rotating $\lambda$ in a
counterclockwise rather than clockwise direction.) Let $\tau$ be
such that, in the initially chosen wedge in the complex $\lambda$
plane, $\phi_\tau$ is maximal among critical points with nonzero
$\n_\tau$. (If there is more than one critical point with maximal
$\phi$, pick the one of smallest $|\I|$.)  Increasing the argument
of $\lambda$ increases all $\phi_\sigma$'s by the same amount.
Increase the argument of $\lambda$ until $\phi_\tau$ exceeds
$3\pi/2$, so that $h_\tau$ becomes positive. In the process, one
may cross Stokes rays, but the fact that $\phi_\tau$ is maximal
and is at least $\pi$ ensures that $\n_\tau$ does not jump in
crossing such rays. (See fig. \ref{Crossing}; if
$\phi_\tau\geq\phi_\sigma$ and $\im\,\I_\tau=\im\,\I_\sigma$, then
$h_\tau\geq h_\sigma$ so $\n_\tau$ does not jump.) Once $h_\tau$
becomes positive, $Q_\lambda$ grows exponentially with $\lambda$,
ensuring that it has an essential singularity at $\lambda=\infty$.

\subsubsection{A Special Situation}\label{specialcase}

Now we will consider a special situation that was  described in
section \ref{analogs}.  We assume that $n=2w$, and that the
polynomial $f(x)$ is actually the imaginary part of a polynomial
$g$ in $w$ complex variables $z^1,\dots,z^w$; we suppose that $g$
is generic enough to have only finitely many critical points, which are all nondegenerate. Then
we consider the integral
\begin{equation}\label{circh} Q_\lambda = \int
\d^wz\, \d^w\bar z\,\exp(\lambda(g(z)-\bar{g(z)}),\end{equation} initially for real
$\lambda$.

\def\CCR{{{\mathcal C}_\R}}
Actually, we can generalize the problem slightly without any
difficulty; the generalization is relevant to Chern-Simons theory.
As $\lambda$ is initially real, we can absorb it in the definition
of $g$. Then instead of simply analytically continuing with
respect to $\lambda$, we can analytically continue with respect to
all of the coefficients in $\bar g$.  In other words, the
following integral
\begin{equation}\label{circo} Q_{g,\tilde g}= \int_\CCR
\d^wz\, \d^w\tilde z\,\exp(g(z)-\tilde g(\tilde z)),\end{equation} is
a convergent oscillatory integral if the polynomial $\tilde g$ is
the complex conjugate of the polynomial $g$ (meaning that $\tilde
g$ is obtained from $g$ by complex conjugating all coefficients)
and the integration cycle $\CCR$ is defined by $\tilde z^i=\bar
z^i$.  We would like to analytically continue with respect to the
coefficients in $g$ and $\tilde g$ to get a holomorphic function
of those coefficients. (A similar generalization could be made in
section \ref{genn}; we could absorb the real number $\lambda$ in
the definition of the polynomial $f$, and then analytically
continue with respect to all of the coefficients in $f$, not just
$\lambda$.)

In our previous analysis, we had to avoid $\lambda=0$, and
similarly in the present context, we must impose a condition on
how $g$ and $\tilde g$ are allowed to vary.  The condition we want
is that as $g$ and $\tilde g$ vary,  the critical points of the
polynomials $g$ and $\tilde g$ should remain nondegenerate and
their number should remain fixed. For example, we do not allow
leading coefficients in $g$ or $\tilde g$ to vanish in a way that
would cause the number of critical points to become infinite or
would cause some critical points to go to infinity. This
restriction is assumed in what follows.

The basic idea behind the analytic continuation with respect to
$g$ and $\tilde g$ is the same as we have already described. For $\tilde g\not= \bar g$,
we will have to relax the condition $\tilde z=\bar z$ and integrate
over a more general cycle $\CC$:
\begin{equation}\label{circop} Q_{\J;g,\tilde g}= \int_\CC
\d^wz\, \d^w\tilde z\,\,\exp(g(z)-\tilde g(\tilde
z)).\end{equation} Of course, the integration cycle $\CC$ must be
chosen so that the integral converges.  By now we know how to
describe cycles of the appropriate form. We set $\I=g(z)-\tilde
g(\tilde z)$ and $h=\re\,\I$. We treat $h$ as a Morse function. To
each critical point $p_\sigma$ of $h$, we attach a Lefschetz
thimble $\J_\sigma$. These cycles form a basis for the appropriate
relative homology group. In the original case $\tilde g=\bar g$,
we express the original real integration cycle $\CCR$  as a linear combination of the
$\J_\sigma$'s,  by the usual formula
$\CCR=\sum_\sigma\,\n_\sigma\J_\sigma$.  The integrals
$Q_{\J_\sigma;g,\tilde g}$ are convergent for all $\sigma$, $g$
and $\tilde g$, and the desired analytic continuation $Q_{g,\tilde
g}$ of our original integral is given by
\begin{equation}\label{ovoc}Q_{g,\tilde g}=\sum_\sigma \n_\sigma
Q_{\J_\sigma;g,\tilde g}.\end{equation}

The only subtlety is the usual one. As one varies $g$ and $\tilde
g$ away from the locus $\tilde g=\bar g$, one may cross surfaces
on which the cycles $\J_\sigma$ and therefore the coefficients
$\n_\sigma$ jump.  This happens when there are flows between
different critical points, as a result of which the downward flow
from the critical point with greater $h$ fails to define a
relative homology cycle.  Since $\im\,\I$ is a conserved quantity
in the flow equation, the jumping can only occur when two distinct
critical points  have the same value of $\im\,\I$.  The jumping
takes the form $\J_\sigma\to \J_\sigma+\m_{\sigma\tau}\J_\tau$,
where $h_\sigma>h_\tau$ and $\im\,\I_\sigma=\im\,\I_\tau$. We will
refer to surfaces on which distinct critical points have the same
value of $\im\,\I$ as Stokes surfaces.  We also use the term
Stokes curve for a Stokes surface of real dimension 1 (thus, a
Stokes ray is a special case of a Stokes curve).

What is rather special about this problem is that $\I=g(z)-\tilde
g(\tilde z)$ is the sum of a function on one space (which we will
call $\C^w$, parametrized by the $z^i$) and a function on another
space (which we will call $\tilde\C^w$, a second copy of $\C^w$
parametrized by the $\tilde z^i$).   $\I$ is defined on the
product space $X=\C^{2w}=\C^w\times\tilde\C^w$. The fact that $\I$
is the sum of functions on the two factors  has obvious
implications for the critical points.

Let $q_\sigma$, $\sigma\in\Sigma_g$ be the critical points of $g$,
as a function on $\C^w$, and let $r_{\tau}$, $\tau\in
\Sigma_{\tilde g}$, be the critical points of $\tilde g$, as a
function on $\tilde \C^w$. Then the critical points of $\I$ are
the points $p_{\sigma,\tau}=q_\sigma\times r_\tau\in \C^w\times
\tilde\C^w=X$. Similarly, the Lefschetz thimble
$\J_{\sigma,\tau}$ associated to a critical point
$p_{\sigma,\tau}$ in $X$ is the  product of cycles
$\J^g_\sigma\in\C^w$ and $\J^{\tilde g}_\tau\in\tilde\C^w$.  Here
$\J^g_\sigma$ is the Lefschetz thimble in $\C^w$ associated to
the critical point $q_\sigma$ of the Morse function $\re\,g$, and
similarly $\J^{\tilde g}_\tau$ is the Lefschetz thimble in
$\tilde\C^w$ associated to the critical point $r_\tau$ of the
Morse function $\re\,(-\tilde g)$.

We can therefore rewrite (\ref{ovoc}) as follows:
\begin{equation}\label{norvoc}Q_{g,\tilde
g}=\sum_{\sigma\in\Sigma_g}\sum_{\tau\in\Sigma_{\tilde
g}}\n_{\sigma,\tau}\int_{\J^g_\sigma}\d ^wz\,\exp(g(z))\,
\int_{\J^{\tilde g}_\tau}\d^w\tilde z\,\exp(-\tilde g(\tilde
z)).\end{equation} Thus $Q_{g,\tilde g}$ is everywhere a finite
sum of functions of $g$ times functions of $\tilde g$, obtained
from integrals over the $z$'s and the $\tilde z$'s, respectively.
The coefficients in the sum, however, can jump as one crosses
Stokes surfaces.

As was explained in section \ref{overview}, the problem that we have just
described is a prototype for analytic continuation of Chern-Simons
gauge theory with complex gauge group.

\subsection{Symmetries}\label{symmetries}

Since gauge symmetry is an essential ingredient in Chern-Simons
theory, including symmetries in our finite-dimensional models will
improve their analogy with Chern-Simons.  Before taking this step,
we consider a few preliminaries.

\subsubsection{Oscillatory Integrals On More General
Manifolds}\label{oscmore}

One generalization is to consider, instead of an oscillatory
integral on $\R^n$, an oscillatory integral on some other space
$Y$.   $Y$ must have a suitably  nice  complexification $X$, but
we will not attempt to be precise about this. Instead we just
describe a typical example.  We take $Y$ to be a real affine
variety such as the one defined by the equation
\begin{equation}\label{purkey} x^2+y^2=z^2 +1.\end{equation} Then
we introduce a sufficiently generic polynomial $f(x,y,z)$, and consider the
oscillatory integral $Z_\lambda=\int_Y \d \theta\,\exp(i\lambda
f)$, where $\d\theta=\d x\,\d y/z$ is the natural volume form on
$Y$.

Analytic continuation in $\lambda$ can now be carried out by an
obvious generalization of what we have said for $\R^n$.  We embed
$Y$ in the complex manifold $X$ defined by the same equation
(\ref{purkey}) but now with $x,y$, and $z$ regarded as complex
variables.  The top degree form $\d\theta$ on $Y$ extends to a
holomorphic form of top degree on $X$ (given by the same formula
$\d x\,\d y/z$),
 which can be integrated over any middle-dimensional
cycle in $X$.   Introducing the Morse function $h=\re\,(i\lambda
f)$, where $f$ is regarded now as a holomorphic function on $X$,
we associate to each critical point a relative homology cycle
obtained by downward flow.  The analytic continuation is made as usual  by
expressing the original integration cycle $Y$ in terms of these cycles, and then taking Stokes phenomena into
account.

\subsubsection{Manifolds Of Critical Points}\label{mancrit}

Another generalization is to consider a polynomial $f$ on $\R^n$
(or on some more general space, as just described) that has not
isolated critical points but a manifold $\N$ of critical points.
  The analytic
continuation of $f$ to a holomorphic function on $\C^n$ likewise
has a manifold $\M$ of critical points; $\M$ is a complex
manifold, of course, and its intersection with $\R^n$ coincides
with $\N$. (We will consider only the case that $\M$ is smooth,
and we assume that the matrix of second derivatives of $f$ in the normal
direction to $\M$ is invertible.) We want to use Morse theory on
$\C^n$ to study the analytic continuation of the usual integral
$Z_\lambda=\int\d^n x \,\exp(i\lambda f)$.

Suppose that $\M$ is a union of components $\M_\alpha$ of complex
dimension $t_\alpha$. $\I=i\lambda f$ has a constant value
$\I_\alpha$ on $\M_\alpha$, since $\M_\alpha$ is a critical point
set.  As usual, we consider $h=\re\,\I$ as a Morse function. Then
$\M_\alpha$ has Morse index $n-t_\alpha$. (This is shown by
essentially the same argument used in section  \ref{morset} for
the case $t_\alpha=0$; one can choose local coordinates in the
normal direction to $\M_\alpha$ so that
$\I=c+\sum_{i=1}^{n-t_\alpha}z_i^2$.)  The explanation in section
\ref{morset} for why $h$ is a perfect Morse function assumed that
$t_\alpha=0$ for all components, so that all critical points have
the same Morse index. Here is an alternative argument that does
not assume $t_\alpha=0$. In general, a Morse function is perfect
if there are no flows between distinct critical points. (This is
only possible for Morse functions that are unbounded below, like
the ones we consider here, since otherwise any downward flow from one critical point
ends at another critical point.) As we have
found in eqn. (\ref{irc}), $\im\,\I$ is a conserved quantity for
the flow equation. Moreover, there never are flows between different critical points with the
same value of $\I$.  For generic $\lambda$, no two distinct
components of $\M_\alpha$ with different values of $\I_\alpha$
have the same value of $\im\,\I_\alpha$,
so generically there are no flows between distinct critical points and $h$
is a  perfect Morse function.

There is a general recipe, given a perfect Morse function on any
manifold $X$, for describing the relative homology of $X$ in terms
of downwards flows from critical points.  Let $\mathcal S$ be a
component  of the critical point set that has Morse index $s$. Let
$\W\subset \mathcal S$ be a cycle of dimension $r$. For any point
$w\in \W$, the points in $X$ that lie on downwards flows from $w$
form an $s$-dimensional subspace of $X$. So the points in $X$ that
can be reached by downward flow from somewhere on $\W$ are a
family of dimension $r+s$.  This family is a cycle $\J_\W$ of
dimension $r+s$ in the relative homology. These cycles, which we
view as generalized Lefschetz thimbles,
 generate the relative homology of $X$
(subject only to the relations that hold in the homology of the
critical point set).

Applying this to our problem, if $\M_\alpha$ is a component of the
critical point set that has complex dimension $t_\alpha$ and Morse
index $n-t_\alpha$, then to get an integration cycle on $X$ --
which should be a cycle of dimension $n$ -- we need to start with
a cycle  on $\M_\alpha$ of real dimension $t_\alpha$, that is, a
middle-dimensional cycle.

So the analytic continuation of the integral $Z_\lambda$ can be
described using relative homology cycles attached to the
components $\M_\alpha$ of the critical point set.  But in general
to a component $\M_\alpha$ we attach not just one cycle but a
number equal to the rank of the middle-dimensional homology of
$\M_\alpha$.

This construction may have various applications to Chern-Simons
theory, since in general the moduli space of representations of
the fundamental group of a three-manifold has components of
positive dimension.   However, the only application we will make
in this paper concerns reducible critical points, that is, flat
connections that preserve some gauge symmetry. Moreover, this
application concerns a situation in which every relevant component
of $\M_\alpha$ has middle-dimensional cohomology of rank 1, and
hence contributes only one cycle to the middle-dimensional
cohomology. In a sense, therefore, our application involves a
situation that is as simple as the case of only isolated critical
points.

\def\O{{\mathcal O}}
\subsubsection{Group Actions}\label{gaction}

Why might critical points not be isolated? One simple answer to
this question will be important in our study of gauge theory. In
the presence of a group action, the critical points will form
orbits of the group, and this may force them to be non-isolated.

We consider a manifold $Y$  with a chosen action of a compact connected Lie
group $H$. We also assume that $Y$  has a good complexification $X$, which
possesses a real involution that leaves $Y$ fixed. 
We endow $Y$
with an $H$-invariant function $f$ whose critical points consist of
finitely many $H$ orbits $\O_\sigma$, $\sigma\in\Sigma$. 

As usual,
we want to study the analytic continuation of the integral
$Z_\lambda=\int_Y \d\theta \,\exp(i\lambda f)$, where $\d\theta$
is the restriction to $Y$ of a holomorphic top form on $X$. Assuming that $f$ can be
analytically continued to a holomorphic function on $Y$ (which we also denote as $f$),
we analytically continue the integral $Z_\lambda$  by embedding $Y$ in $X$ and using the Morse function
$h=\re\,(i\lambda f)$ on $X$.  We assume that the $H$ action on $Y$ analytically
continues to an action on $X$ of a complex Lie group $G$ that is
the complexification of $H$; in this case, the critical points of $h$ are
orbits of $G$.  We assume that there are only finitely many such
critical orbits, near each of which $h$ is nondegenerate in the normal directions.
For clarity, we will write $\O^G$ or $\O^H$ for,
respectively, a critical orbit of $G$ acting on $X$ or a critical
orbit of $H$ acting on $Y$.

To orient ourselves, let us first consider the case that $H$ acts freely on $Y$, and $G$
acts freely on $X$.  Then we can write $Z_f$ as an integral over $Y'=Y/H$:
\begin{equation}\label{helme}Z_f=\int_{Y'}\d\theta' \,\exp(i\lambda f),\end{equation}
where the volume form $\d\theta'$ is obtained by integrating
$\d\theta$ over the fibers of $Y\to Y'$.  From a ``downstairs''
point of view, to analytically continue $Z_f$, we must embed $Y'$
in a suitable complex manifold $X'$. Roughly speaking, the
appropriate choice is $X'=X/G$, which -- as $G$ is a
complexification of $H$ and $X$ is a complexification of $Y$ -- is a
complexification of $Y'=Y/H$.    (Unstable orbits -- the notion is discussed below --
should be discarded in defining
the quotient $X/G$.)

Since $f$ has only finitely many critical orbits as a function on
$X$, it follows that as a function on $X'$, $f$ has only finitely
many critical points.  We are thus in the familiar situation of
isolated critical points only, and the relative homology of $X'$
therefore has a basis consisting of a Lefschetz thimble
$\J'_\sigma$ for every critical point $p'_\sigma$. We can
analytically continue the integral $Z_f$ in the usual way by
expressing the real cycle $\CC_\R=Y'\subset X'$ as a linear
combination of the $\J'_\sigma$, and keeping track of Stokes
phenomena.

How would we get an equivalent result by computing upstairs on $Y$
and $X$ rather than downstairs on $Y'$ and $X'$?  Every critical
point $p'_\sigma$ on $X'$ corresponds to a critical orbit
$\O^X_\sigma$ on $X$.  Since we have assumed that $G$ acts freely
on $X$, the orbits are all copies of $G$.  Topologically, the
complex Lie group $G$ is isomorphic to the cotangent bundle $T^*H$
of its maximal compact subgroup $H$.  The middle-dimensional
homology of $T^*H$ is of rank 1, generated by the zero section of
the cotangent bundle.  Hence, from an upstairs point of view,
the middle-dimensional relative homology of $X$  has a basis with one generator for
every critical orbit.  These generators correspond in an obvious way to the generators
of the relative homology of $X'$.  The associated Lefschetz thimbles also have an obvious
correspondence, and  the Stokes surfaces are  the same
whether one computes upstairs or downstairs.  So we can perform
more or less equivalent computations upstairs or downstairs.

In the downstairs description, the critical points are
isolated, so we do not really need the theory sketched in section
\ref{mancrit} as long as $H$ acts freely.
  Now let us consider a more general case in which
$H$ does {\it not} act freely on $Y$. This being so, the quotient
$Y'=Y/H$ is singular, and the general theory would not readily
apply to it. So instead we will calculate on $Y$.  We still assume
that the critical point set of $f$ on $Y$ is a finite union of $H$
orbits $\O_\sigma^H$, and that the critical point set of $f$ on
$X$ is an analogous (but perhaps larger\footnote{ This is possible
even in the absence of a group action.  For example, the function
$f(x)=x^3/3+x$ has complex critical points that do not lie on
the real axis.}) finite union of $G$ orbits $\O_\tau^G$.

Any critical orbit $\O_\sigma^H$ in $Y$ has a complexification
that is one of the $\O_\sigma^G$.   The complexification of an $H$
orbit $\O^H\subset Y$ is, from a topological point of view, simply
$T^*\O^H$.  Its middle-dimensional homology is of rank 1,
generated by the zero section of the cotangent bundle.  Each
critical orbit in $Y$ therefore contributes one generator of the
middle-dimensional relative homology of $X$, just as in the case of a free action.

What about critical orbits in $X$ that do not intersect $Y$?  We will show that not
all critical orbits need to be taken into account, but only
certain ``semistable'' ones.  The semistable orbits are all topologically
of the form $T^*\O^H$ for some $H$ orbit $\O^H$, and again they
have middle-dimensional homology of rank 1.  (Every critical orbit
that intersects $Y$ is semistable.)  So in short, the analytic
continuation of $Z_f$ is made using a middle-dimensional relative
homology of $X$ that has one generator for every semistable critical
$G$ orbit.

Now we will explain how semistability comes in.  First of all, to
construct the downward flowing cycles associated to a critical
point, we need to pick a metric on $X$.  We generally cannot pick
this metric to be $G$-invariant, but it is commonly possible to pick an
$H$-invariant Kahler metric, such that the Kahler form $\omega$ is odd
under the real involution of $X$ that leaves $Y$ fixed.  (For example, if $Y=\R^n$ with
$H$-invariant metric $\sum_i(\d x^i)^2$, then we take on $X=\C^n$
the $H$-invariant Kahler metric $\sum_i |\d x^i|^2$.  In this case, $\omega=-i\sum_k \d x^k\wedge
\d \bar {x^k}$, and is odd under $x^k\leftrightarrow \bar{x^k}$.)
We assume the existence of such a metric.
$H$ acts on $X$ preserving the symplectic form $\omega$, and it is possible
to define a moment map $\vec\mu$ for the action of $H$ on $X$.
Here $\vec\mu$ maps a vector field $V$ that generates a
one-parameter subgroup of $H$ to a function $\mu_V$ that obeys
\begin{equation}\label{zorme}\d\mu_V=\iota_V \omega,\end{equation}
where $\iota_V$ is the operation of contraction with $V$.  In
indices
\begin{equation}\label{orme}\frac{\partial\mu_V}{\partial
x^i}=V^{\bar j}\omega_{\bar j i},~~ \frac{\partial \mu}{\partial\bar{ x^{ i}}}= V^j\omega_{j\bar i}.
\end{equation}
(The map from $V$ to $\mu_V$ is a linear map from the Lie algebra
of $H$ to the space of real-valued smooth functions on $X$.) We
fix possible additive constants in the $\mu_V$ by requiring that
these functions vanish identically on $Y$.  This also ensures that
the $\mu_V$ are odd under complex conjugation.  This follows from
(\ref{zorme}), since the vector fields $V$ are even under complex
conjugation and $\omega$ is odd.

The downward flow equations of Morse theory have the beautiful
property that $\vec\mu$ is conserved, in other words $\mu_V$ is
conserved for all $V$.  This follows from a straightforward
computation. We have
\begin{equation} \label{torme} \frac{\d \mu_V}{\d
t}=\frac{\partial
\mu_V}{\partial x^i}\frac{\d x^i}{\d t}+\frac{\partial\mu_V}{\partial \bar {x^i}}\frac{\d \bar{x^i}}{\d t}
      = -V^{\bar j}\omega_{\bar j i} g^{i\bar k}\frac{\partial h}{\partial \bar{x^k}}
      - V^j\omega_{j\bar i}g^{\bar i k}\frac{\partial h}{\partial x^k}.\end{equation}
Since $h$ is the real part of a holomorphic function $\I$, we have
$\partial h/\partial x^i=i\partial \im\,\I/\partial x^i$,
$\partial h/\partial \bar{x^i}=-i\partial\,\im\,\I/\partial
\bar{x^i}$.  On the other hand, the Kahler form $\omega$ and
Kahler metric $g$ obey $\omega_{\bar j i} g^{i\bar
k}=i\delta^{\bar k}_{\bar j}$, $\omega_{j\bar i}g^{\bar i
k}=-i\delta ^k_j$.  So
\begin{equation}\label{loopmex}\frac{\d\mu_V}{\d t} =-V^{\bar
j}\frac{\partial \,\im\,\I}{\partial \bar
{x^j}}-V^j\frac{\partial\,\im\,I}{\partial x^j}=-\iota_V\d\,\im\,\I
.\end{equation} The right hand side vanishes, since $\I$ is $H$-invariant.

A less computational explanation of this result can be found by observing that the
gradient flow with respect to the Morse function
$\mathrm{Re}\,\I$ coincides with the Hamiltonian flow
with Hamiltonian  $\mathrm{Im}\,\I$.  (Here we regard $X$ as a symplectic manifold with
symplectic form $\omega$.)  As usual, the moment map for $H$ is a conserved quantity for  the
flow generated by an $H$-invariant Hamiltonian.

We will call a critical $G$-orbit semistable if it contains at least
one point with $\vec\mu=0$ and unstable otherwise.  (If a semistable orbit is free or has
only a finite stabilizer, we call it stable.)
 Since $\vec\mu$ is conserved along
flows, and vanishes identically on $Y$, any critical $G$-orbit
that is connected to $Y$ by a flow is semistable.  So when we
express the real integration cycle $\CC_\R=Y$ in terms of
generalized Lefschetz thimbles associated to critical orbits, only
semistable critical orbits enter.  Similarly, as we explain
momentarily, in crossing a Stokes surface, a cycle in the relative
homology of $X$ associated to a semistable critical orbit is
always re-expressed in terms of other cycles associated to
semistable critical orbits. So the analytic continuation of the
integral $Z_f$ can always be described using the semistable
critical orbits only.

It remains to explain why Stokes phenomena connect semistable critical
orbits only to each other. Suppose that $\O^G$ is a semistable
critical $G$-orbit.  It contains a distinguished $H$ orbit, namely
the orbit $\W=\O^H$ with $\vec\mu=0$. $\O^G$ is the
complexification of $\O^H$ and is topologically  the cotangent
bundle $T^*\O^H$. The middle-dimensional homology of $\O^H$ is
generated by  $\W$. Since $\vec\mu$ is conserved along flows, any
critical orbit to which $\W$ can be connected by a flow has a
point with $\vec\mu=0$ and therefore is semistable.

In contrast to the  middle-dimensional cohomology of a semistable
$G$ orbit, which is always of rank 1, an unstable $G$ orbit might
have vanishing or more complicated middle-dimensional cohomology.
For $G=SL(2,\C)$, an example of an unstable orbit is $\Bbb{CP}^1$,
whose middle-dimensional cohomology vanishes.  (Since $\Bbb{CP}^1$
is a homogeneous space for $H=SU(2)$, and the condition
$\vec\mu=0$ is $H$-invariant, $\vec\mu$ would have to vanish
everywhere on $\Bbb{CP}^1$ if it vanishes anywhere; but if
$\vec\mu$ is identically zero, then (\ref{zorme}) implies that
$V=0$ for all generators of $H$, and the $H$ action is trivial.)
More generally, for any $G$, the flag manifold, which in general
can have a middle-dimensional cohomology of high rank, is an
example of an unstable orbit.

\subsubsection{More on Group Actions}\label{moreg}

Continuing the discussion of the last subsection, let $\J_\sigma$ be the Lefschetz
thimble associated to a semistable
critical orbit $\O^G_\sigma$.  $\J_\sigma$ consists of the points
that can be reached by downward flow from $\O_\sigma^H$, the points in $\O_\sigma^G$
with $\vec\mu=0$.

Suppose that we are given an integration cycle $\CC$ and we wish
to express it in terms of the $\J_\sigma$.  $\CC$ is a
middle-dimensional, $H$-invariant element of the relative
homology, and in practice one is usually interested in the case
that $\CC$ lies in $\vec\mu^{-1}(0)$.

To find the coefficients in an expansion
\begin{equation}\label{turm}\CC=\sum_\sigma\n_\sigma\J_\sigma,\end{equation}
one wants to intersect $\CC$ with a dual set of cycles
$\K_\sigma$.    The dual cycles should have a natural pairing with
the $\J_\sigma$, such that
\begin{equation}\langle\J_\sigma,\K_\tau\rangle
=\delta_{\sigma\tau}.\end{equation} As we have defined $\J_\sigma$
by downward flow from $\O^H_\sigma\subset\O^G_\sigma$, one's first
thought is to define $\K_\sigma$ by upward flow from
$\O^H_\sigma$.  This does not work well, since in this case the
intersection of $\J_\sigma$ with $\K_\sigma$, rather than a single
point, would be the orbit $\O^H_\sigma$.  (The algebraic
intersection number would equal the Euler characteristic of this
orbit; for instance, it would vanish for the case of a free
orbit.)  What one must do instead is to define $\K_\sigma$ by
upward flow starting from a suitable dual to $\O_\sigma^H\subset
\O_\sigma^G$.  The appropriate dual is a fiber of the cotangent
bundle.  In other words, identifying $\O_\sigma^G$ as
$T^*\O_\sigma^H$, we pick an arbitrary point $q_\sigma\in
\O^H_\sigma$ and write $\tilde\O_\sigma^H$ for the fiber of
$T^*\O_\sigma^H$ at $q_\sigma$.  Then $\K_\sigma$, defined by
upward flow from $\tilde\O_\sigma^H$, meets $\J_\sigma$ precisely
at $q_\sigma$. ($\K_\sigma$ lies in a dual version of the relative
homology, spanned by cycles that are allowed to go to infinity
along $G$ orbits as well as in the region where the Morse function
becomes large.) Moreover, away from a Stokes curve, $\K_\sigma$
meets $\J_\tau$ not at all if $\sigma\not=\tau$.  So
$\langle\J_\sigma,\K_\tau\rangle=\delta_{\sigma\tau}$, and the
coefficients in (\ref{turm}) are
$\n_\sigma=\langle\CC,\K_\sigma\rangle$.

Similar questions arise when one considers the jumping phenomena
$\J_\sigma\to\J_\sigma +\sum_\tau\m_{\sigma\tau}\J_\tau$ in
crossing a Stokes curve.  The coefficients $\m_{\sigma\tau}$ is
computed by counting downward flow lines from $\O_\sigma^H$ to
$\tilde\O_\tau^H$. (Each such flow line is weighted by the sign of
a certain determinant, as described in section \ref{apflow}.) The
asymmetry between the definitions of downwards and upwards flowing
cycles has an interesting consequence, which we will describe for
the special case that there are two types of critical orbits --
trivial orbits and free orbits. Suppose that the critical point
set of an $H$-invariant Morse function $h=\re\,\I$ contains a
critical point $p$ (located at $\vec\mu=0$) and a  free stable $G$ orbit
$\O^G$.  In general (when allowed by conservation of $\im\,I$),
there may be flows from $p$ to $\tilde \O^H$, and the coefficient
$\m_{p\O}$ that appears in a jumping formula
$\J_p\to\J_p+\m_{p\O}\J_\O$ may be nonzero. However, the
coefficient $\m_{\O p}$ that counts flows from $\O^H$ to $p$
always vanishes. A quick way to show this is to observe that by
replacing $\O^H$ by an everywhere nonzero section of the cotangent
bundle $\O^G=T^*\O^H\to\O^H$, such as the section
$\vec\mu^{-1}(c)$, where $c$ is a nonzero constant, one can
eliminate all flows from $\O^H$ to $p$.  So the algebraic
invariant counting such flows must vanish.

In a more general case in which the critical point set consists of
several isolated points $p_\sigma$ and several free orbits
$\O_\alpha$, the coefficients $\m_{\O_\alpha,p_\sigma}$ will
vanish by the argument just indicated. (All other coefficients may
be nonzero in general.)   This means that the subspace of the
relative homology generated by the cycles $\J_{\O_\alpha}$ is
invariant under all jumping phenomena.  The subspace generated by
the cycles $\J_{p_\sigma}$ is not invariant under jumping. This
perhaps surprising fact will be demonstrated in an explicit and
useful example in section \ref{sotwo}.

\def\k{\kappa}
\subsection{Singularities}\label{singularities}

In studying an oscillatory integral $\int \d^n x\,\exp(i\lambda
f)$, for a single and fairly generic polynomial $f$, it is
reasonable to expect that the critical points of $f$ are isolated
and irreducible, as we have assumed so far.  But as soon as $f$
depends on one or more parameters, one should expect to meet
singularities, where two or more critical points meet.

Singularities play an important role because Stokes surfaces
always pass through the locus (in the parameter space of the
polynomial $f$) at which two or more critical points meet.  If we
vary a parameter in $f$ so that two critical points $p_\sigma$ and
$p_\tau$ coincide, then $f(p_\sigma)=f(p_\tau)$, which certainly
ensures that $\mathrm{Im}\,(i\lambda
f(p_\sigma))=\mathrm{Im}\,(i\lambda f(p_\tau))$.  So a Stokes
surface always passes through the singular locus with
$p_\sigma=p_\tau$.

Moreover, the flow equations simplify at a singularity.  To
understand what happens to the Lefschetz thimbles in crossing a
Stokes surface, in general we must solve the flow equation in $n$
complex dimensions.  However, near a singularity, most of the $n$
complex variables are unimportant.  If a singularity can be
modeled by the behavior of a complex polynomial in $k$ variables
-- where in practice $k$ will often be 1 or 2 -- then near the
singularity, the flow equations reduce effectively to equations
for those $k$ variables; the other modes are ``massive'' and decay
quickly in solving the flow equations.

The relation between singularities and Stokes phenomena is
therefore very useful in practice.  Indeed, the subject reviewed
in this section was developed partly for analyzing singularities and their
relation to oscillatory integrals
\cite{Arnold}.

In section \ref{anajones}, where
we study the analytic continuation of the colored Jones
polynomial, we will vary one parameter (the ratio of the highest weight
of a representation to the Chern-Simons level) and we will encounter three
rather generic types of singularity.  These will be described here.

\subsubsection{More on the Airy Function}\label{twocrit}

The only singularity that arises generically in varying a single complex parameter
in the absence of any symmetries
 is a meeting of two critical points.  Even
if this phenomenon  happens in complex dimension $n$, only one complex variable is
essential in describing it.
We simply consider a cubic polynomial that depends on a parameter
$\epsilon$:
\begin{equation}\label{zormo} f(x)=\frac{x^3}{3}-\epsilon
x.\end{equation} There are two critical points at $x=\pm
\sqrt\epsilon$; they meet at $\epsilon=0$.

The corresponding oscillatory integral $I(\lambda,\epsilon)=\int
\d x\exp(i\lambda f)$ is essentially the Airy function.  We can use the scaling $x\to x\lambda^{-1/3}$
to express this integral in terms of a function of one variable:
\begin{equation}\label{urtz}I(\lambda,\epsilon)=\lambda^{-1/3}I(1,\lambda^{2/3}\epsilon)=\lambda^{-1/3}
\int\d x\,\exp(i(x^3-\lambda^{2/3}\epsilon x)).\end{equation}
Setting $\epsilon=0$, a special case is  $I(\lambda,0)\sim\lambda^{-1/3}$.
Comparing this to a Gaussian integral $\tilde I(\lambda)=\int\d x
\exp(i\lambda x^2)\sim \lambda^{-1/2}$, the ratio is
\begin{equation}\label{porkoff} \frac{I(\lambda,0)}{\tilde I(\lambda)}
\sim \lambda^{1/6},~~\lambda\to\infty.\end{equation}
 These facts will be useful in section \ref{mmod}.

For our remaining remarks, we use the above scaling to reduce to the case $\lambda=1$. The good regions at infinity
in the complex $x$-plane where $\exp(i f(x))$ decays are then independent of $\epsilon$ and were
sketched in fig. \ref{Airy}.  The Lefschetz thimble associated
to a critical point always connects two of the good regions.

Since the values of $\I=i f$ at the critical points are
$\mp 2 \epsilon^{3/2}/3$, the condition for a Stokes curve
is that $\epsilon^{3/2}$ should be imaginary; thus ${\mathrm
{Arg}}\,\epsilon$ should equal $\pi$ or $\pm \pi/3$.  In particular, several
Stokes curves meet at $\epsilon=0$.  It is clear why this is so:
at $\epsilon=0$, the two critical points meet, and therefore have
the same value of $\I$ and in particular the same value of
$\im\,\I$.

Our application will involve the case that $\epsilon$ is real. If
$\epsilon$ is positive, we write $p_\pm$ for the critical points
at $x=\pm\sqrt \epsilon$, and $\J_\pm$ for the corresponding Lefschetz thimbles.  If $\epsilon$ is negative, we write $\tilde
p_\pm$ for the critical points at which the action is positive or negative, respectively.
Thus the points $\tilde p_\pm$ are located at $x=\mp
i\sqrt{-\epsilon}$, and
the action takes the values
$\I_\pm
=\pm(2/3)(-\epsilon)^{3/2}$ at those points.   We write $\tilde \J_-$ for the Lefschetz thimble associated to
 $\tilde p_-$.  The Lefschetz thimble $\tilde \J_+$ associated to
$\tilde p_+$ is ill-defined if $\epsilon<0$, because the ray
$\epsilon<0$ is a Stokes ray.  We define $\tilde \J_+$ by taking the limit as $\epsilon$
approaches the negative axis from above.

\begin{figure}
 \begin{center}
   \includegraphics[width=5in]{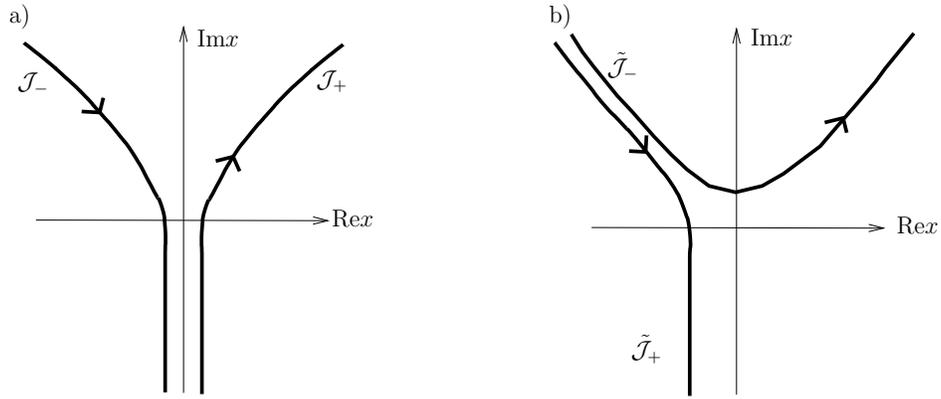}
 \end{center}
\caption{\small Qualitative behavior of Lefschetz thimbles for the Airy function for $\epsilon>0$
(a) and $\epsilon<0$ (b).  In case $(b)$, $\epsilon$ has been given a small imaginary
part, since the negative $\epsilon $ axis is a Stokes curve.}
 \label{Airycycles}
\end{figure}

It is not difficult to describe the  cycles $\J_\pm$ and $\tilde \J_\pm$  qualitatively as curves
in the complex $x$-plane (fig. \ref{Airycycles}).
From the figure, we see that the relation between these two bases
for the relative homology is
\begin{align}\label{helign} \notag \J_+ & =   -\tilde \J_++\tilde \J_-  \\
                             \J_- & =   \tilde\J_+   \end{align}
or equivalently
\begin{align}\label{zelign}\notag \tilde\J_+ & =   \J_-  \\
                           \tilde\J_- & =  \J_++\J_-    . \end{align}

\subsubsection{An Example With $SO(2)$ Symmetry}\label{sotwo}

Now we will consider a singularity that can arise by varying one
parameter in the presence of $SO(2)$ symmetry. A minimal case is a
singularity that can be modeled by two variables $x,y$, with
$SO(2)$ acting on the $xy$ plane by rotations.  The basic invariant
is $x^2+y^2$. The origin $x=y=0$ is a critical point of any
$SO(2)$-invariant polynomial. There may also be critical orbits
characterized by nonzero values of $x^2+y^2$. A meeting of two
critical orbits both with $x^2+y^2\not=0$ can be modeled again by
the Airy integral -- the $SO(2)$ action just factors out.  The new
case is a collision between a critical orbit with $x^2+y^2\not=0$
and the critical point at the origin.

\begin{figure}
 \begin{center}
   \includegraphics[width=3in]{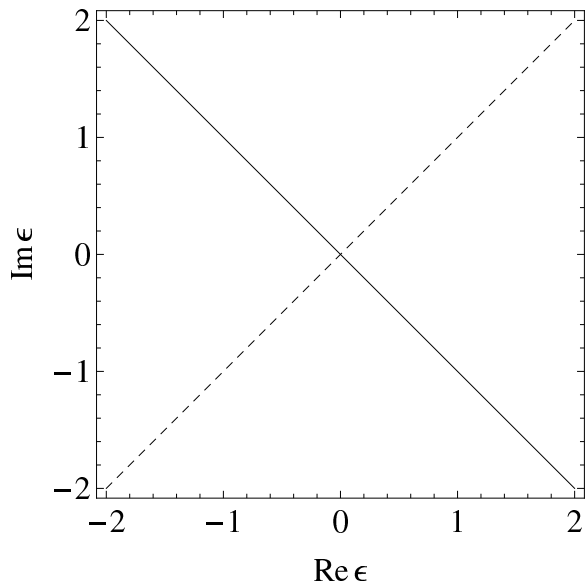}
 \end{center}
\caption{\small The two lines $\epsilon=r\exp(\pm\pi i/2)$, with real $r$, are Stokes lines in the
sense that flows between the critical point $p$ and the critical orbit $\O$ are possible on these lines.
Jumping, however, occurs only in crossing the line $\epsilon=r\exp(-\pi i/2)$, shown here as a solid
line. There is no jumping of Lefschetz thimbles in crossing the dotted line.}
 \label{moresing}
\end{figure}

Like a collision of two critical points in the absence of any
symmetry, this can happen naturally as a result of adjusting one
parameter. We simply consider the polynomial
$f(x,y)=(x^2+y^2)^2/2-\epsilon(x^2+y^2)$.  There is a critical
point  at $x=y=0$ and a critical orbit $x^2+y^2=\epsilon$. We
denote these as $p$ and $\O$ respectively. $\O$ contains real
points if $\epsilon>0$ but not if $\epsilon<0$.  As a real
critical orbit, $\O$ thus disappears in passing through
$\epsilon=0$.   Setting $\I=i\lambda f$ with $\lambda>0$, the
values of $\I$ on the critical orbits are $\I_p=0$ and
$\I_\O=-i\lambda\epsilon^2/2$. So the Stokes curves are given by
$\re\,\epsilon^2=0$, which corresponds to the pair of lines
$\epsilon=r\exp(\pm \pi i/2)$, with real $r$ (fig.
\ref{moresing}). As expected, they meet at the singularity at
$\epsilon=0$.

A simple scaling reduces  $I(\lambda,\epsilon)
=\int \d x\,\d y \,\exp\biggl(i\lambda\bigl((x^2+y^2)^2-\epsilon(x^2+y^2)\bigr)\biggr)$ to a function of one
variable.  Setting $x,y\to \lambda^{-1/4}x,\lambda^{-1/4}y$, we get
\begin{equation}\label{bombo} I(\lambda,\epsilon)=\lambda^{-1/2}I(1,\lambda^{1/2}\epsilon)
=\lambda^{-1/2}\int\d x\,\d y \,\exp\biggl(i\bigl((x^2+y^2)^2-\lambda^{1/2}\epsilon(x^2+y^2)\bigr)\biggr).
\end{equation}    
Setting $\epsilon=0$, a special case is $I(\lambda,0)\sim \lambda^{-1/2}$.  Let us compare
this to the large $\lambda$ behavior for $\epsilon\not=0$.  Assuming
that (as in our application in eqn. (\ref{omsky})), the dominant contribution comes from
the isolated, nondegenerate critical point at $x=y=0$, a simple
scaling shows that $I(\lambda,\epsilon)\sim 1/\lambda$ for $\epsilon\not=0$, so the analog of (\ref{porkoff}) is
\begin{equation}\label{orkoff} \frac{I(\lambda,0)}{I(\lambda,\epsilon)|_{\epsilon\not=0}}
\sim \lambda^{1/2},~~\lambda\to\infty.\end{equation}
By introducing polar coordinates in the $xy$ plane, $I(\lambda,\epsilon)$ can be expressed
in terms of the error function.

The Lefschetz thimbles associated to critical orbits can be explicitly described
in this example,  because the symmetry can be used
to reduce everything to complex dimension 1.  Picking the Kahler metric $\d s^2=|\d x|^2+
|\d y|^2$, the moment map for the action of $SO(2)$ becomes $\mu=-i(x\bar y-y\bar x)$.
Setting $\mu=0$, we find that by an $SO(2)$ rotation we can set $y=0$.  The Lefschetz
thimbles that lie at $\mu=0$ are therefore surfaces of revolution that can be constructed by acting with the
group $SO(2)$ on a real curve in the complex  $x$ plane.  For the rest of our analysis, we scale 
$x$ to set $\lambda=1$.

\begin{figure}
 \begin{center}
   \includegraphics[width=5.5in]{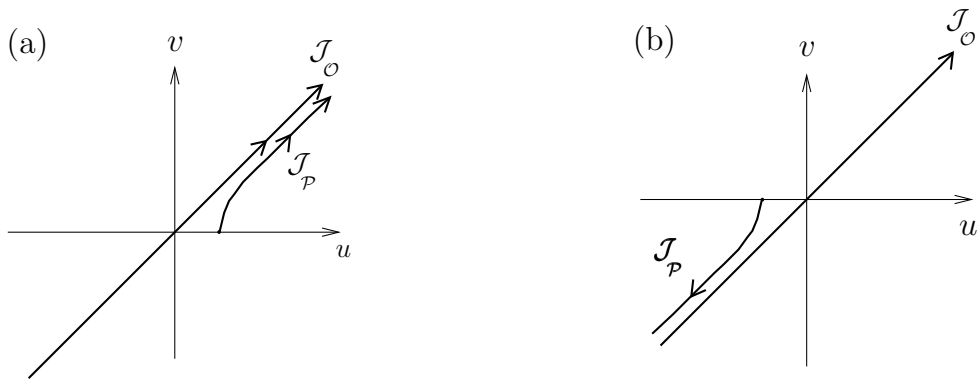}
 \end{center}
\caption{\small
For real $\epsilon$, the cycle $\J_p$ is derived from a curve in the $uv$ plane that starts at $(u,v)=(-\epsilon,0)$
and consists of half of one branch of the hyperboloid $u^2-v^2=\epsilon^2$.  We pick the half
that is asymptotic to the curve $u=v$.  For $\epsilon<0$, the relevant half is asymptotic to
$u=v=\infty$, as sketched in (a).  For $\epsilon>0$, it is asymptotic to $u=v=-\infty$, as sketched in
(b).  Accordingly, the cycle $\J_p$ jumps when $\epsilon$ changes sign.
The jumping is by $\J_p\to\J_p+\J_\O$, where $\J_\O$ is associated to the line $u=v$ in the
$uv$ plane, irrespective of the sign of $\epsilon$.
 In other words, $\J_p$ in (a) or $\J_p+\J_\O$ in
(b) is a path from $(u,v)=(-\epsilon,0)$ to $u=v=\infty$.
 }
 \label{directions}
\end{figure}

The flow equation has a conserved quantity $\mathrm{Im}\,(i f)$,
so to find the right curve in the $x$-plane, we do not need to
solve flow equations; it suffices to set $\mathrm{Im}\,(i f)$ to a constant.  Equivalently,
we must set
$\mathrm{Re}\,(x^2-\epsilon)^2$ to a constant $c$.  Setting
$x^2=\epsilon+u+iv$, the condition becomes $u^2-v^2=c$.  For the
Lefschetz thimble $\J_\O$ that flows down from the critical orbit
$x^2+y^2=\epsilon$, we set $c=0$, and choose the branch $u=v$ to
ensure that the real part of $if(x,y)$ is negative.   The
Lefschetz thimble $\J_p$ that flows down from the critical point
$x=y=0$ passes through $\epsilon+u+iv=0$, so the constant $u^2-v^2=\re((u+iv)^2)$ is
equal to $\re(\epsilon^2)$.
 The hyperboloid
$u^2-v^2=\mathrm{Re}\,(\epsilon^2)$ has two branches.  We pick the
branch that passes through $(u,v)=(-\epsilon,0)$ (in other words, $x=0$),
and then $\J_p$ is derived from the half of this branch that is
asymptotic to $u=v$.  As shown in fig. \ref{directions}, this
branch is asymptotic to $u,v>>0$ or $u,v<<0$ depending on
$\epsilon$.  In particular, if $\epsilon$ is real, $\J_p$ has
different asymptotic behavior depending on the sign of $\epsilon$.

From the figure, we can read off that between $\epsilon<0$ and
$\epsilon>0$, the Lefschetz thimbles change by
\begin{align}\label{klp} \J_p & \to \J_p+\J_\O   \cr \J_\O&\to\J_\O.  \end{align}
The meaning of the second statement is that the curve $\J_\O$,
which is the diagonal $u=v$, is unchanged in going from fig.
\ref{directions}(a) to fig. \ref{directions}(b).  But $\J_p$ in
fig. \ref{directions}(a) is a path from $(u,v)=(-\epsilon,0)$ to
$u=v=+\infty$, while in fig. \ref{directions}, it runs from
$(-\epsilon,0)$ to $u=v=-\infty$. In fig. \ref{directions}(b), to
get a path from $(-\epsilon,0)$ to $(+\infty,+\infty)$, we need to
take the sum $\J_p+\J_\O$. So $\J_p$ jumps in passing through
$\epsilon=0$, though $\J_\O$ does not.  Invariance of $\J_\O$ has
a more general explanation that was described in section
\ref{moreg}.

To recover the Lefschetz thimbles $\J_p$ and $\J_\O$, we have to ``rotate'' (via $SO(2)$)
the curves  that are sketched in fig. \ref{directions}.  In the
case of $\J_p$, the relevant curve is topologically a half-line whose endpoint is $SO(2)$-invariant.
Rotating it by $SO(2)$, we find that
as a manifold with $SO(2)$ action, $\J_p$ is equivalent to $\R^2$;
$SO(2)$ acts on $\J_p$ with a fixed point at the origin that corresponds to the endpoint of the half-line.  On the other hand, in the case of $\J_\O$, the relevant curve is a copy of $\R$, with
its ends at infinity.  Rotating it by $SO(2)$, we find that as a manifold with $SO(2)$ action, $\J_\O$ is equivalent
to $\R\times S^1$ with a free action of $SO(2)$ on the second factor.

It is convenient to express (\ref{klp}) in a dual language. Given a general cycle
$\CC=\n_p \J_p+\n_\O\J_\O$, to compensate for the jumping (\ref{klp}) and ensure
that $\CC$ varies smoothly, the coefficients jump by
\begin{align}\label{zlp}\n_p& \to \n_p\cr
                                              \n_\O& \to -\n_p+\n_\O.\end{align}

Finally, let us return to the Stokes lines of fig. \ref{moresing}.
We write $l_+$ and $l_-$ for the two Stokes lines given by
$\epsilon=r\exp(\pm i\pi/2)$ with real $r$. Scaling to
$\lambda=1$, the Morse function $\re\,(if)$ vanishes at $p$ and
equals $\pm r^2/2$ on $l_\pm$.  So on $l_+$, there can be a
downward flow from $\O$ to $p$, and on $l_-$ there can be a
downward flow from $p$ to $\O$.  Flows of each type actually
exist, as the reader can verify by slightly extending the analysis
of fig. \ref{directions}.  ($\J_\O$ always corresponds to the line
$u=v$, and $\J_p$ to part of a hyperboloid through
$u+iv=-\epsilon$.)    In general, the jumping in eqn. (\ref{klp})
occurs in crossing $l_-$, but there is no such jumping in crossing
$l_+$. The algebraic invariant that ``counts'' the downward flows
from $\O$ to $p$ along $l_+$ vanishes (even though such flows
exist), for a reason explained in section \ref{moreg}.

\subsubsection{One More Case}\label{onemore}

The last singularity that we will encounter is more surprising.  We will describe it
first in a language with
$SO(1,1)$ rather than $SO(2)$ symmetry.  We consider $SO(1,1)$ acting on real variables $u,v$
by $u\to ru$, $v\to r^{-1}v$, with $r$ real.  The basic invariant
is $uv$.  We consider an integral $\int\d u\,\d v\exp(i\lambda f(uv))$ where $f$ is a function of
$uv$.  The obvious case is that $f$ is a quadratic function.  However, and surprisingly, in the world of knots in $S^3$,
there can be a natural reason for $f$ to be an odd function of $uv$.

Under these conditions, the basic example of a function exhibiting critical behavior is
\begin{equation}\label{corn}f(u,v)=(uv)^3/3-\epsilon (uv).\end{equation}
Apart from the critical point $p$ at $u=v=0$, there are two critical orbits $\O_\pm$
at $uv=\pm\epsilon^{1/2}$.  For real $\epsilon$, these critical orbits are real
or imaginary depending on the sign of $\epsilon$.   (As real orbits, these orbits are hyperboloids
with two topological components.)

A closely related problem is the case of $SO(2)$ action on the $xy$ plane with
$f(x,y)=(x^2+y^2)^3/3-\epsilon(x^2+y^2)$.   The relation becomes apparent upon
complexification; one problem can be mapped into the other by taking $u=x+iy$, $v=x-iy$.
The problem of $SO(1,1)$ acting on the real $uv$ plane and the problem of $SO(2)$ acting
on the real $xy$ plane correspond to two real slices of the same situation with complex
variables.  One real slice is defined by  $u=\bar u$, $v=\bar v$ and the other by
$u=\bar v$.  To study analytic continuation, as usual we must complexify the
variables $u$ and $v$, and then the $SO(1,1)$ and $SO(2)$
problems become equivalent. It is convenient to work with $x$ and
$y$.  (We introduced $u$ and $v$ because the real slice with $SO(1,1)$ symmetry
will appear naturally in our application.)

First, we will determine the $\lambda$ dependence of $I(\lambda,\epsilon)=\int\d x\,\,\d y
\,\exp(i\lambda f(x,y))$ using scaling symmetry.
An obvious scaling of $x$ and $y$ shows that
\begin{equation}\label{donko}I(\lambda,\epsilon)=\lambda^{-1/3}I(1,\lambda^{2/3}\epsilon)
=\lambda^{-1/3}\int\d x\,\d y\exp(i((x^2+y^2)^3/3-\lambda^{2/3}\epsilon(x^2+y^2)).\end{equation}
A special case is that $I(\lambda,0)\sim\lambda^{-1/3}$.  
  On the other hand, for
$\epsilon\not=0$, and choosing $\epsilon<0$ so that the only real
critical point is the one at $x=y=0$, a similar scaling gives
$I(\lambda,\epsilon)\sim \lambda^{-1}$.  So
\begin{equation}\label{zorkoff} \frac{I(\lambda,0)}{I(\lambda,\epsilon)|_{\epsilon<0}}
\sim \lambda^{2/3},~~\lambda\to\infty.\end{equation}
By introducing polar coordinates in the $xy$ plane, one can express $I(\lambda,\epsilon)$
as a multiple of the Airy function.  The main difference is that there is an extra possible
integration cycle for $I(\lambda,\epsilon)$, as we now explain.

To study the analytic continuation systematically, we consider the integration cycles
associated to critical orbits.
 Let $\J_0$, $\J_+$, and $\J_-$ be the Lefschetz thimbles
associated, respectively, to the critical point at $x=y=0$ and to
the critical orbits $x^2+y^2=\pm\epsilon^{1/2}$. The same
reasoning as in section \ref{sotwo} can be used to describe them
by applying an $SO(2)$ rotation to suitable curves in the complex
$x^2$-plane. In fact, apart from a substitution $x\to x^2$, the
polynomial $f(x,0)=(x^2)^3/3-\epsilon x^2$ is familiar from our
study of the Airy function. Accordingly, the Lefschetz thimbles
$\J_\pm$ obtained by flowing from the critical orbits $\O_\pm$,
can be derived from the curves that were depicted in fig.
\ref{Airycycles}, but now in the $x^2$ plane.  On the other hand,
the third Lefschetz thimble $\J_0$ is associated to a similar
curve connecting the point $x^2=0$ to infinity.  By an analysis
similar to that of section \ref{sotwo}, one can determine the
jumping that occurs between $\epsilon>0$ and $\epsilon<0$.    The
change in $\J_\pm$ is precisely the same as it is for the Airy
function; what is new is that $\J_0$ picks up a linear combination
of $\J_\pm$.   We leave the details to the reader.

\subsection{Analytic Continuation In $k$: A Prototype}\label{anak}

We have discussed at great length the analytic continuation of the
integral of an exponential $\exp(i\lambda f)$ from real to complex
values of $\lambda$.  For a more precise analog of Chern-Simons
theory, we should consider analytic continuation of an {\it
integer} parameter (such as the Chern-Simons level $k$) to complex
values.

For an example, we consider the integral  representation of the Bessel function:
\begin{equation}\label{besself}I(\k,\lambda)=\frac{1}{2\pi i}\oint\frac{\d z}{z}
z^{\k} \exp(\lambda
(z-z^{-1})),\end{equation}
with an integer $\k$ that in our analysis of section \ref{anajones} will correspond roughly\footnote{In
that analysis, $\lambda$ corresponds roughly to the index $n$ of the colored Jones
polynomial $J_n(q)$, so the analogy would be closer if also $\lambda$ were constrained
to be an integer.}
to the Chern-Simons level $k$.

\begin{figure}
 \begin{center}
   \includegraphics[width=3in]{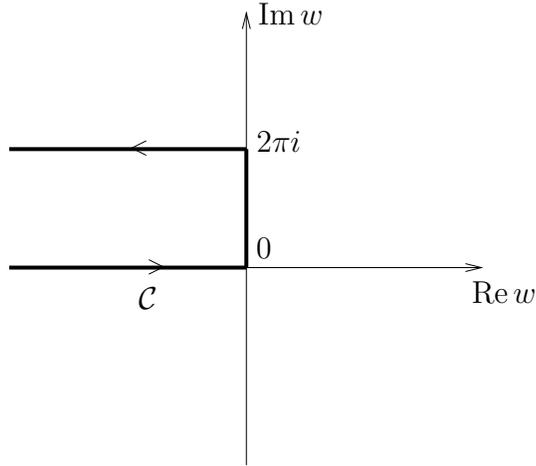}
 \end{center}
\caption{\small The contour $\CC$ in the complex $t$-plane that is used in analytic continuation
of the Bessel function.}
 \label{Contour}
\end{figure}

We take the integration cycle to be the circle $|z|=1$.    Alternatively, we can write
\begin{equation}\label{nessel}I(\k,\lambda)=\frac{1}{2\pi i}\oint\frac{\d z}{z}  \exp(\I),\end{equation}
with
\begin{equation}\I=\lambda
(z-z^{-1})+\k\ln z.\end{equation} For real $\lambda$ and integral
$\k$, this integral is an oscillatory integral of the general type
that we have been considering, except that the integration cycle
is compact, so that  analytic continuation in $\lambda$ poses no
problem.  However, we would like to analytically continue the
integral to complex values of $\k$.   This is a rough analog of
analytically continuing Chern-Simons gauge theory with respect to
the level $k$.

Let us recall the steps that were described somewhat abstractly in section \ref{moredown}.
The holomorphic form that we are trying to integrate is naturally defined in the punctured
$z$-plane, with the point $z=0$ omitted.  To analytically continue in $\k$, we want to lift
the integral from the punctured $z$-plane to its universal cover, which is the complex $w$-plane
where $z=e^w$.  The integration cycle $|z|=1$ can be lifted to the line segment in the $w$-plane  from $\alpha i$ to $(2\pi+\alpha)i$, for an arbitrary  $\alpha$.
Picking $\alpha=0$,  the integral
becomes
\begin{equation}\label{essel} I(\k,\lambda)=\frac{1}{2\pi i}\int_0^{2\pi i }\d w\,\exp(2\lambda
\sinh w +\k w).\end{equation} The integration is now over an open
contour whose ends are the two points $w=0$ and $w=2\pi i$. If we
stop here, Bessel's equation (whose proof requires integration by
parts), will fail when $\kappa\notin\Z$.   An analogous procedure
in quantum field theory would lead to a failure of the Ward
identities. Instead, following a classical procedure, and assuming that
$\re\,\lambda>0$, we add to the contour two more half lines given
(fig.\ref{Contour}) by the negative real axis and its translate
by $2\pi i$, suitably oriented to give
 a closed although noncompact integration contour $\CC$, which represents a cycle in the
 appropriate relative homology.  The integral over this
 cycle \begin{equation}\label{wessel} I(\k,\lambda)=\frac{1}{2\pi i}\int_\CC\d w\,\exp(2\lambda
\sinh w +\k w)\end{equation} converges (if $\re\,\lambda>0$), and in this integral,
 $\kappa$
can now be treated as a complex variable. This gives an analytic
continuation of the Bessel function to complex values of $\k$,
since  if $\k$ is an integer,  the two half lines that have
been added to the integration contour cancel each other and the
extended integral reduces to the original one. To analytically
continue beyond the region $\re\,\lambda>0$, one must shift $\CC$
vertically in the $w$-plane as the argument of $\lambda$ varies.
But we will mainly study the dependence on $\k$ with $\re\,\lambda
>0$.

If $\lambda$ and $\kappa$ are real and obey suitable inequalities
($\lambda$ and $2\lambda+\kappa$ should be positive), the two
half-lines that we added to complete the integration contour are
precisely as described in section \ref{moredown}: they are the
downward flow lines from the endpoints $w=0$ and $w=2\pi i$ of the
naive integration contour in the complex $w$ plane.  If we vary
$\lambda$ and $\kappa$ slightly, the flow lines move but without
changing the relative homology class of the integration cycle
$\CC$.

We have described a simple procedure for analytic continuation away from integer
values of  $\k$, but there
is no avoiding a basic fact discussed  in
section \ref{gench}: analytic continuation away from integer values is not unique.   For an illustrative special
case of this, note that we could have begun by lifting the circle
$|z|=1$ to a line segment in the $w$-plane connecting the points
$w=\pm \pi i$. This would be analogous to a different choice of
$\alpha$ in section \ref{moredown}.  Provided $\lambda$ and
$\kappa$ are real and obey suitable inequalities ($\lambda$ and
$2\lambda+\kappa$ should be negative), the downward flow lines
from $w=\pm i$ are again half-lines parallel to the real axis, but
they extend in the opposite direction to $\re\,w>>0$.   The two
analytic continuations are qualitatively different. The original
Bessel integral (\ref{besself}) obeys
$I(-\kappa,\lambda)=(-1)^\kappa I(\kappa,\lambda)$, as follows
from the change of variables $z\leftrightarrow -1/z$.   This
symmetry is broken by the choice of whether to extend the contour
to $\re\,w<<0$, as in fig. \ref{Contour}, or to $\re\,w>>0$.  With
either choice, it turns out that when $\kappa$ is not an integer,
the large $\kappa$ behavior depends very much on the sign of
$\re\,\kappa$.

In the rest of our analysis, we will stick with the analytic continuation corresponding to the
contour $\CC$ of fig. \ref{Contour}.
We want to study the behavior  of the integral (\ref{wessel}) as $\kappa$  and $\lambda$ become
large with a fixed value of $\rho=\k/2\lambda$.  First we will
learn what we can by inspection, taking $\kappa$ real and
$\lambda$ positive, and then we will reconsider the matter using
Morse theory.

The critical point equation is $z^2+2\rho z+1=0$ or $\cosh w+\rho
= 0$ and the critical points are at
\begin{equation}\label{orf}z=-\rho\pm i\sqrt{1-\rho^2}. \end{equation}
or
\begin{equation}\label{potoc}w=\pm{\mathrm{arccosh}}
(-\rho)+2\pi i n=\ln(-\rho\pm\sqrt{\rho^2-1})+2\pi i n,~~
n\in\Z.  \end{equation}

First let us suppose that $\k$ is an integer, so that
$\I_{\k,\lambda}$ can be defined by the original integral on the
contour $|z|=1$, and we can use (\ref{orf}) to identify the
critical points. If $|\rho|\leq 1$, the critical points are at
$|z|=1$, on the integration contour.  They make oscillatory
contributions to the integral, and this gives the behavior of the
integral for large $\k$ and fixed $\rho$ in this region.

Now, still with integer $\kappa$, suppose that $|\rho|>1$.  The
critical points move off the integration contour $|z|=1$, and  the
integral therefore vanishes exponentially.

Next, suppose that $\k$ is not an integer.  We must consider the
effects of the two half-lines that have been added to the
integration contour.  If $|\rho|\leq 1$, the addition of these
half-lines does not lead to a qualitative change; the integral is
still dominated by oscillatory contributions from critical points
at $|z|=1$. We must be more careful if $|\rho|>1$, and the two
cases $\rho>1$ and $\rho<-1$ turn out to be completely
different.\footnote{This difference results entirely from the
choice of how to analytically continue the function
$I(\k,\lambda)$ to nonintegral values of $\k$, since for $\kappa\in\Z$, the  Bessel function
has  a symmetry $I(-\kappa,\lambda)=(-1)^\kappa I(\kappa,\lambda)$, as already noted.} If
$\rho>1$, there is no critical point on the integration contour
and $I(\kappa,\lambda)$ vanishes exponentially whether $\k$ is an
integer or not. (The most precise analysis is by a saddle point
analysis that we describe shortly.)   For $\rho<-1$ (which means
negative $\k$, since we have taken $\lambda>0$), the behavior is
more delicate.  There are two critical points on our chosen integration
cycle $\CC$, namely at $w=\ln(-\rho-\sqrt{\rho^2-1})$ and
$w=\ln(-\rho-\sqrt{\rho^2-1})+2\pi i$.  The two critical points
have the same positive value of $h=\re\,\I$, and they both make
contributions that are exponentially large for $\k\to-\infty$.
(Indeed, $h\sim -\k f(\rho)$ for  $-\k$ large, where the function
$f(\rho)=(\rho^2-1)^{1/2}/\rho+\ln(-\rho+(\rho^2-1)^{1/2})$ is
positive for $\rho<-1$.) On the other hand, $\im\,\I$ vanishes for
one critical point and is $\k\,\im \,w=2\pi \k$ at the other.   Allowing for
the opposite orientations of the two half-lines that are part of
the contour $\CC$, the two real critical points contribute
\begin{equation}\label{hosedd} \exp(-\kappa f(\rho))\left(1-\exp(2\pi i \k)\right) \end{equation}
times the usual sort of asymptotic series.  For integer $\kappa$,
these contributions cancel, but for generic $\k<<0$, their sum
grows exponentially with $|\k|$.

\subsubsection{Description By Lefschetz Thimbles}\label{deflef}

Finally, we will seek a more precise description by expressing the cycle $\CC$ as a sum
of Lefschetz thimbles $\J_\sigma$ associated to critical points $p_\sigma$.   Let us
formulate the most parsimonious hypothesis that is consistent with the observations above:

(1) For $\rho>1$, it is reasonable to guess that  $\CC$
is equivalent to the Lefschetz thimble associated to the critical
point at $w=\ln(-\rho+\sqrt{\rho^2-1})$ ($w$ has negative real
part and we pick the branch of the logarithm such that
$\mathrm{Im}\,w=\pi$). This is the unique critical point enclosed
by $\CC$ if $\rho>1$, and as $h$ is negative at this critical
point, the associated Lefschetz thimble  makes an exponentially small contribution to the
integral, as expected.

(2) For $1>\rho>-1$, the two critical points at $|z|=1$ must certainly be included.   There
is no obvious need for more.

(3) For $\rho<-1$, the situation is more tricky.  The two real
critical points that account for the exponentially growing
asymptotics in (\ref{hosedd}) must certainly contribute when $\CC$
is expressed in terms of Lefschetz thimbles.  But as they cancel
when $\k$ is an integer, while $I(\kappa,\lambda)$ is not zero in
that case, the expression for $\CC$ in terms of Lefschetz thimbles
must receive a contribution from at least one more critical point.
This critical point must have $h<0$, since $I(\kappa,\lambda)$ is
exponentially small for integer $\kappa$ and $\rho<-1$.

\begin{figure}
 \begin{center}
   \includegraphics[width=5in]{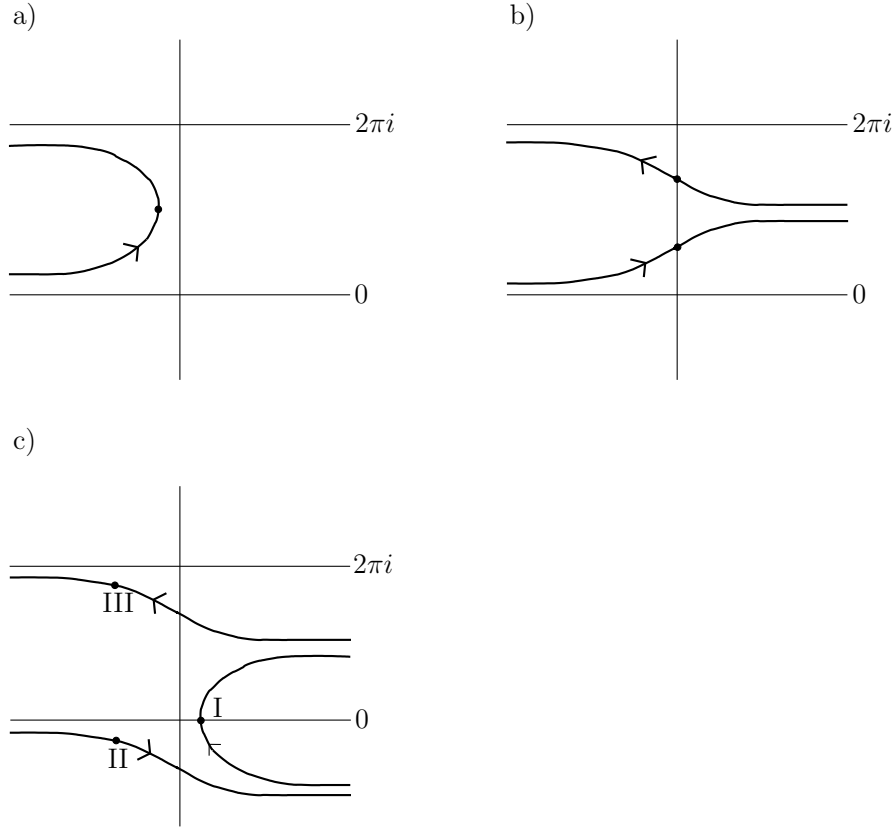}
 \end{center}
\caption{\small   Qualitative behavior of Lefschetz thimbles in analytic continuation of the
Bessel function, for (a) $\rho>1$ , (b) $1>\rho>-1$, and (c) $-1>\rho$.    In each case, the critical
points are marked by black dots.  In (a), the integration cycle of fig. \ref{Contour} is
equivalent to a single Lefschetz thimble associated to a complex critical point that
makes an exponentially decaying contribution to the integral.   In (b), it is equivalent to the
sum of two Lefschetz thimbles that make oscillatory contributions.  And in (c), it is equivalent
to the sum of a Lefschetz thimble associated to a critical point (labeled as I) that makes an exponentially
decaying contribution and two critical points (labeled II and III) whose contributions are exponentially
growing, but cancel if $\k$ is an integer.  In (c), $\rho$ has
been given a small imaginary part to avoid a Stokes line.}
 \label{bessel}
\end{figure}

In summary, in regions (1), (2), or (3), $\CC$ must be a sum of at
least 1, 2, or 3 critical point contours, respectively.  This most
parsimonious interpretation is actually correct, as one can see
from fig.\ref{bessel}, where the relevant Lefschetz thimbles are
sketched.

\subsubsection{Singularities At $\rho=\pm 1$}\label{singrho}

The most interesting values of $\rho$ are at $\rho =\pm 1$, where critical points meet.
Let us first examine the behavior at these points on the $z$-plane, and then on the $w$-plane.
The analysis on the $z$-plane is simpler, but it only tells us what happens when $\kappa$ is
an integer.  To analyze the behavior for non-integral $\kappa$, we need to work on the $w$-plane.

There are precisely two critical points on the $z$-plane, located at $z=-\rho\pm i\sqrt{1-\rho^2}$.
Precisely for $\rho=\pm 1$, the two critical points meet.  This is reminiscent of the most basic
example of critical behavior, the cubic polynomial $f(x)=x^3/3-\epsilon x$ related to the Airy
function, whose two critical
points meet at $\epsilon =0$.
The critical points of $f$ are real for $\epsilon>0$ and form a complex conjugate pair
for $\epsilon<0$.  The consequences of the singularity at $\epsilon=0$ were explored in
section \ref{twocrit}.

Our problem on the $z$-plane is exactly analogous.  We endow the punctured $z$-plane
with the antiholomorphic involution $z\to 1/\bar z$, so that the real points (the fixed points of the
involution) are the circle $|z|=1$.  The two critical points at $z=-\rho\pm i\sqrt{1-\rho^2}$
are real for $\rho$ real and $|\rho|<1$ but form a complex conjugate pair for $\rho$ real and
 $|\rho|>1$.  The singularities at
$\rho=\pm 1$ are  just analogous to the Airy function singularity at $\epsilon=0$.

As long as $\kappa$ is an integer and we work on the $z$-plane,
the behavior of the Bessel integral  that we found in section \ref{deflef} precisely mirrors the behavior of the Airy
integral
\begin{equation}\label{dolf}I(\epsilon)=\int_{-\infty}^\infty\d x\,\exp\left(i(x^3/3-\epsilon x)\right).
\end{equation}
  Region (2) of the Bessel
integral  corresponds to the case $\epsilon>0$ for the Airy integral: the integration cycle
is equivalent to the sum of two Lefschetz thimbles associated to two real critical points.
Regions (1) and (3) of the Bessel integral correspond to the case $\epsilon<0$ for the Airy
integral: the integral is equivalent to an integral over a single Lefschetz thimble that makes
an exponentially small contribution.  In section \ref{deflef}, we found that three critical points
are needed in region (3) to describe the integral on the $w$-plane, but two of them (the ones that make exponentially growing contributions) cancel upon projection to the $z$-plane.

The analysis on the $w$-plane is more complicated because each critical point on the $z$-plane
lifts to infinitely many critical points on the $w$-plane (differing by $w\to w+2\pi i n$, $n\in\Z$).
Nevertheless, it is very illuminating to repeat the analysis of the Bessel function emphasizing
its interpretation in terms of Airy function singularities:

$(1')$  We start for $\rho>1$ with a single critical point
at $\tilde w_-=\ln\left(-\rho+\sqrt{\rho^2-1}\right)$,
 picking the branch
 of the logarithm so that $\tilde w_-=\pi i$ at $\rho=1$.    (Choosing a different branch would multiply
 the analytically continued Bessel integral by an integer power of $\exp(2\pi i \kappa)$.)
The name $\tilde w_-$ is motivated by the fact that in section \ref{twocrit}, a complex critical point that makes an exponentially small contribution to the Airy function
 for $\epsilon<0$ was called $\tilde p_-$.  As in section \ref{twocrit},
 we write  $\tilde\J_-$ for  the
 Lefschetz thimble corresponding to this critical point.  For $\rho>1$, as we have learned in section \ref{deflef},
 the integration cycle $\CC$ that
 we have used to analytically continue the Bessel function is equivalent to $\tilde\J_-$.

$ (2')$ The $\rho>1$ critical point at $\tilde w_-=\ln(-\rho+\sqrt{\rho^2-1})$, when continued
to $\rho=1$, meets
 two real critical points that for $\rho<1$ are located at $w_\pm=\ln(-\rho\pm i\sqrt{1-\rho^2})$.
 Here branches of the logarithm are chosen so that at $\rho=1$, we have $w_+=w_-=\tilde w_-=\pi i$.
  As in the Airy case, we denote
 as $\J_+$ and $\J_-$ the Lefschetz thimbles that correspond to the real critical points $w_\pm$.
We can use the Airy relation (\ref{zelign}) to express the region (1) integration cycle
$\tilde \J_-$ in terms of Lefschetz thimbles $\J_\pm$ appropriate to region (2).
  The relation is $\tilde\J_-=\J_++\J_-$.  So the right integration cycle in region (2) is
  $\J_++\J_-$.

  $(3')$ So far working on the $w$-plane instead of the $z$-plane has not given anything essentially new.  Now we study how the critical points $w_\pm$ evolve as we vary
  $\rho$ from 1 to $-1$.  We have chosen branches of the logarithm so that
  $w_\pm = \ln(-\rho\pm i\sqrt{1-\rho^2})$ are both equal to $\pi i$ at $\rho=1$.  However, when continued
  (along the real $\rho$ axis)
  to $\rho=-1$, they are not equal.  Rather,
  \begin{equation}\label{zury}  w_+(\rho=-1)=0,~w_-(\rho=-1)=2\pi i.\end{equation}

  Hence, we cannot analyze what is happening at $\rho=-1$ in terms of a single Airy singularity.
We must consider two different Airy singularities, located at
$w=0$ and at $w=2\pi i$.   This will make the analysis a little longer,
though the basic idea is just to use what we know about the Airy integral.

Near $w=0$, let us write $\J'_\pm$ for the
Lefschetz thimbles associated to real critical points at
$\rho>-1$, and $\tilde\J'_\pm$ for the Lefschetz thimbles
associated to complex critical points at $\rho<-1$.  We write
$\J''_\pm$ and $\tilde \J''_\pm$ for Lefschetz thimbles associated to the analogous real and
complex critical points near $w=2\pi i$. (The reason for the
notation $\J'$ and $\J''$ is to distinguish these cycles from the
ones we have already defined near $\rho=1$.)

We have already found that for $\rho$ just less than 1, the
integration cycle is $\J_++\J_-$.
 Continuing along the real axis to $\rho$ just greater than $-1$ with the aid of eqn. (\ref{zury}), we see that
$\J_++\J_-$ evolves to $\J_+'+\J_-''$.       To continue past $\rho=-1$, we use the Airy relations
(\ref{helign}), which give $\J_+'=-\tilde\J_+'+\tilde\J_-'$, and  $\J_-''=\tilde\J_+''$.  So
$\J_+'+\J_-''=-\tilde\J_+'+\tilde\J_+''+\tilde\J_-'$, and this is the integration cycle for $\rho<-1$.
  For integer $\kappa$, $-\tilde\J_+'$ and $\tilde\J_+''$ cancel
and the integration cycle is equivalent to $\tilde\J_-'$.  For general $\kappa$,  the contribution
from $-\tilde\J_+'+\tilde\J_+''$ dominates and is  described qualitatively in eqn. (\ref{hosedd}).

\begin{figure}
 \begin{center}
   \includegraphics[width=1.5in]{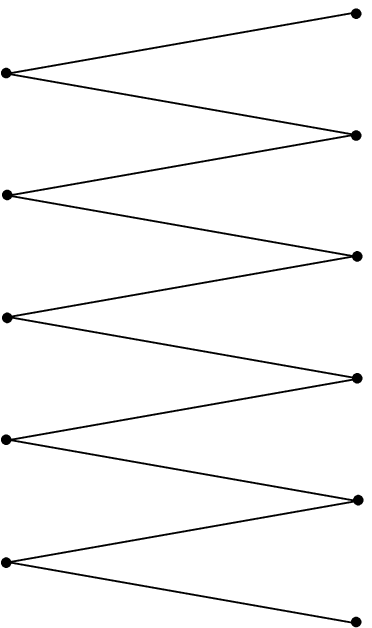}
 \end{center}
\caption{\small  Qualitative behavior of the conserved quantity $\im\,I/\k$ for  critical
points in the region $-1\leq \rho\leq 1$.  $\rho$ is plotted horizontally and $\im\,\I/\k$ vertically.
A black dot represents a pair of real critical points that become coincident at $\rho=1$ or
$\rho=-1$. At $\rho=1$ (right), there are a pair of real critical
points for every value of $\im\,\I/\k$ of the form $(2w+1)\pi$, $w\in\Z$.  Reducing $\rho$, $\im\,\I/\k$
increases for one of these critical points and reduces for the other.  At $\rho=-1$, the critical
points recombine in pairs, but now with $\im\,\I/\k$ of the form $2\pi w$, $w\in\Z$.      In all cases,
these critical points are found by solving the equation $\cosh w=-\rho$. }
 \label{critflow}
\end{figure}

With a view to our application in section \ref{piecesofeight}, the
key point to remember is that the two real critical points on the
$z$-plane, if lifted to the $w$-plane so that they coincide at
$\rho=1$, do not coincide at $\rho=-1$.   The flow of critical points is sketched in fig. \ref{critflow}.

\section{Application To Chern-Simons Theory}\label{backtocs}

Can the framework of section \ref{stokes} be applied in the
infinite-dimensional context of Chern-Simons gauge theory?  The
critical points $p_\sigma$ certainly make sense. The
Euler-Lagrange equations simply say that the gauge field should be
flat (perhaps when restricted to the complement of a knot).
Moreover, by standard quantum field theory methods, one certainly
knows how to expand the path integral around any given critical
point, and hence in effect the Lefschetz thimbles $\J_\sigma$
are known at least perturbatively.

The aspects of this framework that are not standard in quantum field theory are the use of the
flow equation to define the critical point cycles nonperturbatively, to determine the coefficients
when the real integration cycle is expressed in terms of critical point cycles, making analytic
continuation possible, and to describe the Stokes jumping phenomena that are essential
for the consistency of the framework, once one aims for nonperturbative results.  All of these
questions involve the flow equations, so the question of whether the Morse theory
machinery can be applied to Chern-Simons theory is largely the question of whether
the flow equation for Chern-Simons makes sense and has properties similar
to what it has in finite dimensions.

This  is  the topic of section \ref{flowchern}.  The flow
equations in ordinary Morse theory are ordinary differential
equations that describe flow in an auxiliary ``time'' direction.
Chern-Simons theory is three-dimensional even before introducing
an auxiliary direction, so the flow equations become partial
differential equations in four dimensions. Prototypes for such
equations are the Yang-Mills instanton equations, which are
interpreted in Floer theory as flow equations for a real
connection $A$ with the Chern-Simons function $W(A)$ as a Morse
function. In that example, the flow equation has a
four-dimensional symmetry, and this is a hint of what we will find
in section \ref{flowchern}.

Our problem is analogous to the case just mentioned but is more
elaborate because we want  the flow for a complex connection $\A$
with a Morse function derived from the complex Chern-Simons
function $W(\A)$. Also our flow equations depend on a coupling
parameter. It turns out that the flow equations in our problem are
essentially equations that have appeared before, in the
application of supersymmetric gauge theory to the geometric
Langlands program \cite{KW}. The reason for this is not yet fully
clear, but at any rate the elliptic nature of the flow equations
for Chern-Simons theory means that they have properties similar to
the Morse theory flow equations with finitely  many variables.

\subsection{The Flow Equation For Chern-Simons}\label{flowchern}

\def\T{\mathcal T}
Let $M$ be a three-manifold.  Let $H$ be a compact Lie group and
set $G=H_\C$. Let $\T$ be the space of connections $A$ on a fixed
$H$-bundle $E\to M$, and let $\T_\C$ be the space of
complex-valued connections $\A$ on the complexification of $E$. We
can write $\A=A+i\phi$ where $A$ and $\phi$ are the real and
imaginary parts of $\A$; $A$ is an $\frak h$-valued connection,
and $\phi$ is a one-form on $M$ with values in the adjoint bundle
$\mathrm{ad}(E)$.  The curvature of $A$ will be denoted as $F=\d A+A\wedge A$,
and that of $\A$ as $\F=\d\A+\A\wedge \A$.

$\T_\C$ is a complex manifold; to define a flow equation, we need
to endow it with a Kahler metric. We pick a Riemannian metric $g$
on $M$, which determines a Hodge star
operator\footnote{\label{pink} If $\d x^1,\d x^2,\d x^3$ is an
oriented orthonormal frame, we pick the sign of
$\star_M$ so that $\star_M(\d x^1)=\d x^2 \wedge  \d x^3$, etc.
Similarly, later the star operator of $I\times M$ is defined so
that $\star(\d t\wedge\d x^1)=\d x^2\wedge \d x^3$, etc.}
$\star_M$. We define the metric on $\T$ by
\begin{equation}\label{kurof} \d
s^2=-\int_M\,\Tr\,\delta\,\bar\A\wedge \star_M \delta\A.\end{equation}
This is a Kahler metric, with Kahler form
\begin{equation}\label{urof}\omega=\int_M\Tr
\,\delta\phi\wedge\star_M \delta A.\end{equation} The metric $\d
s^2$ is invariant under $H$-valued local gauge transformations,
though not under $G$-valued ones.  The moment map for the
$H$-valued local gauge transformations is
\begin{equation}\label{zurof}\mu= \d_A\star_M\phi,\end{equation}
where $\d_A=\d+[A,\cdot]$ is the covariant derivative with respect
to $A$.

Critical ``points'' of the Chern-Simons function $W(\A)$, understood as a function
on the space of connections, actually form orbits
under the group of complex gauge transformations.
For reasons described in section \ref{gaction}, we will only consider critical orbits
that are semistable, meaning that they admit a locus with $\mu=0$.  The condition for a flat
connection $\A$, that is a connection obeying $0=\F=\d\A+\A\wedge\A$, to
be gauge-equivalent to a connection for which $\mu=0$ is \cite{Corlette} that the
holonomy of the flat connection $\A$ is not strictly triangular.  In other words, if the holonomies
of $\A$ around one-cycles in $M$ can be simultaneously put in upper triangular form
\begin{equation}\label{gonfry}\begin{pmatrix}\alpha & \beta\cr 0 & \gamma\end{pmatrix},
\end{equation}
then the upper right elements $\beta$ all vanish.  Strictly triangular flat connections
are unstable and we omit them.  (A block diagonal flat connection with $\beta=0$ is considered
semistable, and one that cannot be put in a triangular form at all is stable.)

The flow equation for a Morse function that is the real part of a
holomorphic function $\I$ is
\begin{equation}\label{purof}\frac{\d\A}{\d
t}=-\star_M\frac{\delta\,\I}{\delta\bar\A}.\end{equation} We are
interested in the case that $\I=\lambda\,W(\A)$ where $W$ is the
Chern-Simons functional and $\lambda$ is a complex number.  By
rescaling $t$, we can reduce to the case that $\lambda=2\pi\exp(-i\alpha)$ with
real $\alpha$.  The flow equation then becomes
\begin{equation}\label{kurtof}\frac{\d\A}{\d
t}=-\exp(-i\alpha)\star_M\bar\F,\end{equation}
where $\bar\F$ is the complex conjugate of the curvature
$\F=\d\A+\A\wedge\A$. More explicitly, this is
\begin{align}\label{urtof}\frac{\d A}{\d
t}&=-\star_M\bigl(\cos\alpha(F-\phi\wedge\phi)-\sin\alpha\,d_A\phi\bigr)\cr
\frac{\d\phi}{\d t}&=\star_M\bigl(\sin\alpha(F-\phi\wedge\phi)+\cos\alpha\,\d_A\phi\bigr).\end{align}

The time $t$ parametrizes a real one-manifold $I$ (which is the
real line or a portion of it). We pull back $E$ to a bundle over
$I\times M$ and view $A$ as a connection on this pullback.  As
such, $A$ has vanishing component in the $I$ direction. We view this as a gauge condition.  Relaxing this gauge condition, we
can write $\d A_i/\d t$ ($i$ is an index tangent to $M$) in
gauge-invariant language as a component $F_{ti}$ of the curvature
$F=\d A+A\wedge A$.  Similarly, we view $\phi$ as an
$\mathrm{ad}(E)$-valued one-form on $I\times M$; as such, it is
not generic, since its component $\phi_t$ in the $I$ direction
vanishes.  We endow $I\times M$ with a Riemannian metric that is the
sum of the metric $\d t^2 $ on $I$ with the metric $g$ on $M$, and
we write simply $\star$ for the Hodge star operator on $I\times M$
(normalized as in footnote \ref{pink}). After taking linear
combinations of the equations in (\ref{urtof}), one finds that
these equations can be put in a form with four-dimensional
symmetry
\begin{align}\label{kweqns} (F-\phi\wedge\phi)^+&=u D\phi^+ \cr
             (F-\phi\wedge\phi)^-&=-u^{-1}D\phi^- ,\end{align}
with
\begin{equation}\label{tdef}
u=\frac{1-\cos\alpha}{\sin\alpha},~u^{-1}=\frac{1+\cos\alpha}{\sin\alpha}.\end{equation}
(For a two-form $v$, we define the projections $v^\pm=(v\pm \star v)/2$.) To be more precise, the four-dimensional symmetry is broken in
our context by the condition $\phi_t=0$.  But if we relax this
condition, the equations have full four-dimensional symmetry.

The moment map $\mu=\d_A\star_M\phi$ is as usual a conserved
quantity for the flow equations.  As explained in section
\ref{gaction}, we are primarily interested in solutions of the
flow equations with $\mu=0$.  Bearing in mind that we anyway  have
the condition $\phi_t=0$, $\d_A\star_M\phi$ can be written in
four-dimensional language as $\d_A\star \phi$.  So we want to
supplement the  equations (\ref{kweqns})  with an additional condition
\begin{equation}\label{zweq}\d_A\star\phi=0.  \end{equation}
Henceforth, the phrase ``flow equations'' will refer to this combined system of equations.

\def\V{{\mathcal V}}
The flow equations have been first encountered \cite{KW} in the gauge
theory approach to geometric Langlands.\footnote{The parameter $u$
is called $t$ in \cite{KW}, but here we prefer to avoid confusion
with the time coordinate.} These are elliptic equations, modulo
the gauge symmetry, and this means that their general properties
are much like those of flow equations for finitely many variables in ordinary Morse theory. The relevant properties
are stated in section (\ref{apflow}).

\def\Tr{\mathrm{Tr}}
The condition $\phi_t=0$ can be reinterpreted in terms of a
vanishing theorem similar to those of section 3.3 of \cite{KW}.
This argument proceeds as follows.  Set $\V^+(u)=(F-\phi\wedge\phi)^+-u D\phi^+$,
 $\V^-(u)=    (F-\phi\wedge\phi)^-+u^{-1}D\phi^-$, $\V_0=\d_A\star\phi$, so
 the flow equations are $\V^+(u)=\V^-(u)=\V_0=0$.  Now consider the following sum
 of squares of these equations:
 \begin{align}\label{ohw}&-\int_{I\times M}\,\Tr\left(\frac{u^{-1}}{u+u^{-1}}\V^+(u)\wedge\V^+(u)
 +\frac{u}{u+u^{-1}}\V^-(u)\wedge\V^-(u)+\V_0\wedge\star\V_0\right)
\cr &~~~ =-\int_{I\times M}\d^4x\sqrt g\,\Tr\left(\frac{1}{2}F_{\mu\nu}F^{\mu\nu}+D_\mu\phi_\nu D^\mu\phi^\nu
 +R_{\mu\nu}\phi^\mu\phi^\nu+\frac{1}{2}[\phi_\mu,\phi_\nu]^2\right)\cr
 &~~~~~~+\frac{u-u^{-1}}{u+u^{-1}}\int_{I\times M}\Tr\,F\wedge F.\end{align}
 (Indices $\mu,\nu$ are tangent to $I\times M$, and $R_{\mu\nu}$ is the Ricci tensor of
 $I\times M$, which is a pullback from $M$.)
 Clearly, the left hand side of (\ref{ohw}) is stationary when the flow equations are obeyed.
 Therefore the right hand side is also stationary.  The part of the right hand side that depends on
 $\phi_t$ is simply
 \begin{equation}\label{polyz}Z(\phi_t)=
 -\int_{I\times M}d^4x\sqrt g\,\Tr\left(\sum_\mu D_\mu\phi_t D^\mu\phi_t
+\sum_{\mu\not=t}[\phi_t,\phi_\mu]^2\right).\end{equation} The flow equations
therefore imply that $\phi_t$ obeys the Euler-Lagrange equations
derived from (\ref{polyz}), or
\begin{equation}\label{olyz}-\sum_\mu D_\mu D^\mu \phi_t+\sum_{\mu\not=t}[\phi_\mu,
[\phi_\mu,\phi_t]]=0.\end{equation}
(Of course, this can also be proved by differentiating the equations (\ref{kweqns}) and (\ref{zweq}).)
Provided that one can integrate by parts, which is true if $\phi_t$
obeys either Dirichlet or Neumann boundary conditions at each end of $I$,  (\ref{olyz}) implies
that $Z(\phi_t)=0$.  Given the definition of $Z(\phi_t)$ as a sum of squares,
this immediately implies
that $\phi_t$ is covariantly constant and commutes with all $\phi_\mu$.  Thus, $\phi_t$
generates a symmetry of the solution.

We will always choose boundary conditions such that $\phi_t$
vanishes at (at least) one end  of $I$, and integration by parts
is possible at the other end. The above argument shows that in a
solution of the flow equations obeying such boundary conditions,
$\phi_t$ is identically zero. There is no need to spoil the
four-dimensional symmetry, or the elliptic nature of the
equations, by postulating in advance that $\phi_t$ is identically
zero.  This is a consequence of the equations plus boundary
conditions.

\subsubsection{The Index}\label{napflow}

\def\rm{\mathrm}
\def\ZX{(X,\rm{ad}(E))}
\def\ade{\mathrm{ad}(E)}

For applications, we need to know the expected value of the
dimension of the moduli space $\M$ of solutions of the above
equations, with appropriate boundary conditions  and modulo gauge
transformations. This equals the index of the linear elliptic
operator $L$ that arises from those equations, after linearization
and gauge fixing.   We will show that on a four-manifold of the form
$I\times M$,  with certain types of boundary condition at the ends of $I$,
the index vanishes.   This  will
lead in section \ref{apflow} to behavior similar to that of Morse
theory with finitely many variables.

The index is independent of $u$, so it suffices to set $u=1$. The
relevant linear system  can be described as follows. We expand
around a classical solution $(A_0,\phi_0)$ by setting
$(A,\phi)=(A_0+\delta A,\phi_0+\delta\phi)$.  A convenient
gauge-fixing condition is $\d_{A_0}\star\delta A=0$. For a
four-manifold $X$, let  $\Omega^q_X(\ade)$ be the space of $q$-forms
on $X$ with values in $\mathrm{ad}(E)$.

After acting on $\delta\phi$ with the $\star$ operator, we can
think of $\delta A$ and $\delta\phi$ as elements of
$\Omega^1_X(\ade)$ and $\Omega^3_X(\ade)$, respectively.  We can view
the gauge-fixing condition as asserting the vanishing of the
element $\star\d_{A_0}\star \delta A$ of $\Omega^0_X(\ade)$, while
the equations $\V^+=\V^-=0$ assert the vanishing of the element
$\V^++\V^-$ of $\Omega^2_X(\ade)$, and the equation $\V_0=0$ asserts
the vanishing of an element of $\Omega^4_X(\ade).$

All told, our linear operator $L$ maps $\Omega^1_X(\ade)\oplus
\Omega^3_X(\ade)$ to $\Omega^0_X(\ade)\oplus\Omega^2_X(\ade)\oplus
\Omega^4_X(\ade)$. A familiar operator that does this is the operator
$L_0=\d_A+\d_A^*$ mapping $\rm{ad}(E)$-valued differential
forms of odd degree to those of even degree. $L$ is actually equal
to $L_0$ modulo zeroth order terms that do not affect the index.
As we explain momentarily,
 the index of the operator $L_0$, or of $L$,
is unaffected by the twisting by the gauge bundle $E$.  If $X$ is
compact, this index
 is simply the product $-\chi(X)\,{\mathrm{dim}}(H)$, where $\chi(X)$ is the
Euler characteristic of $X$ and $\mathrm{dim}(H)$ is the dimension
of the Lie group $H$. For $X=I\times M$, the topological Euler
characteristic of $X$ vanishes, and boundary contributions to the
index are also zero, so the index of $L$ vanishes.

The value of the index for compact $X$ and its vanishing for
$X=I\times M$ can be understood as follows.  Let $S_+$ and $S_-$
be the two spin bundles of $X$; each is of rank 2. It is
convenient to express the operator $L_0$ as a  Dirac operator
mapping sections of $S_+\otimes V$ to sections of $S_-\otimes V$
for some vector bundle or difference of vector bundles $V$;  the
index of $L_0$ can then be computed from the index theorem for the
Dirac operator. To put $L_0=\d_A+\d_A^*$ in the desired form, we
must take $V$ to be a formal difference of vector bundles
$V=(S_+\ominus S_-)\otimes \mathrm{ad}(E)$. Taking the formal
difference $S_+\ominus S_-$ where $S_+$ and $S_-$ have the same
rank causes all bulk and boundary contributions to the index that
depend on $E$ to cancel.  (As $X$ is a four-manifold and $E$ has vanishing first Chern
class, the only bulk $E$-dependent contribution is a multiple of the second Chern class of 
$E$; it cancels when we take the difference $S_+\ominus S_-$.  Boundary contributions
are further discussed in the next paragraph.)   Hence $\mathrm{ad}(E)$ can be
 replaced by a trivial bundle of the same rank,  namely
 $\mathrm{dim}(H)$, and $L_0$ then reduces to the sum of
 that number of  copies of the standard operator $\d+\d^*$ mapping
 ordinary differential forms of odd  degree to those of even
 degree. The index of this operator on a four-manifold without boundary is $-\chi(X)$.

Since the vanishing of boundary corrections to the index is
important, let us spell it out in more detail.  If $I$ has a
boundary at a finite distance -- for example the right boundary of
the half-line $(-\infty,0]$ -- then we will pick local boundary
conditions.  With local boundary conditions, there is no
$E$-dependent boundary contribution to the index, and as the metric of
$I\times M$ is a product, there is also no boundary contribution
involving the extrinsic curvature of the boundary.  Alternatively,
if $I$ has an ``end'' at infinity -- for example, the left end of
the half-line -- then we want normalizable solutions of the
equation $L\Psi=0$. Asking for normalizable solutions is
equivalent to cutting off the half-line at a finite distance and
imposing global Atiyah-Patodi-Singer boundary conditions.  In
general, in such a situation there is a boundary contribution to
the index of a Dirac-like operator, but it cancels when we take
the formal difference $(S_+\ominus S_-)\otimes \mathrm{ad}(E)$
because $S_+$ and $S_-$ are equivalent when restricted to $M$.

The operator $L$ is real, so it has a real determinant line bundle.  This determinant
line bundle is topologically trivial\footnote{In Donaldson theory, it is shown
that the determinant line bundle of the Dirac operator twisted by $S_+\otimes \mathrm{ad}(E)$
is trivial under a mild topological restriction that is satisfied in our applications.
In our case, because we are twisting by the difference $(S_+\ominus S_-)\otimes \mathrm{ad}(E)$,
the topological condition in question is not needed.}
        and after fixing a trivialization, $L$ has a natural real-valued determinant
function $\det\,L$.

\subsubsection{Applications Of The Flow Equation}\label{apflow}

The flow equation has two primary applications.  The first is to
determine the coefficients when a given cycle $\CC$ in the
relative homology is expanded as a sum of Lefschetz thimbles
$\J_\sigma$ in the usual way,
\begin{equation}\label{totto}\CC=\sum_\sigma \n_\sigma \J_\sigma.\end{equation}   We
are only interested in cycles $\CC$ that lie at $\mu=0$.

The coefficient $\n_\sigma$ is obtained by counting, with suitable
signs, the solutions of the upward flow equation on the half-line
$\R^-=(-\infty,0]$. As usual, the upward flow equations are
obtained from the downward ones by reversing the sign of the time.
The boundary conditions are that the flow starts at a flat
connection $\A_\sigma$ (corresponding to the cycle $\J_\sigma$)
at $t=-\infty$ and ends on $\CC$ at $t=0$. Equivalently, we can
consider downward flow equations on the half-line
$\R^+=[0,\infty)$.  In this case, we consider flows that start on
$\CC$ at $t=0$ and end at $\A_\sigma$ at $t=\infty$.

We write $\A_M=A_M+i\phi_M$ for the restriction of  $\A=A+i\phi$
to $ M$, at a fixed $t$. The boundary condition that we want at
$t=0$ is  that $\A_M$ should lie in $\CC$. Since we assume that
$\CC$ lies at $\mu=0$, the condition that $\A_M$ lies in $\CC$
implies in particular that $\mu=\d_A\star_M\phi_M=0$ at $t=0$.
Subtracting this from  the bulk equation $\d_A\star\phi=0$, which is one of the flow
equations, we
find that the covariant normal derivative of $\phi_t$ vanishes at
$t=0$: $D_t\phi_t=0$.  This is enough to justify integration by
parts in showing that $Z(\phi_t)$, defined in (\ref{polyz}),
vanishes.  So it will lead to global vanishing of $\phi_t$ after
we also pick a suitable boundary condition at $t=-\infty$.

The most basic example is the case that $\CC$ is the real integration
cycle of Chern-Simons theory with compact gauge group $H$.
In this case,  the condition for $\A_M$ to lie in $\CC$ is simply $\phi_M=0$.  This certainly implies
that $\d_A\star\phi_M=0$.

At $t=-\infty$ (or $t=+\infty$ if we formulate the flow on
$\R^+$), we ask that $\A_M$ approaches a specified flat connection
$\A_\sigma$ and that $\phi_t$ vanishes. Given this vanishing, and
the fact that we have chosen a boundary condition that allows
integration by parts at $t=0$, the vanishing theorem implies that
$\phi_t$ will vanish everywhere.

 Let $\M$ be the moduli
space of solutions of the flow equations with boundary conditions
just described. Since the index in the linearized problem
vanishes, the expected dimension of $\M$ is zero and for a generic
metric on $M$, $\M$ consists of a finite set of points.  We count
these points with a weight that is 1 or $-1$ depending on the
sign\footnote{The overall sign of the function $\det\,L$ depends
on a choice of orientation of the relevant determinant line
bundle.  This amounts to a choice of relative orientations of the
cycles $\CC$ and $\J_\sigma$ that we are comparing.  Such choices of
sign and orientation are discussed in, for example,  \cite{KM} in the context of finite-dimensional
Morse theory and Floer cohomology.}
 of $\det\,L$.
 Summing the contributions of all critical points,
 we get the integer $\n_\sigma$ that appears in the expansion $\CC=\sum_\sigma \n_\sigma\J_\sigma$.

Let us compare the definition of $\n_\sigma$ to the definition of
the most basic invariant in Donaldson theory of smooth four-manifolds.  This
 invariant is defined in a situation in which the index of the operator $L_-$ that arises in linearizing
 the Yang-Mills instanton equation on  $X$  is zero.   Under this circumstance, for a generic metric on $X$, the
 instanton equation has finitely many solutions,
 and Donaldson's invariant is defined by summing over all such solutions with a weight
 given by the sign of $\det\,L_-$.   Obviously, $\n_\sigma$ is a close cousin of
 this, with the usual instanton equations replaced by the flow
 equations for a complex-valued connection.

So far, the elliptic nature of the flow equations has enabled us
to express a given cycle $\CC$ in terms of the Lefschetz thimbles
$\J_\sigma$, according to eqn. (\ref{totto}).
 The other important application of the flow equation is to describe Stokes phenomena.
 Here we must consider complete flow lines between two critical points corresponding
 to semistable flat connections $\A_\sigma$ and $\A_\tau$ on $M$.    We must define
 an invariant $\m_{\sigma\tau}$ that controls what happens in crossing a Stokes ray
 at which there are flows from $\A_\sigma$ to $\A_\tau$.

\def\TT{{\mathbf T}}
  We consider the flow
 equations on $X=\R\times M$, with the requirement that $\A_M\to\A_\sigma$ for $t\to-\infty$
 and $\A_M\to\A_\tau$ for $t\to+\infty$.  Further we require that $\phi_t\to 0$ at both ends.
 The index of $L$ still vanishes, but now there is a new ingredient: the flow
 equations
 are invariant under the group $\TT\cong\R$ of translations in the $\R$ direction, but a solution with $\sigma\not=\tau$ cannot
 be translation-invariant.  So any solution with $\sigma\not=\tau$ will have to come in a
 one-parameter family, and hence it is impossible to have an isolated solution.  This clashes with
 the fact that the index is 0, and means that generically there are no flows from $\A_\sigma$ to
 $\A_\tau$.

 We know an independent reason for this: the flow equation has a conserved quantity,
 namely the imaginary part of the holomorphic function $\I=\lambda W(\A)$.   For generic
 $\lambda$, conservation of $\mathrm{Im}\,\I$ makes a flow from $\A_\sigma$ to $\A_\tau$
 impossible.   The condition for a Stokes ray is precisely that a flow from $\A_\sigma$ to $\A_\tau$
 is consistent with the conservation law.   Taking $\lambda=2\pi \exp(-i\alpha)$, a Stokes
 ray is characterized by some condition $\alpha=\alpha_0$.  Instead of regarding $\alpha$
 as a constant in our flow equations (\ref{kweqns}) and (\ref{zweq}), regard it as a variable
 that takes values in a small interval near $\alpha=\alpha_0$.  This adds one variable
 to the problem without adding any new equations, and the index of the linearized problem is now 1.  Thus the moduli space
 of solutions should generically be one-dimensional -- corresponding to families of
 solutions, obtained from each other by the action of $\TT$, that appear at
 $\alpha=\alpha_0$.  To get back to a problem with zero index, we can identify two
 field configurations that differ by the action of $\TT$.  This removes one variable, and the number of solutions is now
 generically finite.  Weighting the solutions by the sign of the determinant of the operator $\tilde L$
 that appears in the linearization around a solution, we arrive at the desired invariant
 $\m_{\sigma\tau}$.

Concretely, adding one variable and removing another  and replacing
$L$ by $\tilde L$ has the following effect.  Suppose we set $\alpha=\alpha_0$ and find
a solution of the flow equations.  The linearization $L$ around a solution of the flow equations
will have a zero mode $\psi$,
generated by the action of $\TT$ on the given solution.  Hence $\det L=0$.
Now vary $\alpha$ away from $\alpha_0$.  The eigenvector $\psi$ of $L$ can be deformed
with $\alpha$ so as to remain an eigenvector, but its eigenvalue $\lambda(\alpha)$ generically vanishes
only at $\alpha=\alpha_0$.  Replacing $L$ by $\tilde L$ means that in evaluating the determinant
at $\alpha=\alpha_0$,
the factor that comes from  the eigenvector $\psi$ is not $\lambda(\alpha_0)=0$ but
$\left.({\d\lambda}/{\d\alpha})\right|_{\alpha=\alpha_0}$, which is generically not zero.
So $\det\,\tilde L$ is generically nonzero and its sign is the contribution of
a given $\TT$-orbit of flow lines to $\m_{\sigma\tau}$.

We have defined an invariant $\n_\sigma$ associated to the flow
equations on a half-line, and another invariant $\m_{\sigma\tau}$
associated to the same equations on the full real line. The
relation between these invariants  is that in crossing a Stokes
ray at which there is a net number $\m_{\sigma\tau}$ of downward
flow lines from $\A_\sigma$ to $\A_\tau$, the invariant $\n_\tau$
jumps by $\n_\tau\to \n_\tau -\n_\sigma \m_{\sigma\tau}$. As
explained in the discussion of eqn. (\ref{orkey}),  where $\m_{+-}$ is
taken to be $\pm 1$, this jumping of  the coefficients $\n_\tau$ compensates
for the jumping in the cycles $\J_\sigma$ and ensures that the
sum $\sum_\sigma\n_\sigma\J_\sigma$ is invariant.   In infinite
dimensions, it seems most concise to describe the jumping
phenomenon in terms of the behavior of the coefficients.

Just as in Morse theory with finitely many variables, the jumping
happens because at $\alpha=\alpha_0$ the set of points that can be
reached  by downward flow from $\A_\sigma$ ceases to
be closed -- its closure contains $\A_\tau$. A downward flow from
$\CC$ to $\A_\tau$ can split into a pair of flows, first from
$\CC$ to $\A_\sigma$ and then from $\A_\sigma$ to $\A_\tau$.  This
leads to the ``wall-crossing'' behavior $\n_\tau\to \n_\tau
-\n_\sigma \m_{\sigma\tau}$.  See for example \cite{CV} in the
context of flows with finitely many variables. Ellipticity means
that such statements about Morse theory carry over to the
infinite-dimensional case.  This is an important fact in the
context of Floer theory (of three-manifolds and of loop spaces of
symplectic manifolds).  See for example \cite{KM,MH} for  reviews.
Hopefully the usual proofs in the context of Floer theory can be
adapted to the present situation.

\subsection{Knots And Morse Theory}\label{placeknots}

\def\L{\mathcal L}
Beginning in section \ref{spurz}, we described the analytic continuation of Chern-Simons knot
invariants in a rather informal way.  To make this more precise, we need to incorporate
in the language of oscillatory integrals
the holonomy functions that are included in the path integral to define knot invariants.
How to do this has been briefly explained in \cite{Wi}, section 3.3, and in
\cite{wilecture}, pp. 1218-9, and in much more detail in \cite{Beasley}.
We consider first the case of a compact gauge group $H$ with connection $A$.

\def\Tr{\mathrm{Tr}}
For a knot $K$ labeled by a representation $R$, the basic factor is
\begin{equation}\label{kurf}\Tr_R\,P\exp\left(-\oint_KA\right).\end{equation}
According to the Borel-Weil-Bott theorem, letting $T$ denote a maximal torus in $H$, the representation
$R$ can be constructed
as the space of holomorphic sections of a certain holomorphic line bundle $\L_R$
over the flag manifold $H/T$.    This means that there is a quantum mechanical
problem with $H$ symmetry for which the space of physical states is the representation
$R$.  Writing a path integral representation of this system and coupling it to the gauge
field $A$ will give the representation that we need of the holonomy function.

The line bundle $\L_R\to H/T$ admits an essentially unique $H$-invariant
 connection that we will call $a_R$.
We can describe as follows a classical problem with $R$ symmetry
whose quantization leads,  in view of the Borel-Weil-Bott theorem,
to the representation $R$. Let $S$ be a one-manifold and let
$\rho:S\to H/T$ be a map.  On the space of such maps, we define an
action $I_R(\rho)=\int_S\rho^*(a_R)$. For $S$ a circle, this is
defined mod $2\pi$, so that $\exp(iI_R(\rho))$ is well-defined.
(If instead $S$ has boundaries, $\exp(iI_R(\rho))$ is not a
complex number but a transition amplitude between quantum
wavefunctions;
 we will not need the details here.)   Since $I_R(\rho)$ is invariant under
reparametrizations of $S$, this system is a topological field theory in the one-dimensional
sense; the Hamiltonian vanishes.

The basic path integral for this system is the integral over
maps $\rho:S\to H/T$:
\begin{equation}\int D\rho\,\exp(iI_R(\rho)).\end{equation}
For $S$ a circle, the value of the path integral can be determined by observing that the path
integral determines a trace in the physical Hilbert space.  The physical Hilbert space
is simply the representation $R$, and as the Hamiltonian vanishes, the trace is simply
the trace of the identity operator.  So the value of the path integral is $d_R$, the dimension
of the representation $R$.

We want to generalize this to couple to a gauge field $A$ and to compute the trace not
of the identity operator but of the holonomy operator  $\Tr_RP\exp\left(-\oint_SA\right)$.  To
do this, we simply couple the gauge field $A$ to the $\rho$ field in a gauge-invariant fashion.
Quantum mechanically, we would couple $A$ to the matrices that generate the action of $H$
in the representation $R$.  The classical limit of these matrices is the moment function
$\mu_R$ associated to the action of $H$ on $H/T$.  $\mu_R$ is a map from $H/T$
to the dual of the Lie algebra $\frak h$.  So for $h\in\frak h$, we have a real valued function
$\mu_R(h)$ on $H/T$.  It obeys
$\d\mu_R(h)=\iota_h\d a_R$ (where  $\iota_h$ is contraction with the vector
field on $H/T$ that corresponds to $h$).
As $A$ is $\frak h$-valued, there is a natural
pairing $(A,\mu_R)$.  The gauge-invariant extension of the action $I_R(\rho)$
is $I_R(\rho,A)=I_R(\rho)-\int_S (A,\mu_R)$.  (A more intrinsic description of the geometry
involved in defining this action can be found in \cite{wilecture}; for more detail see also
\cite{Beasley}.)  Thus we arrive
at a path integral representation of the holonomy:
\begin{equation}\label{polics}\Tr_RP\exp\left(-\oint_SA\right)= \int D\rho\,\exp\left(i I_R(\rho,A)\right).
\end{equation}

Now we can rewrite the basic formula (\ref{mydef}) for the
Chern-Simons invariant of a knot in a convenient form:
\begin{equation}\label{ydef}Z_H(K;k)=\int_\U DA\int D\rho\,\exp(i {k} W(A)+iI_R(\rho,A)).\end{equation}
What we have gained is that now, although we are integrating over a larger number of
variables, the integrand is simpler: it is just the exponential of  a local
functional of the fields.  In other words, we have reduced the problem to an oscillatory
integral (with more variables), and we can apply the Morse theory machinery
rather as if no knot were present.

As usual, the main step in analytic continuation is to extend the action to a holomorphic
function of complex variables.  We analytically continue
the connection $A$ to a complex-valued connection $\A$ and we replace the flag
manifold $H/T$ with $H_\C/T_\C$, where $T_\C$ is a complex maximal torus in $H_\C$.
The analytically continued integral thus takes the form
\begin{equation}\label{yydef}Z_H(K;k)=\int_\CC D\A\int D\rho\,\exp(i {k} W(\A)+iI_R(\rho,\A)),
\end{equation}
where $\CC$ is an integration cycle in the space of fields $\A,\rho$.

Any integration cycle $\CC$, including the original real integration cycle of Chern-Simons
theory in the presence of a knot, can be expressed in terms of Lefschetz thimbles in the usual
way.   Such  cycles   are convenient for
several reasons.  One reason, discussed in section \ref{asyc}, is that it is straightforward
to determine the asymptotic value of the integral over such a cycle, in the semiclassical
region in which the action is large.  What is this region?  In the absence of a knot, the action is
proportional
to $k$, so the semiclassical region is the region of large $k$.  In the presence of a knot
labeled by a representation $R$ with highest weight $\lambda_R$,
the action $I_R(\rho,A)$ is proportional to $\lambda_R$, since it is linear in $a_R$ and $\mu_R$
which are both proportional to $\lambda_R$.
So in the presence of a knot, the natural semiclassical limit is obtained by taking $k$ and
$\lambda_R$ to infinity with a fixed ratio.   The Morse theory
machinery will enable us to understand the asymptotic behavior of the analytically continued
Chern-Simons path integral  in this limit.

\subsubsection{Classical Equations In The Presence Of A Knot}

In the presence of a knot  $K$, the classical equation for $A$ receives an extra contribution
and no longer asserts simply that $F$ is flat.  Rather, the equation receives an extra
contribution that is supported on $K$:
\begin{equation}\label{orork}\frac{kF}{2\pi}=\delta_K\mu_R.   \end{equation}
(Here $\delta_K$ is a delta function that is Poincar\'e dual to $K$, and we use the linear
form $\Tr$ to identify the Lie algebra $\frak h$ with its dual; this is also done in eqn.
 (\ref{zorkk})
below.)    The fact that the curvature is
a delta function supported on $K$ means that the connection is flat away from $K$ and
has a monodromy around $K$.
Exponentiating the delta function in the curvature, this monodromy is
\begin{equation}\label{zome}U=\exp(2\pi \mu_R/ k).\end{equation}
To understand this result, recall that
 $H/T$ is a homogeneous space, and that the conjugacy class
of $\mu_R$ (as an element of $\frak h^*$) is constant and depends only on $R$.  So
eqn. (\ref{zome}) really defines a conjugacy class in $H$; the monodromy around the
knot must lie in this conjugacy class.  We note that the conjugacy class in question
depends only on the ratio $\mu_R/k$.

Let us make this more explicit for $H=SU(2)$.  Take $R$ to be the $n$-dimensional
irreducible representation of $SU(2)$.    Its highest weight is
\begin{equation}\label{zorkk} \frac{n-1}{2}\begin{pmatrix}i & 0 \cr 0 & -i\end{pmatrix}.\end{equation}
(To keep the structure constants real, we represent a real Lie algebra such as $\frak{su}(2)$
by anti-hermitian matrices.)  Up to conjugacy, this is $\mu_R$ for the representation $R$.
So the monodromy around a knot labeled by this representation is
\begin{equation}\label{spork}\exp\left(\frac{i\pi (n-1)}{k}
\begin{pmatrix}1 & 0 \cr 0 & -1\end{pmatrix}\right).\end{equation}

We will now make a remark on quantum corrections to this classical formula.
In physics in general, quantum corrections cannot be entirely understood by shifting
 the values of classical parameters, but many quantum results
in Chern-Simons
theory can be understood more readily from a semiclassical point of view if one shifts $n-1$ to
$n$ and $k$ to $k+2$ in (\ref{spork}).  (The shift in $k$ was discussed in section \ref{littlemore},
and the shift in $n$ has a similar interpretation.  For the analogous shifts for noncompact
Lie groups, see \cite{DW}.)  For our purposes, it is not necessary
to take these shifts into account explicitly.  Our goal is to determine the correct integration
cycle for the path integral in the limit of large $k$ and $n$ with fixed ratio $n/k$.  The
answer to this question is not affected by shifts $n\to n+1$ or $k\to k+2$.  Those
shifts enter when, after having found the right integration cycle for the path integral,
one carries out the integral including one-loop effects.
This will generate the shifts in $n$ and $k$ in the usual way.

\subsubsection{More On The Classical Equations}\label{further}

 There are some important
subtleties concerning the gauge fields obeying (\ref{orork}),
especially after we analytically continue to a complex gauge field
$\A$.  In that case, the equation becomes
\begin{equation}\label{zurob} \frac{k
\F}{2\pi}=\delta_K\mu_R,\end{equation} where $\F=\d\A+\A\wedge\A$.
We want to supplement this equation with a condition for vanishing
of the moment map,
\begin{equation}\label{urob}\d_A\star_M\phi=0,\end{equation}
where $\A=A+i\phi$.  To describe what sort of gauge fields obey
this equation, it suffices to work in a two-plane normal to $K$.
We take polar coordinates $r,\theta$ on this two-plane, with $K$
understood to lie at $r=0$.  To obey (\ref{zurob}) and
(\ref{urob}), we can take
\begin{align}\label{hurk}A=&\,\alpha\,\d\theta +\dots \cr  \phi=&
-\gamma\,\d\theta+\dots,\end{align} where $\alpha$ and $\gamma$
are commuting elements of the real Lie algebra $\frak h$ with
$\alpha-i\gamma=\mu_R/2k$, and the ellipses refer to terms less
singular than $1/r$.  (The angular form $\d\theta$ has a $1/r$
singularity at $r=0$.)  This type of singularity in gauge theory
was studied in detail\footnote{In that context, there was another
term in $\phi$ of the form $\beta\,\d r/r$; $\beta$ can be removed
by a  gauge transformation (which is singular at $r=0$), and tentatively it seems that
in the present problem we can just as well take $\beta=0$.} in
\cite{OGW}, starting in section 2.1. The most natural way to
proceed is to fix a gauge in which $\mu_R$ lies in the Lie algebra
of a specified complex maximal torus $T_\C$.  This can always be
done globally along $K$. Then we require $\A$ to have a
singularity of the sort described in (\ref{hurk}), and we only
allow gauge transformations that preserve the gauge condition --
in other words, gauge transformations that are $T_\C$-valued when
restricted to $K$.

One important point, discussed in \cite{OGW}, is that by a gauge
transformation that is singular at $r=0$, one can shift $\alpha$
by a coroot of $H$.  This means that the formalism is invariant
under shifting $\mu_R$ by $k$ times a coroot.  For $H=SU(2)$, this
corresponds to the symmetry
\begin{equation}\label{helf}n\to n+2k,\end{equation}
which actually undergoes a quantum correction to $n\to n+2(k+2)$.

There is an important subtlety, which we will describe first for
$H=SU(2)$.  Suppose that we vary $\mu_R$ and $k$ so that $\mu_R/k$
goes to zero, or more generally so that it approaches a
cocharacter (which we then remove by  a singular gauge
transformation, as mentioned in the last paragraph).  This means
that $\alpha$ and $\gamma$ go to zero, and a naive look at
(\ref{hurk}) might make one believe that in the limit, $A$ and
$\phi$ become nonsingular. The correct conclusion, however,
because of the ellipses in (\ref{hurk}), which represent terms
less singular than $1/r$, is only that in such a case, $A$ and
$\phi$ are less singular than $1/r$. The equations $\F=\d_A\phi=0$
have a solution with a singularity at $r=0$ that is (slightly)
milder than $1/r$, and this less singular solution may be  the
result if one varies a family of solutions with the $1/r$
singularity (\ref{hurk}) so that the coefficient of $1/r$
approaches zero. The singular solution in question, which was
essentially first studied by Nahm \cite{N}, takes the form
\begin{align}\label{fork}A & = \frac{t_1\,\d\theta}{\ln r}\cr
                        \phi & =\frac{t_2 \,\d r}{r\ln r}-\frac{t_3\,\d\theta}{\ln r},\end{align}
where $[t_1,t_2]=t_3$, and cyclic permutations thereof.  This
condition means that the $t_i$ define a homomorphism from
$\frak{su}(2)$ to $\frak h$.  For $H=SU(2)$, there is up to
conjugation only one nontrivial homomorphism and we pick this one.
(For other groups, all possible homomorphisms can arise, in
general.)  For this solution, the monodromy of $\A$ around the
singularity at $r=0$ is not trivial but is in the conjugacy class
that contains the non-diagonalizable, unipotent element
\begin{equation}\label{ozork}V=\begin{pmatrix}1 & 1 \cr 0 &
1\end{pmatrix}.\end{equation} This is demonstrated by directly
showing that, with $A,\phi$ given in (\ref{fork}), the holonomy of
$\A=A+i\phi$ along any path that loops around $r=0$ is in the
stated conjugacy class.  For example, if the path is the circle
$r=r_0$ for some $r_0$, the monodromy is $\exp(2\pi(t_1-it_3)/\ln
r_0)$, which is in the stated conjugacy class.  (For more detail,
see section 3.8 of \cite{OGW}.)

What is happening here is that in the complex Lie group $G=H_\C$, as $\mu_R/k\to 0$,
the conjugacy class that contains the element
$U=\exp(2\pi\mu_R/k)$ does not approach the conjugacy class of the
identity; rather it approaches the conjugacy
class of the element $V$. This may become clear if one considers
the family of $SL(2,\C)$ elements
\begin{equation}\begin{pmatrix}\lambda & 1\cr 0 &
\lambda^{-1}\end{pmatrix},\end{equation} which are conjugate to
the diagonal matrix ${\mathrm {diag}}(\lambda,\lambda^{-1})$ for
generic $\lambda$, but to $V$ if $\lambda=1$. So if we consider a
family of connections $\A$ representing Chern-Simons critical
points, parametrized by $\mu_R/k$, with monodromy
$U=\exp(2\pi\mu_R/k)$, then as $\mu_R/k\to 0$, the monodromy of
$\A$ around the knot might approach 1, but it also might approach
a unipotent element conjugate to $V$. The latter behavior is generic, since the conjugacy
class of $V$ has  dimension two, while that of the identity  consists of a single point.

Modulo a singular gauge transformation that shifts $\alpha$ to
zero,\footnote{Such a gauge transformation is single-valued in
$SO(3)$, but not in $SU(2)$.  It is not quite a symmetry of
$SU(2)$ Chern-Simons theory, but it is useful for analyzing the
classical equations and describing critical points.  For more on
this, see section \ref{tensoring}.} the behavior is exactly the
same if $\mu_R/k$ approaches a value for which $U\to -1$. Given a
family of flat connections $\A$ of monodromy $U$, consider a limit
in which $U\to -1$.  The monodromy of $\A$ may approach $-1$, but
instead it may approach
\begin{equation}\label{ozorkox}-V=-\begin{pmatrix}1 & 1 \cr 0 &
1\end{pmatrix}.\end{equation} We will encounter both cases in section \ref{anajones}.

We have described this exceptional behavior for $SL(2,\C)$, but it
has a close analog for any complex Lie group $G$.  As long as
$U=\exp(2\pi\mu_R/k)$ is regular (that is, at long as the subgroup
of $G$ that it commutes with is a copy of $T_\C$), Chern-Simons
critical points in the presence of a knot correspond to flat
connections whose monodromy around the knot is conjugate to $U$.
But in general, the monodromy may be in any conjugacy class whose
closure contains $U$.  For more on this, see section 3.8 of
\cite{OGW}.

\bigskip\noindent{\it The Flow Equations In The Presence Of A
Knot}

Now let us briefly describe the flow equations in the presence of
a knot.  As usual, to formulate the flow equations, we introduce
an extra ``time'' coordinate.  So to study Chern-Simons theory on
a three-manifold $M$ in the presence of a knot $K$, we consider
flow equations on the four-manifold $X=\R\times M$, with the knot
now filling out the two-manifold $\R\times K$.  The flow equations
are the same as before, but now we must consider solutions with a
singularity along $\R\times K$.  The singularity is given by
(\ref{fork}) modulo terms less singular than $1/r$.  In
particular, for special values of $\mu_R/k$, we must allow a
singularity that (perhaps after a singular gauge transformation to
set $\alpha$ to zero) takes the form (\ref{fork}).  This seemingly
exceptional case is important in applications, because the flat
$SL(2,\C)$ connection associated to a hyperbolic metric on the
knot complement has this sort of monodromy.

In effect, the flow equations have a codimension two singularity
in four dimensions that can be modeled by the same surface
operators studied in \cite{OGW}.

\subsubsection{Analog For Noncompact Lie Groups}\label{humf}

For the case of a compact Lie group $H$, we have described how to incorporate
a knot labeled by a representation $R$ in the framework of oscillatory integrals and Morse
theory.

What happens if the gauge group is not compact, for example if it is $G=H_\C$?  If
the representation $R$ is unitary and arises by quantization of a coadjoint orbit
$\mathcal O$ of $G$, then there is a close analog of (\ref{polics}), with $H/T$ replaced by
$\mathcal O$.  The resulting knot invariants, however, have not been seriously studied.

  Let us consider instead
the case that $R$ is a finite-dimensional holomorphic
representation of $G$, obtained by analytic continuation of an
irreducible representation of $H$, as described in section
\ref{knots}.  It is straightforward to  bring such a
representation into the Morse theory framework. We represent the
holonomy around the knot  in such a representation by the same
factor as in (\ref{yydef}), namely $\int
D\rho\,\exp(iI_R(\rho,\A))$. So the path integral definition
(\ref{onk}) of knot invariants associated to such a representation
can be rewritten
\begin{equation}\label{konk}Z_G(K;t,\tilde t)=\int_\CC D\A\,D\tilde \A \,D\rho\,\exp\left(itW(\A)/2+i\tilde t \,W(\tilde \A\,)/2\right)\,
     \exp(iI_R(\rho,\A)).\end{equation}
 Morse theory can now be used in the usual way to describe possible integration cycles for this
 integral. After analytic continuation, $\A$ and $\tilde\A$ are independent.  Since $I_R(\rho,\A)$
 is independent of $\tilde \A$, the presence of the knot does not affect the Euler-Lagrange
 equation for $\tilde A$.  The
  equation for a critical point thus tells us simply that $\tilde\A$ is flat,
 while $\A$ is a flat connection on the complement of $K$ whose monodromy around
 $K$ is in the conjugacy class of the  element $U=\exp(2\pi\mu_R/ k)$ described in eqn.
 (\ref{zome}). (When $U$ is not regular, this statement must be refined in a way described
 in section
 \ref{further}.)  Of course, now we are supposed to take the conjugacy class of $U$ in
 $G$ rather than in $H$.

 One point deserves note.  As $I_R(\rho,\A)$ is not real,
the exponential $\exp(i I_R(\rho,\A))$
 grows exponentially in some regions.  As a result, even if $t$ and $\tilde t$ are complex conjugates,
 the path integral (\ref{konk})
  does not converge  if evaluated on the
 usual real cycle $\tilde\A=\bar\A$.  Perhaps it is possible to slightly displace the real
 cycle to make the integral converge, or perhaps the integral can only be defined using Lefschetz
 thimbles.  As explained in section \ref{knots}, in the case of knots in $S^3$, a suitable
 integration cycle can essentially be borrowed from the case of a compact gauge group.

A similar analysis applies if the gauge group is a noncompact real form $H'$ of
$G=H_\C$.  If $R$ is a unitary representation, one can imitate (\ref{polics}) using
the appropriate coadjoint orbit.  If $R$ is a finite-dimensional representation of $H'$ (obtained
by analytic continuation from $H$ to $G=H_\C$ and then restriction to $H'$), one can represent
the knot by the now familiar factor $\int D\rho\exp(iI_R(\rho,A'))$, where now $A'$ is
$\frak h'$-valued.  Prior to analytic continuation, $\rho$ is a map to $H/T$, which admits
an action of $G$ and hence of $H'$.  After analytic continuation, the fields and the integrand
of the path integral are the same as they are in the $H$-valued theory, but the integration
cycle is different.  This last statement holds exactly as it does in the absence of the knot.

\subsubsection{Classification Of Critical Points}\label{gumf}

To study the Chern-Simons path integral via the Morse theory machinery, a first step is to
classify the critical points and evaluate the classical action at a critical point.   As long
as one considers only physical values of the coupling parameters, critical points
correspond to flat connections modulo gauge transformation (perhaps with monodromy
around a knot), and the Chern-Simons
invariant $W(A)$ of such a flat connection is understood as an element of $\R/2\pi\Z$
(or $\C/2\pi\Z$ in the case of a complex Lie group).

Upon analytic continuation away from physical values of the coupling, one must be more
careful.  Let us consider first the case that the gauge group is a compact Lie group $H$
and one wishes to continue away from integer values of $k$.
In this case, the integrand
$\exp(ikW(A))$ is invariant only under the subgroup of gauge transformations that
leave $W(A)$ fixed, rather than shifting it by a multiple of $2\pi$.  If $H$ is connected and
simply-connected, this subgroup is the connected component of the group of gauge transformations.   At any rate, $\exp(ikW(A))$ should be understood as a  function not on the space
$\U$ of gauge fields modulo gauge transformations, but on a cover $\hat \U$ on which
the Chern-Simons invariant is well-defined.  For connected and simply-connected $H$,
$\hat \U$ is the universal cover of $\U$.

So we need to consider critical points of the Chern-Simons function on $\hat \U$.
Such a critical point is simply a flat connection together with a lift of its Chern-Simons invariant
to $\R$ (or $\C$, if we consider complex critical points).

As an example, let us consider Chern-Simons theory on $S^3$ in the
presence of a knot $K$.   Typically there are only finitely many
critical orbits in the usual sense.  In other words, up to gauge
transformation, there are typically only finitely many flat
connections on the knot complement $S^3\backslash K$ with
prescribed monodromy about $K$.  However, when we analytically
continue away from integer values of $k$, we need a real-valued
Chern-Simons function, meaning that a critical point must be
understood as a pair consisting of a flat connection and a lift of
its Chern-Simons function to $\R$.  This combined data determines
a critical point of the function $W(A)$ on $\hat\U$. There always
are infinitely many of these critical points, as the Chern-Simons
invariant of any given flat connection can be lifted to $\R$ in
infinitely many ways.   In section \ref{anajones}, we will study
in detail the analytic continuation of the Chern-Simons path
integral for some specific knots (the trefoil and the figure-eight
 knot) and we will see the important role played by distinct
critical points that differ only in the value of their
Chern-Simons invariant. This should come as no surprise, since in
eqn. (\ref{hosedd}) we have already seen similar behavior in the
analytic continuation of the Bessel function away from integer
values of $\kappa$.

The same ideas hold if the gauge group is a non-compact real form $H'$ of $G=H_\C$.
However, the
 case that the gauge group is the complex Lie group $G$ involves a new wrinkle.  The
Chern-Simons
theory depends naturally on an integer-valued parameter $\ell$ and a real parameter $s$,
as shown in eqn. (\ref{mork}).    As explained in section \ref{overview}, to analytically continue
to complex $s$,
we view $\tilde\A$, which begins life as the complex conjugate of the gauge field $\A$,
as an independent $\frak g$-valued connection in its own right.  The action is a linear
combination of the two Chern-Simons actions $W(\A)$ and $W(\tilde\A)$,
\begin{equation}\label{hoagie}I=\frac{t\,W(\A)}{2}+\frac{\tilde t\,W(\tilde\A)}{2},~~t=\ell+is,~
\tilde t=\ell-is.\end{equation}
If we impose invariance under all gauge transformations, a critical point is a pair of
flat connections $\A,$ $\tilde\A$, up to gauge equivalence.  In that case,
$W(\A)$ and $W(\tilde\A)$ are each naturally defined as elements of  $\C/2\pi\Z$.
But then (for complex $s$) the  indeterminacy of $I$ is not just an integer multiple of $2\pi$
and the integrand $\exp(iI)$ of
the path integral is not defined.

 One might think that to make sense of the path integral, we would have to lift
 both $W(\A)$ and $W(\tilde\A\,)$ from $\C/2\pi\Z$ to $\C$.  We indeed have to do
 this if the aim is to analytically continue to generic values of both $\ell$ and $s$.
 However, there is a fairly natural problem, described in section \ref{overview}, in which
 $\ell$ remains an integer and only $s$ is continued to complex values.  In this case,
 $(t+\tilde t\,)/2$     remains an integer, so adding $2\pi$ to both $W(\A)$ and $W(\tilde A)$
 causes no change in $\exp(i I)$.   The most natural way to proceed is to regard $\exp(iI)$
 as a function on the smallest cover of $\U_\C\times\U_\C $ where it can be defined.
 In practice, this means that a critical point is a pair of flat connections $\A,\tilde\A$ together
 with a lift of the pair $W(\A)$, $W(\tilde\A)$ to $(\C\times \C)/2\pi \Z$, where $\Z$
 is diagonally embedded.   In other words, we can shift both Chern-Simons invariants
 by a common multiple of $2\pi$, but we cannot shift them independently.   We can use
 this freedom to arbitrarily specify how we want to lift the Chern-Simons
 invariant of either $\A$ or $\tilde\A$, but then we must consider all lifts of the other.

It is instructive to consider the usual example of a knot $K$ in
$S^3$.  In this case, as explained in section \ref{humf}, a
critical point in the usual sense corresponds to an ordinary  flat
connection $\tilde \A$ on $S^3$ plus a flat connection $\A$ on
$S^3\backslash K$ whose monodromy around $K$ is in a prescribed
conjugacy class. As $S^3$ is simply connected, $\tilde\A$ is
gauge-equivalent to a trivial flat connection, and $W(\tilde\A\,)$
vanishes modulo $2\pi$. It is natural to fix the complex lift of
$W(\tilde \A)$ to be zero.  Once we do this, we have exhausted our
freedom to shift the Chern-Simons invariants, and we must consider
all possible complex lifts of $W(\A)$.

The possible critical points, then, are classified by a flat
connection $\A$ on $S^3\backslash K$ with appropriate monodromy
and with all possible choices for its Chern-Simons invariant.  But
this is the same set of critical points that we encounter in
analytically continuing the theory on $S^3$ with compact gauge
group $H$ in the presence of the same knot $K$ labeled by the same
representation $R$.   What we have just described is a
nonperturbative aspect of the relation between theories with gauge
groups $H$ or $G=H_\C$ for the special case of knots in $S^3$.
This relation was already discussed in section \ref{relation}.

\subsubsection{Variation Of The Action}\label{zoknot}

\def\m{\mathfrak m}
\def\l{\mathfrak l}
To use the Morse theory formalism, we need some information about
the value of the Chern-Simons function $W(\A)$ for a classical
solution.  We take the gauge group to be a complex Lie group
$G=H_\C$.

\begin{figure}
 \begin{center}
   \includegraphics[width=3in]{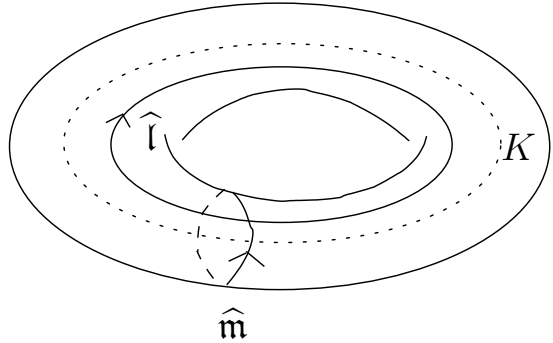}
 \end{center}
\caption{\small A knot $K$ (dotted line)
surrounded by a two-torus $Y$.  On $Y$ are drawn the meridian $\hat\m$,
a circle whose linking number with $K$ is 1, and the longitude $\hat\l$, whose linking  number with
$K$ vanishes.}
 \label{meridian}
\end{figure}

The following ideas prove to be  useful.  First of all, as
sketched in fig. \ref{meridian}, the boundary of a small
neighborhood of a knot $K$ is a two-torus $Y$. On $Y$, we draw two
circles -- the meridian $\hat\m$, which goes the short way around $Y$
and has linking number 1 with $K$, and the longitude $\hat\l$, which
goes the long way around $Y$ and which we choose to have linking
number 0 with $K$.

Since the fundamental group of $Y$ is abelian, if $\A$ is a flat
connection, its monodromies around $\hat\m$ and $\hat\l$ (which we will
denote by the same symbols) are commuting
elements of $G$.  Generically, we can conjugate them to a common
maximal torus  $T_\C\subset G$.  According to (\ref{zome}), for a
knot labeled by a representation $R$, $\hat\m$ is conjugate to
$U=\exp(2\pi \mu_R/k)$.  (More exactly, as explained in section
\ref{further}, this is so if $U$ is regular.  In general, $U$ is
contained in the closure of the conjugacy class that contains
$\hat \m$.)

Let us consider in detail the case $G=SL(2,\C)$.  The maximal
torus of $G$ is a copy of $\C^*$, consisting of diagonal matrices
\begin{equation}\label{ofto}\begin{pmatrix}\lambda & 0 \cr
0 & \lambda^{-1}\end{pmatrix}.\end{equation} An important fact is
that in a classical solution of Chern-Simons theory in the
presence of a knot $K$, we have a little more structure than just
a flat connection on the knot complement.
 In putting knots in the Morse theory
framework, we introduced an $H/T$ or $G/T_\C$-valued field $\rho$
along $K$. In a classical solution, $\rho$ is invariant under
$\hat\m$ and $\hat\l$, and once $\hat\m$ and $\hat\l$ are given,
the choice of $\rho$ amounts to a choice of one of the two
eigenvalues of $\hat\m$ and $\hat\l$, say the first eigenvalue
(corresponding to $\lambda$ as opposed to $\lambda^{-1}$ in
(\ref{ofto})).
 We write  $\m$ and $\l$ for the
eigenvalues  of  the loops $\hat\m$ and
$\hat\l$ that correspond to $\rho$.
 According to (\ref{spork}), if $R$ is the
$n$-dimensional representation of $SU(2)$, then
\begin{equation}\label{forsk}\m =\exp(i\pi(n-1)/k). \end{equation}
In section \ref{anajones}, we will use this formula in the
semiclassical limit of large $n,k$ with fixed ratio $n/k$. (In
this limit, the difference between $n$ and $n-1$ is irrelevant.)

\def\AA{{\mathbf A}}
Once $K$ is embedded in a three-manifold -- which we will take to
be simply a three-sphere -- $\l$ and $\m$ are no longer
independent. They are constrained by a requirement that an
$SL(2,\C)$ flat connection on $T$ with monodromies $\hat\l$ and $\hat\m$
should actually extend over the knot complement $S^3\backslash K$.

One simple class of $SL(2,\C)$ flat connections are the abelian
ones.  These are the flat connections whose structure group
reduces to $\C^*$, the group of diagonal matrices (\ref{ofto}).
They are classified very simply.  Although the fundamental group
of $S^3\backslash K$ can be very complicated, its first homology
group is simply isomorphic to $\Z$, generated by $\m$.  This means
that there is up to gauge transformation one flat $\C^*$-valued
connection on $S^3\backslash K$ for every choice of $\m$.  Such a
flat connection always has $\l=1$.

What about flat connections whose structure group does not reduce
to $\C^*$?  Flat connections with triangular structure group
\begin{equation}\label{unstable}\begin{pmatrix} \alpha & \beta \cr 0 &
\gamma
\end{pmatrix}\end{equation}
 are unstable and need not be considered as critical points.  What
 this really means is that to any strictly triangular flat connection there
 is a corresponding abelian one obtained by deleting the
 off-diagonal part of $\A$, and this should be considered instead.  An
 important point which follows directly from the definition of the
 Chern-Simons function $W(\A)$ is that this function is invariant
 under this operation:
\begin{equation}\label{diagonalize}\begin{pmatrix} \alpha & \beta \cr 0 &
\gamma
\end{pmatrix}\to\begin{pmatrix} \alpha & 0 \cr 0 & \gamma
\end{pmatrix}.\end{equation}

There remain the irreducible $SL(2,\C)$ flat connections.  (A flat
connection is called irreducible if the only elements of
$SL(2,\C)$ that commute with all monodromies are the central
elements $\pm 1$.)  An  irreducible flat connection does not exist
for all values of $\l,\m$.  Rather, for each knot $K$, there is a
Laurent polynomial $\AA(\l,\m)$ (a polynomial in $\l,$ $\l^{-1}$,
$\m$, $\m^{-1}$) such that the moduli space of irreducible
$SL(2,\C)$ flat connections on $S^3\backslash K$ is described by
an equation
\begin{equation}\label{omurk}\AA(\l,\m)=0 \end{equation}
which parametrizes values of $\l$ and $\m$ for which there exists
such a connection.  $\AA(\l,\m)$, which is often called the $\AA$
polynomial, is effectively computable for any given knot.  We will
make use of some examples in section \ref{anajones}.  The $\AA$
polynomial is invariant under the exchange
\begin{equation}\label{zurk}\l\to\l^{-1},~~\m\to\m^{-1},\end{equation}
which comes from a Weyl transformation exchanging the two
eigenvalues in (\ref{ofto}).  Such a Weyl transformation does not
leave invariant the field $\rho$ that is defined along the knot,
but the definition of the $\AA$ polynomial does not ``know'' about
$\rho$.  Another symmetry of the $\AA$ polynomial is
\begin{equation}\label{purk}\l\to\l,~~\m\to -\m.\end{equation}
This symmetry reflects the fact that on $S^3\backslash K$, there
exists a flat complex line bundle $\L$ with monodromies $\m=-1$,
$\l=1$. If $E\to S^3\backslash K$ denotes a flat complex vector
bundle of rank 2 and determinant 1, corresponding to a flat
$SL(2,\C)$ connection, then the operation $E\to E\otimes \L$
reverses the sign of $\m$ without changing $\l$.    It is not
immediately apparent whether this operation leaves fixed the
Chern-Simons function $W(\A)$ and therefore whether it is a
symmetry of Chern-Simons theory.  We return to this issue in
section \ref{tensoring}.

To describe the function $W(\A)$ once the $\AA$ polynomial is
known (see \cite{kk} for a mathematical treatment), we simply use
the variational principle for $W(\A)$, which follows directly from
its definition:
\begin{equation}\label{omicro}\delta
W(\A)=\frac{1}{2\pi}\int_{S^3}\Tr\,\delta \A\wedge
\F.\end{equation} On the other hand, for a flat connection on the
knot complement, we have $\F=(2\pi\mu_R/k)\delta_K=(\ln \hat\m)
\delta_K$.  In the last step, we used the fact that
$\hat\m=\exp(2\pi\mu_R/k)$.  So the right hand side of
(\ref{omicro}) reduces to $\Tr\,\oint_K \delta\A \ln\hat\m$. Here,
once we diagonalize $\m$, $\A$ and $\delta\A$ are also diagonal
along $K$, and $\oint_K\delta\A=\delta\ln\hat\l$.  So the formula
becomes $\delta W=\Tr\,(\delta\ln\hat\l) \ln \hat\m/2\pi$. In
expressing this result in terms of the eigenvalues
 $\l$ and $\m$ of the Lie algebra elements $\hat\l$ and $\hat\m$,  we get a
factor of 2 from the trace:
\begin{equation}\label{csvalue}\delta
W(\A)=\frac{1}{\pi}\delta\ln\l\cdot\ln \m.\end{equation}

In practice, once the relation between $\l$ and $\m$ is known, the
important results about $W(\A)$ can usually be obtained by
integrating this formula. Some slightly subtle examples will be
considered in section \ref{anajones}.  Here we will practice with
some easy cases.

 For an abelian flat connection, $\l$ is always 1,
so $0=\delta\ln\l=\delta W(\A)$.  So $W(\A)$ is constant as a
function on the moduli space of abelian flat connections on
$S^3\backslash K$.  It vanishes for a trivial flat connection
($\l=\m=1$) so it vanishes for any abelian flat connection. For
another example, consider a unitary flat $SL(2,\C)$ connection --
in other words, one whose holonomy lies in $SU(2)$. In this case,
$\l$ and $\m$ are both of absolute value 1, so $\ln\m$ and $\ln\l$
are both imaginary. Hence, $\delta W$ is real on this locus,
consistent with the fact that $W$ is real for an $SU(2)$ flat
connection.

Apart from the factor $\exp(ikW(\A))$, the integrand of the path
integral in the presence of a knot also contains an additional
factor $\exp(i I_R(\rho,\A))$, so we need to know how to evaluate
this factor for a classical solution.  In the usual gauge in which
$\rho$ is constant along $K$, this factor reduces to
$\exp(-i\int_K(\mu_R,\A))$.  This depends only on $\l$ and can be
understood as a character of the maximal torus in which $\l$ takes
values.   Schematically we might write it as $\l^{\mu_R}$. For
the $n$-dimensional representation of $SU(2)$, this factor is
$\l^{(n-1)}.$  Equivalently,
\begin{equation}\label{gorl}I_R(\rho,\A)=-i(n-1)\ln\l.\end{equation}

\subsubsection{Branches Of The Logarithm}\label{branches}

An obvious point about this last formula is that $\ln\l$ has many
branches. This reflects the fact that the underlying action
$I_R(\rho,\A)$ is similarly defined only modulo $2\pi$.  There are
two possible ways to proceed.  If we wish to analytically continue
to non-integral values of $n$, we should consider a critical point
to be a pair consisting of a flat connection of suitable monodromy
and a lift of $\ln\l$ to $\C$ (plus a similar lift of $W(\A)$ if
we are analytically continuing in $k$).  This amounts to replacing
the loop space of $H/T$ or $G/T_\C$  with its universal cover.

If we leave $n$ as an integer and analytically continue only with
respect to $k$, we need not consider the choice of a lift of
$\ln\l$ to be part of the definition of a critical point. Instead,
the multivaluedness of the logarithm enters in the following way.
Consider the quantity $\mathrm{Im}(ikW(\A)+iI_R(\rho,\A))$ which
one may expect to be conserved by the flow equations. The
multivaluedness of the logarithm means that this  quantity is
actually only conserved modulo $2\pi (n-1)\Z$. It is conserved in
any flow provided we make suitable choices for the branches of
$\ln\l$, but there may be no way to choose branches of $\ln\l$
that depend only on the choice of a critical point (and not on the
choice of a flow line) such that
$\mathrm{Im}(ikW(\A)+iI_R(\rho,\A))$ is conserved in every flow.

If instead we analytically continue with respect to $n$, we must
consider the choice of branch of $\ln\l$ as part of the definition
of a critical point and then the quantity
$\mathrm{Im}(ikW(\A)+iI_R(\rho,\A))$ really is conserved by all
flows.

See the end of section \ref{sc} for an example in which this issue
becomes relevant.

\subsubsection{Tensoring With A Flat Line Bundle}\label{tensoring}

As a further application of eqn. (\ref{csvalue}), we will return
to the question of whether, for $H=SU(2)$, the operation $E\to
E\times \L$, where $\L$ is a flat line bundle of order 2, is a
symmetry of Chern-Simons theory.

Let $\Delta W(\A)$ denote the change in $W(\A)$ under this
operation.  We can differentiate $\Delta W(\A)$ with respect to
$\A$ by making use of (\ref{csvalue}).  The operation $E\to
E\otimes \L$ leaves $\l$ unchanged and (as it changes the sign of
$\m$) shifts $\ln\m$ to $\ln \m -i\pi$.  So $\delta\Delta
W(\A)=(\delta\ln l/\pi)(-i\pi)=-i\delta \ln\l$.  Integrating, we
have $\Delta W(\A)=-i\ln\l$ up to an additive constant.  In
Chern-Simons theory on a general three-manifold (such as
$\Bbb{RP}^3$) the constant may be nonzero.\footnote{It is always
an integer multiple of $\pi$, though not necessarily of $2\pi$.
This can be proved by considering the Chern-Simons invariant for
$SO(3,\C)=SL(2,\C)/\Z_2$.} But it vanishes for the special case of
a knot  $K\subset S^3$.  This can be proved by considering the
special case that $\A$ is a flat abelian connection (embedded in
$SL(2,\C)$) with appropriate monodromy around $K$.  Tensoring with
$\L$ gives another flat abelian connection $\A'$. The monodromies
of $\A$ and $\A'$ around the longitude and the meridian are
related by $\m'=-\m$, $\l'=\l=1$, and moreover  $W(\A)=W(\A')=0$.
  So for a knot in $S^3$, we have simply $\Delta W(\A)=-i\ln\l$, and, making use of
(\ref{gorl}), we get a symmetry of
$kW(\A)+I_R(\rho,\A)$ if we combine $E\to E\otimes \L$ with $n\to n+k$.
Allowing for a one-loop quantum shift described in section  \ref{littlemore}, the symmetry is
actually
\begin{equation}\label{monkey}E\to E\otimes \L,~~n\to n+k+2.\end{equation}

\begin{figure}
 \begin{center}
   \includegraphics[width=2in]{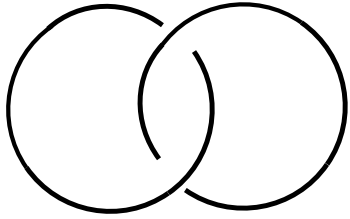}
 \end{center}
\caption{\small Two linked circles in $S^3$.}
 \label{linked}
\end{figure}

This is actually the one place in the present paper that it is
important to distinguish a knot (an embedded circle in $S^3$) from
a link (a disjoint union of embedded circles).  The operation
$E\to E\otimes \L$, where $\L$ has monodromy around a knot, does
not affect the longitude for that knot.  But in the case of two or
more disjoint embedded circles with nonzero linking (fig.
\ref{linked}), if $\L$ has monodromy around one circle, then
tensoring with $\L$ may reverse the sign of the longitude for a
second circle.  So (\ref{monkey}) is a symmetry of Chern-Simons
theory for a knot, but not necessarily for a link.

The square of the operation (\ref{monkey})
\begin{equation}\label{ozzo}n\to n+2(k+2)\end{equation}
is a universal symmetry of Chern-Simons  theory (already mentioned
in eqn. (\ref{helf})).  Indeed, $\L^2$ is completely trivial, and
tensoring with $\L^2$ leaves $\m$ fixed.

\section{Analytic Continuation Of The Colored Jones Polynomial}\label{anajones}

Here we will finally apply all this machinery to some specific
examples.  We will consider $SU(2)$ Chern-Simons theory at level
$k$ on $S^3$, with a knot $K$ labeled by the $n$-dimensional
representation of $SU(2)$.  We denote as $J_n(q)$, where
$q=\exp(2\pi i/(k+2))$, the (unnormalized) Chern-Simons path
integral in the presence of this knot.  $J_n(q)$ is essentially
the colored Jones polynomial, though  our normalization is not the
most common one in the mathematical literature.   (As explained at the end of section
\ref{spurz}, the two normalizations differ by whether one chooses to divide by the Chern-Simons
path integral for the unknot.)

\begin{figure}
 \begin{center}
   \includegraphics[width=3.5in]{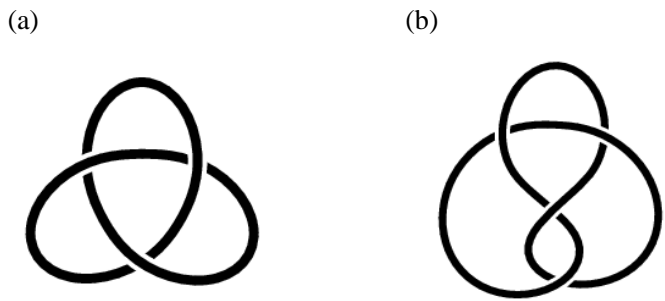}
 \end{center}
\caption{\small The trefoil knot (a) and the figure-eight knot (b).  (From
Wikipedia.)}
 \label{twoknots}
\end{figure}

We consider in detail two illustrative and relatively simple knots
-- the trefoil knot and the figure-eight knot (fig.
\ref{twoknots}). For background on these examples, see
\cite{Mura}.  The trefoil knot is a basic example of a larger
class of knots known as torus knots. They can be understood as
orbits of a $U(1)$ action on $S^3$, and this leads to
exceptionally simple formulas for their quantum invariants, as has
been explained recently via path integrals \cite{Beasley}. We
expect similar behavior for other torus knots, though we consider
only the trefoil in detail. However, torus knots are atypical.  A
more typical knot is one whose complement in $S^3$ admits a
complete hyperbolic metric.\footnote{The relevant metrics look
near the knot like $\d s^2=\d z^2/z^2+z^2(\d x^2+\d y^2)$, where
the knot is located at $z=0$, the ``infrared'' region in
gauge/gravity duality.  The ``ultraviolet'' region for
$z\to\infty$ is cut off by the embedding in $S^3$.} The
figure-eight knot (also called the $4_1$ knot) is possibly the
simplest example of a knot that is generic in this sense, and as a
result its colored Jones polynomial has been studied extensively.

We will
 study the behavior of $J_n(q)$ in the limit of large $n,k$
with fixed ratio $\gamma=n/k$. We call this the semiclassical
limit. For physical values of $n$ and $k$ -- that is, positive
integers -- this limit can be analyzed by standard methods. One
simply sums over contributions of real saddle points of the path
integral; for physical values of $n$ and $k$, the expansion around
such real saddle points suffices for computing $J_n(q)$ to all
orders in an asymptotic expansion in $1/k$ (or in $1/n$ with $\gamma$ fixed).  The virtue of the
Morse theory machinery of this paper is that it enables us to
understand what happens in the semiclassical limit after analytic
continuation away from integer values of $k$ (and of $n$, if one
so chooses).

\subsection{Preliminaries}\label{preliminaries}

To apply the Morse theory machinery, we need to start at some
point in the complex $\gamma$ plane where we know how to express
the integration cycle  of Chern-Simons theory as a sum over
Lefschetz thimbles. With this in hand, one can then vary
$\gamma$ and take Stokes phenomena into account.

\def\ab{\mathrm{ab}}
If $k$ and $n$ are positive integers, so $\gamma>0$, Chern-Simons
theory is defined by an integral over $SU(2)$-valued connections
modulo gauge transformations. In the framework of the present
paper, this means that the integration cycle for $\A=A+i\phi$ is
the real cycle $\phi=0$, which we call $\CC_\R$.  This cycle can be
expressed in terms of Lefschetz thimbles as described most
precisely in eqn. (\ref{genexp}):
\begin{equation}\label{wenexp}\CC_\R=\sum_{\sigma\in
\Sigma_\R}\J_\sigma +\sum_{\sigma\in\Sigma_-}\n_\sigma
\J_\sigma,\end{equation} Here $\Sigma$ is the set of critical
points -- or rather critical gauge orbits, but we will speak
loosely of critical points.  The critical points are flat
connections on $S^3\backslash K$ with appropriate monodromy.
$\Sigma_\R$ consists of the real critical points (the ones with
$\phi=0$), and $\Sigma_-$ consists of complex critical points corresponding
to flat connections $\A$
for which the real part of the exponent of the path integral is
negative.  This exponent, including the contribution from the
knot, is
\begin{equation}\label{pathem}\I=ikW(\A)+(n-1)\ln\l.\end{equation}
The coefficients $\n_\sigma$ must be computed by counting
solutions of flow equations. Because of the difficulty of actually
doing this, our results for the figure-eight knot will be
incomplete.

An important topological fact about $SU(2)$ gauge theory is that
for the case of a knot in $S^3$ and monodromy sufficiently close
to the identity, the only real critical point corresponds to the
abelian flat connection $\A_\ab$. So for sufficiently small values
of $\gamma=n/k$, the contribution to $\CC_\R$ from the sum over
$\Sigma_\R$ is simply the cycle $\J_\ab$ associated to $\A_\ab$.

$\CC_\R$ is initially defined as a real integration cycle in
$\U_\C$, the space of complex-valued connections.  To analytically
continue to complex values of $k$, we must lift and extend
$\CC_\R$ to a cycle  $\CC\subset\hat\U_\C$; recall that
$\hat\U_\C$ is a cover of $\U_\C$ where the Chern-Simons function
$W(\A)$ is well-defined as a complex-valued function.   This
process was described schematically in section \ref{turz}.
Concretely, analytic continuation can be made by choosing a lift
to $\C$ of the Chern-Simons invariant $W(\A_\sigma)$ for each
critical point that appears in (\ref{wenexp}) with a nonzero
coefficient.  In section \ref{spurz}, a geometrical interpretation
was given for a choice in which, for some given real number
$\alpha$, one requires $\alpha< \im\,\I<\alpha+2\pi$ for all
$\sigma$.  The most general possible choice was described in
section \ref{gench}.

The choice that we will make of analytic continuation is guided by
the fact that for fixed $n$, the colored Jones polynomial has a
unique analytic continuation to a function $J_n(q)$ with no
essential singularity at $k=\infty$ (or equivalently at $q=1$).  Moreover \cite{GL}, in the
semiclassical limit, $J_n(q)$ is bounded as $n\to\infty$ for sufficiently small $\gamma$.
We
take these facts to mean that there is a unique choice of $\CC$ such
that, in a sufficiently small neighborhood of $\gamma=0$, the path
integral has no exponentially growing contributions; it receives
contributions only from critical points for which $\re\,\I$ is
zero or negative.   This should hold regardless of the direction
in which one approaches the point $\gamma=0$ in the complex plane.

The Chern-Simons invariant  of the abelian flat connection
$\A_\ab$ vanishes modulo $2\pi$. Avoiding exponential growth tells
us how to lift $\J_\ab$ to $\hat\U_\C$: we must pick the lift such
that the Chern-Simons invariant $W(\A_\ab)$ is 0 (rather than a
non-zero multiple of $2\pi$). Similarly, we pick the branch of the
logarithm such that $\ln\l=0$. Henceforth, when we refer to
$\A_\ab$ as a critical point in $\hat\U_\C$ or to $\J_\ab$ as a
cycle in $\hat\U_\C$, we have in mind this lifting, which we call
the canonical lifting.

For any $K$, the contribution of $\A_\ab$ to the integration cycle
is always precisely $\J_\ab$, regardless of $\gamma$.  In fact,
Stokes phenomena never cause a jumping in the coefficients with
which abelian critical points contribute to the integration cycle.
This is explained in sections \ref{moreg} and \ref{sotwo}. Jumping
definitely can occur for the coefficients of Lefschetz thimbles
associated to nonabelian critical points.

For a typical knot,  nonabelian critical points have nonzero $\I$
near $\gamma=0$; let us assume this to be the case. At first
sight, the argument of section \ref{essential} seems to show that
if flat connections with nonzero $\I$ contribute to the path
integral near $\gamma=0$, then an essential singularity for
$k\to\infty$ is unavoidable.  If this were the case, we would
conclude that in a neighborhood of $\gamma=0$, the desired
integration cycle $\CC$ must precisely coincide with $\J_\ab$ for
all knots. There seem to be two fallacies in this reasoning.

First, the argument in section \ref{essential} assumed that the
flow equations conserve $\im\,\I$.   However, as discussed in
section \ref{branches}, in the presence of a knot, this is only so
if one replaces the loop space of $G_/T_\C$ by its universal
cover, as one should do if one wishes to analytically continue
$J_n(q)$ to non-integral values of $n$.  In this case, a choice of
branch of $\ln\,\l$ is regarded as part of the data specifying a
critical point.   If the concept of a critical point is understood
in this way, then $\im\,\I$ is conserved in the Chern-Simons flow
equations and the argument of section \ref{essential} is
applicable.  Otherwise, we cannot make this argument.

At best (since we note another problem shortly), the conclusion from section \ref{essential} is that if $J_n(q)$, after being
analytically continued to non-integral $n$, has no essential singularity for $k\to\infty$
or $q\to 1$, then the integration cycle for sufficiently small $\gamma$ is simply
$\CC=\J_\ab$.

Unfortunately, although it seems clear in the framework of the
present paper that $J_n(q)$ can be analytically continued away
from integer values of $n$, this does not seem to have been
explored in the literature.  The claim that $J_n(q)$ has no
essential singularity at $k=\infty$ is only known in general for
integer $n$, and we have little guidance about whether this
remains so after analytic continuation in $n$.

The simplest example is the unknot $K_0$.  The explicit formula\footnote{Recall that in our
normalization,  $J_n(q)$ is defined by a path integral rather than a ratio of path integrals; for an
explanation of the relation of this to the usual normalization in the mathematical literature,
 see the last paragraph of section \ref{spurz}.}  is
\begin{equation}\label{zolly}J_n(q;K_0)=\sqrt{\frac{2}{k+2}}\sin\left(\frac{\pi n}{k+2}\right).
\end{equation}
Clearly, this function can be analytically continued in $n$, and
the analytically continued function has no essential singularity
at $k=\infty$.  For the trefoil, we will see at the end of section
\ref{sc} that for $n\not\in \Z$, there is an essential singularity
at infinity,\footnote{I am informed by C. Beasley that this can
also be verified for more general torus knots using formulas
 of \cite{Beasley}.} but the situation for more general knots is very unclear.

 The second problem is that Chern-Simons theory is more complicated than the situation
 considered in section \ref{essential}.  It is assumed there that the Stokes curves are straight
 rays, which is not the case in Chern-Simons theory, and perhaps more important, it is assumed
 that there are only finitely many critical points and Stokes curves.  In Chern-Simons theory,
 once one analytically continues away from integer $k$, there are infinitely many critical orbits
 (corresponding to different liftings of the Chern-Simons invariant of a flat connection) meeting at $\gamma=0$, and
 this can play an important role.  In the case of the trefoil
 knot, we will see that along the positive imaginary $\gamma$
 axis, arbitrarily close to $\gamma=0$, an infinite number of critical points must be included to
 describe the integration cycle.

We do not have a general theory of how to determine the
integration cycle so as to avoid exponential growth of the path
integral for $k\to\infty$, so we will have to take an exploratory
point of view in our examples. In our study of the trefoil, we
will show that $\J_\ab$ is indeed the correct integration cycle if
one approaches $\gamma=0$ along any direction except the positive
imaginary axis.    For the figure-eight knot, we will not make an
equally detailed analysis.

Once one knows the correct integration cycle near $\gamma=0$,
whether it is simply $\J_\ab$ or not, the appropriate integration
cycle for any $\gamma$ can be found in the usual way by varying
$\gamma$ and taking account of Stokes phenomena.

\subsubsection{Local Models}\label{locmod}

\def\ade{{\mathrm{ad}}(E)}
Since studying the behavior near a singularity of the moduli space
of flat connections is an important technique for understanding
Stokes phenomena,  it may be helpful for some readers if we
briefly discuss the deformation theory of a flat bundle. (When we
consider examples later, we do so in a more informal language, and
some readers may omit the following discussion.)

Let $E$ be a flat bundle over a manifold $M$, with structure group
$G$ and connection $\A=\A_0$.  Let $Y$ be the automorphism group
of $E$, and set\footnote{Here $\ade$ is the sheaf of infinitesimal gauge transformations,
leaving fixed the structure along the knot if a knot is present.} $V=H^1(M,\ade)$, $V'=H^2(M,\ade)$.  $V$
parametrizes first order deformations of $E$, and $V'$
parametrizes obstructions to such deformations.  In deformation
theory, it is shown that there is a $Y$-invariant map $\sigma:V\to
V'$ such that the moduli space of deformations of $E$ is
$\sigma^{-1}(0)/Y$. (Rather than dividing by $Y$, it may be more
convenient to think of the moduli space as the space
$\sigma^{-1}(0)$ with an action of $Y$.) We can expand $\sigma$ as
a power series in a variable $v\in V$.  The lowest order term is
quadratic and comes from the cup product. Higher order terms come
from Massey products. (For example, see \cite{JM}.
From a physical point of view, these higher
order terms arise by integrating out massive modes of the gauge
field; see section 4.2 of \cite{WCT}.)

If $M$ is a three-manifold, then $V'$ is dual to $V$ and the
discussion simplifies.  In this case, there is a $Y$-invariant
function $\Phi$ on $V$, and the map $\sigma$ is $\d\Phi$, the
differential of $\Phi$.  Again $\Phi$ can be expanded as a power
series in $v$.  The leading term is cubic and comes by evaluating
the classical action $\I(\A)$ as a function on $V$.  From a
physical point of view, $\Phi$ is the effective action for the
zero modes of the gauge field; it contains cubic couplings that
can be read off from  $\I(\A)$, and higher
order couplings (described mathematically in terms of Massey
products) that arise by integrating out massive modes.

In our application, there will be an additional parameter
$\gamma$, controlling the monodromy around a knot.  Suppose that
the flat bundle $E$ appears for $\gamma=\gamma_0$.  To include
$\gamma$ in the above discussion, simply set
$\epsilon=\gamma-\gamma_0$.  The effective action $\Phi$ can be
extended from a function on $V$ to a function on $V\times
\C_\epsilon$, where $\C_\epsilon$ is a copy of the complex plane
parametrized by $\epsilon$, in such a way that flat connections,
for any $\epsilon$, correspond to points in $V\times\C_\epsilon$
at which the equation $\d_V\Phi=0$ is obeyed.  Here $\d_V$ is the
exterior derivative in the $V$ direction.  In other words, flat
connections for a given $\epsilon$ correspond to critical points
of the restriction of $\Phi$ to the fiber of the projection
$V\times\C_\epsilon\to \C_\epsilon$ that lies above that value of
$\epsilon$.

The question now arises of how many Massey products must be
included to get a good model of the behavior near the flat bundle
$E$ that we started with.  The answer, assuming\footnote{This is
not always true in three-manifold theory, since in general the
moduli space of flat connections may have components of positive
dimension.  It is true in our applications involving knots.  In
general, one needs to include enough Massey products to lift
whatever degeneracies do get lifted.} that for $\epsilon=0$ (and
hence any $\epsilon$ near 0), there are only finitely many flat
bundles near $E$, is that one must go far enough to lift all
degeneracies. If the cubic form obtained by evaluating $W(\A)$ as
a function on $V$ is nondegenerate, this gives a complete answer
for $\Phi$; otherwise, one must carry out a power series to higher
order, including enough Massey products to get a nondegenerate
result.

Singularity theory gives a convenient framework for this
discussion. Pick a transversal $\U$ to the gauge orbit of the flat
bundle $E$.  Consider  $\I(\A)$ as a function on $\U$. This
function has a singularity (that is, a degenerate critical point)
at the point $\A=\A_0$.  Though $\U$ is infinite-dimensional, the
singularity can be modeled in finite dimensions.  In fact, after
making some  choices, one can embed $V$ in $\U$ and write $\U$ as
$V\times N$, where $N$ is an infinite-dimensional space that
parametrizes the nonzero modes of $\A$; we assume that $\A=\A_0$
corresponds to the product of the origin in $V$ and a point in $N$
that we call the origin. The decomposition $\U=V\times N$ can be
made in a $Y$-invariant fashion. Singularity theory says that one
can choose the decomposition so that $\I(\A)$ near $\A=\A_0$ is
the sum of a function on $V$ and a function on $N$, such that the
function on $V$ is the effective action $\Phi$, and the function
on $N$ has a nondegenerate critical point at the origin. The
function $\Phi:V\to\C$ thus gives a finite-dimensional model of
the singularity of the function $\I(\A):\U\to\C$. Because of the
choices that are made in identifying $\U$ as $V\times N$, $\Phi$
is only determined up to a diffeomorphism of $V$ (more exactly, a
$Y$-invariant diffeomorphism that leaves fixed the origin). Within
its equivalence class, one can take $\Phi$ to be a polynomial,
determined by finitely many Massey products. 

The importance of this for Chern-Simons theory is as follows. The
integral over $Y$ behaves in the semiclassical limit as an
infinite-dimensional Gaussian integral (with perturbative
corrections).  It does not contribute to Stokes phenomena near
$\A=\A_0$.  The interesting subtleties in the behavior of the
Chern-Simons path integral near the flat bundle $E$ in the
semiclassical limit are all contained in the integral over $V$.

We will encounter three examples of this construction.
Conceptually, the most simple example arises for the figure-eight
knot at $\gamma=1/3$, where two nonabelian flat connections meet.
The flat bundle $E$ is irreducible, and its automorphism group $Y$
is trivial.   The space $V$ is one-dimensional, and  $\Phi$ is the
function $\Phi=x^3/3-\epsilon x$ related to the Airy integral. As
$\Phi$ is cubic, Massey products play no role.  The other two
examples involve the case that $\A_0=\A_\ab$ is the abelian flat
connection, so the group $Y$ is $\C^*$.   For either the trefoil
or the figure-eight knot, there is a critical value of $\gamma$ at
which $\A_\ab$ meets one or more non-abelian flat connections. In
each case $V$ is two-dimensional and $Y$ acts with weights $1$ and
$-1$.  We write $u,v$ for coordinates on $V$ corresponding to
these weights.  For the trefoil knot, it turns out that the
appropriate local model is $\Phi=(uv)^2/2-\epsilon uv$.  The
quartic term in $\Phi$ comes from a Massey product.  For the
figure-eight knot, one gets instead $\Phi=(uv)^3/3-\epsilon uv$,
where now a Massey product of rather high order is involved in
generating the $(uv)^3$ term.

\subsubsection{Normalization Of The Path Integral}\label{normo}

Instead of simply saying, as we have just done, that the contribution to the Chern-Simons
path integral of a degenerate critical point can be modeled
by a corresponding finite-dimensional integral, it is possible, as we will briefly sketch,
 to be more precise and
to describe the  normalization factors.    The facts we will describe could be used
to sharpen the comparisons that we will describe later between the Chern-Simons path
integral and computations in the mathematical literature that are based on the exact
solution of Chern-Simons theory, obtained by combinatorial methods.  (We will not
do so in this paper and the reader might choose to omit this explanation.)

We first recall the nondegenerate case.  The contribution of a nondegenerate critical point
to the Chern-Simons path integral was evaluated in \cite{Wi},  compared to the exact
solution (and slightly corrected\footnote{A normalization was used in \cite{Wi} in which
$\eta$ and $\log T$ were both two times smaller than usual.}) in \cite{FG}, and further
refined in \cite{LJe,Roza}.  In Chern-Simons theory on a three-manifold
$M$ with compact gauge group $H$, let $E\to M$ be an $H$-bundle and suppose
that $H^0(M,\ade)=H^1(M,\ade)=0$. 
The automorphism group $P$ of $E$ is then a finite
group (in general, the Lie algebra of $P$ is $H^0(M,\ade)$).   Let $\I_E=ikW$ be the action of Chern-Simons
theory, evaluated for the critical point $E$.  The contribution of the flat bundle $E$ to the Chern-Simons path integral is then
\begin{equation}\label{sysmo}
\frac{1}{\# P}\exp(\I_E)\exp(i\pi\eta(\ade)/4)\sqrt{T(\ade)}.\end{equation}
where $\eta(\ade)$ is the Atiyah-Patodi-Singer $\eta$-invariant of the operator $\star\d+\d\star$ acting on odd degree forms with
values in
$\ade$; $T(\ade)$ is the Ray-Singer analytic torsion of this flat bundle; and $\# P$ is the
order of the finite group $P$.  The expression (\ref{sysmo}) arises by performing the
path integral for fluctuations around the flat bundle $E$, in the Gaussian approximation.
Higher order corrections will multiply this expression by an asymptotic series in powers of
$1/k$ (the same is true for eqn. (\ref{tysmo}) below), as explored in \cite{AS1,AS2}.

Now let us assume that $E$ corresponds to a degenerate critical point.  We continue
to assume that $H^0(M,\ade)=0$, but we take $V=H^1(M,\ade)$ to be nonzero.
We can still carry out a Gaussian integral over the nonzero modes of the gauge field
(the modes that in the terminology of section \ref{locmod} parametrize
 $N$).  The result of integrating over the nonzero
modes is still given by the formula (\ref{sysmo}), if properly interpreted.  There is one essential
point: for an acyclic flat connection, the torsion is a number, but in the general case,
as explained in the original construction of analytic torsion \cite{RS}, and used in the
present context in \cite{LJe}, it is a linear
form on  a certain
one-dimensional space -- constructed as a tensor product of the determinants of the cohomology
groups of the flat bundle in question.  In the present situation, the only nonzero cohomology
group is $V=H^1(M,\ade)$, and $\sqrt{T(\ade)}$ is a linear form on $\det V$.  It determines
a translation-invariant measure on $V$ that we will denote as $\d v$.

What we want to integrate over $V$, using this measure,  is essentially $\exp(\Phi)$, where the function $\Phi$ was described in section \ref{locmod}.  The generalization of (\ref{sysmo}) for $V\not=0$
is
\begin{equation}\label{tysmo}
\frac{1}{\# P}\exp(i\pi\eta(\ade)/4)(k/2\pi^2)^{\mathrm{dim}\,V/2}\int_V\d v\exp(\Phi),\end{equation}
where we replace $\sqrt{T(\ade)}$ by the measure $\d v$ and use it to integrate over $V$.
(The factor $\exp(\I_E)$ has disappeared as $\I_E$ is simply the value of $\Phi$
at $v=0$.  If $v=0$ were a nondegenerate critical point of $\Phi$, the factor $k^{\mathrm{dim}\,V/2}$ would be canceled upon doing the Gaussian integral over $V$.
That this factor is  precisely a power of  $k/2\pi^2$ was found for $H=SU(2)$ in \cite{Roza}.
This has apparently not been established in complete generality, though it can plausibly
be done using localization formulas \cite{BW}.)
The idea of (\ref{tysmo}) is that the integrand $\exp(i\pi\eta(\ade)/2)\,\d v\exp(\Phi)/\# P$ comes from the path
integral over $N$ and we complete the path integral by integrating over $V$.

In the general case (and in fact in two of the three examples mentioned at the end of
section \ref{locmod}), we must consider a flat bundle $E$ such that $H^0(M,\ade)\not=0$.
The automorphism group $P$ of $E$ then has dimension
$d_P=\mathrm{dim}\,H^0(M,\ade)$.  
Moreover, $\sqrt{T(\ade)}$ is now the ratio of a measure on $V$ and one on the Lie algebra
$\frak p$ of $P$.  A convenient way to proceed is to pick a metric on $\frak p$ (such as
the usual one in which short coroots have length squared two) and let $\mathrm{Vol}(P)$ be
the volume of $P$ using this metric.  Once we have picked a metric and hence a measure on 
$\frak p$, $\sqrt{T(\ade)}$ can be regarded just as before as a measure $\d v$ on $V$.
The generalization of (\ref{tysmo}) is then
\begin{equation}\label{utysmo}
\frac{1}{\mathrm{Vol}( P)}\exp(i\pi\eta(\ade)/4) (k/2\pi^2)^{(\mathrm{dim}\,V-\mathrm{dim}\,P)/2}\int_V\d v\exp(\Phi).\end{equation}

The factor $\exp(i\pi\eta(\ade)/4)$ depends on the metric of $M$.  The metric dependence can
be removed (by subtracting a ``counterterm'') once a framing (or two-framing) of $M$ is picked; this was the approach in \cite{Wi}.  A canonical two-framing exists \cite{At} and is often used.
The formulas specialized to this two-framing were described in detail in \cite{FG} for $H=SU(2)$.

\subsubsection{Orientation Of The Integration Cycle}\label{orcycle}

 As a last preliminary comment, we should perhaps discuss the fact
that although the invariant $J_n(q;K_0)$ of the unknot -- given in
eqn. (\ref{zolly}) -- has no essential singularity at $k=\infty$,
it does have a square root singularity there. This square root
singularity is common for all knots.\footnote{It cancels in ratios
$J_n(q;K)/J_n(q;K_0)$, which are the functions usually considered
in the mathematical literature.  These ratios are Laurent
polynomials in $q=\exp(2\pi i/(k+2))$.} The meaning of the square
root singularity is that under analytic continuation around
$k=\infty$, the orientation of the integration cycle $\CC$ is
reversed.

This reversal of orientation under analytic continuation is not an
exotic phenomenon. It occurs for the simple Gaussian integral
$\int_{-\infty}^\infty \d x\,\exp(ikx^2)=\sqrt{\pi/ik}$, where the
square root shows that under analytic continuation around
$k=\infty$, the orientation of the integration cycle is reversed.
(Concretely, this happens because if $k$ is rotated by an angle
$\phi$ in the complex plane, the integration cycle in the
$x$-plane should be rotated by an angle $\phi/2$ to maintain
convergence of the integral.  For $\phi=2\pi$, we get back the
original integration cycle with opposite orientation.)  The
existence of this monodromy shows that the treatment of the
orientation of the integration cycle will be subtle in
Chern-Simons theory, but we will phrase our discussion  to
minimize the role of such details.

\subsection{The Trefoil Knot}\label{trefoil}

\def\x{{\mathbf x}}
\def\y{{\mathbf y}}
For
both of our examples, the necessary facts about the representation
of the fundamental group of the knot complement are described in
\cite{Mura}.  We begin with the case that our knot $K$ is the trefoil.

The fundamental group of $S^3\backslash K$ can be described by two generators
$ \x,\y$ with one relation $\x \y \x=\y \x \y$. $\x$ and $\y$ are both conjugate to the meridian $\hat\m$;
for further details, see fig. 11 in \cite{Mura}.  The longitude is $\hat\l=\y \x^2\y \x^{-4}$.

For every choice of the conjugacy class of $\hat\m$ (apart from
certain exceptional values that will be discussed), there are up
to gauge transformation two $SL(2,\C)$ flat connections on
$S^3\backslash K$.  The first is an abelian flat connection
$\A_\ab$ with $\x=\y=\rm{diag}(\m,\m^{-1})$. (As usual $\m$ is the eigenvalue of the meridian $\hat\m$.)  Its Chern-Simons invariant vanishes.

\def\nab{\mathrm{nab}}
There is also a nonabelian flat connection $\A_\nab$ whose
monodromies  are given up to conjugation by\footnote{Our $\m$ is
called $m^{1/2}$ in \cite{Mura}.}
\begin{equation}\label{ozord}\x=\begin{pmatrix}\m & 1\cr 0 & \m^{-1}\end{pmatrix},~~\y=\begin{pmatrix}\m&
0\cr 1-\m^2-\m^{-2}& \m^{-1}\end{pmatrix}.\end{equation} The
eigenvalue of the longitude is
\begin{equation}\label{orff}\l=-\m^{-6}.\end{equation}

According to (\ref{forsk}), in $SU(2)$ Chern-Simons theory, $\m$
is of modulus 1, so we will need to analyze this case carefully.
If $|\m|=1$, the flat connection $\A_\nab$ has a structure group
that reduces to either $SU(2)$ or $SL(2,\R)$. This can be
anticipated as follows. The complex conjugate of a representation
of the fundamental group in $SL(2,\C)$ is still a representation
of the fundamental group in $SL(2,\C)$, in general with
$\m\to\bar\m$. However, if $|\m|=1$, the operation $\m\to\bar\m$
is equivalent to $\m\to\m^{-1}$ and does not affect the conjugacy
class of the meridian, which is
\begin{equation}\label{pozord}\begin{pmatrix}\m & 0 \cr 0 &
\m^{-1}\end{pmatrix}.\end{equation}  So for $|\m|=1$,
representations of the fundamental group in $SL(2,\C)$ occur in
complex conjugate pairs or else are isomorphic to their own
complex conjugates.  A representation with the latter property has
a structure group that reduces to a real form of $SL(2,\C)$ --
either $SU(2)$ or $SL(2,\R)$.

For the trefoil knot, as there is only one nonabelian
representation up to conjugacy, it must be equivalent to its own
complex conjugate.  Concretely, let us try to conjugate the
representation into $SU(2)$.  We set
\begin{equation} C=\begin{pmatrix} 1 & -({\m-\m^{-1}})^{-1}\cr 0
& 1\end{pmatrix},\end{equation} so
\begin{equation}\label{toxo}C \x C^{-1} =\begin{pmatrix} \m & 0 \cr 0 &
\m^{-1}\end{pmatrix}\end{equation} is an element of $SU(2)$. Then
\begin{equation}\label{poxo} C\y C^{-1}= \begin{pmatrix} (\m-\m^3)^{-1} &
(\m-\m^{-1})^{-2}\cr 1-\m^2-\m^{-2}
&(\m^{-1}-\m^{-3})^{-1}\end{pmatrix}.\end{equation} If $\x$ and $\y$
can be conjugated into $SU(2)$, this can be done in a way that
makes $\x$ diagonal. Having put $\x$ in the form (\ref{toxo}), the
only further conjugation that leaves it diagonal is by a diagonal
matrix $\rm{diag}(\lambda,\lambda^{-1})$.  Conjugation by a
diagonal matrix can map a unimodular matrix of the form
\begin{equation}\label{moxo}\begin{pmatrix}a & b\cr c & \bar
a\end{pmatrix}, \end{equation} such as $C\y C^{-1}$, into $SU(2)$ if
and only if $b\bar c<0$. For $|\m|=1$, this is equivalent to
\begin{equation}\label{ofox}1>\m^2+\m^{-2},\end{equation}
and that therefore is the condition that the flat connection
$\A_\nab$ has structure group $SU(2)$.  If instead
\begin{equation}\label{zoofox}1<\m^2+\m^{-2},\end{equation}
the structure group of $\A_\nab$ reduces to $SL(2,\R)$.

\subsubsection{Singularities Of The Moduli Space Of Flat
Connections}\label{singmod}

The remaining case $1=\m^2+\m^{-2}$ requires further study.  This
case corresponds to a singularity of the moduli space of flat
connections, where different branches meet, as we will now
describe.

The condition $1=\m^2+\m^{-2}$ is equivalent to
\begin{equation}\label{bordox}\ln\m=\pm i\pi/6,~~\rm{mod}\,\,\pi i\Z.\end{equation} In this case, the
monodromies of eqn. (\ref{ozord}) are strictly upper triangular,
and the structure group of a flat connection $\A_\nab$ with those
monodromies reduces to the triangular subgroup of $SL(2,\C)$, and
not to either $SU(2)$ or $SL(2,\R)$.  In this exceptional case,
there is also another flat connection $\A'_\nab$ that is
inequivalent to $\A_\nab$; its monodromies are the lower
triangular cousins of eqn. (\ref{ozord}):
\begin{equation}\label{ozordo}\x=\begin{pmatrix}\m & 0\cr 1 & \m^{-1}\end{pmatrix},~~\y=\begin{pmatrix}\m&
0\cr 0& \m^{-1}\end{pmatrix}.\end{equation} $\A'_\nab$ is actually
gauge equivalent to the complex conjugate of $\A_\nab$; indeed,
eqn. (\ref{ozord}), specialized to $\m^2+\m^{-2}=1$, can be
transformed to (\ref{ozordo}) via complex conjugation composed
with conjugation by the matrix $\Omega=\begin{pmatrix}0&1\cr
-1&0\end{pmatrix}$.

Both $\A_\nab$ nor $\A'_\nab$ represent unstable critical points
in Chern-Simons theory, as their structure groups are strictly
triangular. In fact, when $1=\m^2+\m^{-2}$, the only semistable
critical point is the abelian one $\A_\ab$ with
$\x=\y=\rm{diag}(\m,\m^{-1})$.  Let us discuss the behavior of
Chern-Simons theory in the vicinity of $\A_\ab$.  We think of
$\A_\ab$ as a diagonal flat connection with $\C^*$ structure group
\begin{equation}\label{ormok}\begin{pmatrix}* & 0 \cr 0 &
*\end{pmatrix}\end{equation} embedded in $SL(2,\C)$.  If $\m$ is
fixed (by specifying the values of $n$ and $k$ in Chern-Simons
gauge theory), then $\A_\ab$ has no diagonal deformations as a
flat connection.  However, it does have off-diagonal deformations.
There is a one-dimensional space of deformations of upper
triangular form
\begin{equation}\label{rmok}\begin{pmatrix}0 & u\cr 0 & 0
\end{pmatrix}.\end{equation}
A deformation of $\A_\ab$ with $u\not=0$ gives, up to gauge
transformation, the triangular flat connection $\A_\nab$. The
value of $u$ does not matter, as long as it is nonzero, because
$\A_\ab$ has a $\C^*$ group of symmetries consisting of constant
gauge transformations $\rm{diag}(\lambda,\lambda^{-1})$, and these
act on $u$ by $u\to\lambda^2 u$.  Similarly (or by complex
conjugation composed with the gauge transformation $\Omega$),
$\A_\ab$ has a one-dimensional space of lower triangular
deformations
\begin{equation}\label{rmoko}\begin{pmatrix}0 & 0\cr -v & 0
\end{pmatrix}.\end{equation}
$v$ transforms under $\C^*$ by $v\to\lambda^{-2}v$, so if $v$ is
nonzero, its value does not matter.  A lower triangular
deformation of $\A_\ab$ with nonzero $v$ gives $\A'_\nab$.

If we take $u$ and $v$ to be both nonzero, then only their product
$uv$ is $\C^*$-invariant.  It is not possible to take both
$u,v\not=0$ without varying $\m$ away from $\m^2+\m^{-2}=1$.  This
is evident from (\ref{ozord}), according to which a
two-dimensional representation of the fundamental group of
$S^3\backslash K$ that is not triangular has $\m^2+\m^{-2}\not=1$.
Let $\epsilon=1-\m^2-\m^{-2}$; $\epsilon$ is real if $|\m|=1$. In
the space of possible deformations of $\A_\ab$ to an irreducible
flat connection, we can normalize the coordinates $u,v$ so that
the relation between $u,v$, and $\epsilon$ is $uv=\epsilon$.  This
statement simply says that for a non-abelian flat connection $\A_\nab$,  the $\C^*$-invariant $uv$ has a unique
value for all $\epsilon$ (since $\A_\nab$ is unique up to gauge
transformation), with $uv=0$ at $\epsilon=0$ (as we know from eqn.
(\ref{ozord})), and finally the zero of $uv$ at $\epsilon=0$ is
simple (the monodromies (\ref{ozord}) show that it is possible to
have $u\sim 1$, $v\sim \epsilon$, and so $uv\sim \epsilon$; this
should be no surprise as there is no symmetry that would force
$uv$ to have a zero of higher order).

Now let us specialize this to $SU(2)$ Chern-Simons theory.  In
this case, the gauge field $\A$ should be anti-hermitian, so we
want $v=\bar  u$. Also, the symmetry group of the abelian flat
connection reduces to $U(1)=SO(2)$. If we set $u=x+iy$, $v=x-iy$,
where $x,y$ are real coordinates on the space of first order
deformations of the flat $SU(2)$ connection $\A_\ab$, then the
relation between $x,y$ and $\epsilon$ is
\begin{equation}\label{zonkomo}x^2+y^2=\epsilon.\end{equation}
This equation has real solutions only for $\epsilon\geq 0$; this encodes the fact that
$\A_\nab$ can be conjugated to $SU(2)$ only under that condition.
Let us consider the exponent $\I=ik W(\A)+iI_R(\rho,\A)$ of the
path integral of Chern-Simons theory for small $\epsilon$ and near
the abelian flat connection $\A_\ab$. It has critical orbits with
$x=y=0$ and with $x^2+y^2=\epsilon$.  Moreover, $\I$ vanishes for the abelian flat
connection $\A_\ab$, corresponding to $x=y=0$, and in general is of order $n$ in
the semiclassical region of large $n$ with fixed $\gamma=n/k$.  Finally it has one nonzero
critical orbit at $x^2+y^2=\epsilon$.
These facts suggest that $-i\I/n$ can be modeled near $x=y=\epsilon=0$ by the simple $SO(2)$-invariant
function
\begin{equation}\label{polyx}F(x,y)=c\left(\frac{1}{2}(x^2+y^2)^2-\epsilon(x^2+y^2
)\right)\end{equation}
with a real constant $c$.   We will verify this shortly using (\ref{morkop}), but for
the moment let us discuss the implications.

Generically, in expanding around a flat connection, the
Chern-Simons function is a nondegenerate Morse function in the
directions transverse to the gauge orbits.  When this is so, after
gauge fixing, the expansion around a given critical point can be
approximated for large $n$ and $k$ as an infinite-dimensional
(oscillatory) Gaussian integral.  Nondegeneracy is visible in
(\ref{polyx}) for $\epsilon\not=0$.  As long as $\epsilon$ does
not vanish, $x=y=0$ is a nondegenerate critical point of $F(x,y)$,
and near the critical orbit $x^2+y^2=\epsilon$, the function
$F(x,y)$ is nondegenerate in the directions transverse to the
$SO(2)$ action.   However, this nondegeneracy and the Gaussian
nature of the Chern-Simons path integral break down at
$\epsilon=0$, where  the two critical orbits meet.

The expansion of the Chern-Simons path integral around a critical point can then no
longer be approximated by a Gaussian, but the Gaussian approximation fails only for the
two variables $x$ and $y$.  To improve upon the Gaussian approximation, we need
only a suitable approximation to the dependence of $\I$ on $x$ and $y$ near
the singularity.  Assuming
that (\ref{polyx}) is an appropriate approximation, we can model the integral over the
variables $x$ and $y$ by
\begin{equation}\label{mooro}\int \d x\,\d y\exp\left(in F(x,y)\right).\end{equation}
(For the constant multiplying this integral, see (\ref{tysmo}).)
  The integral (\ref{mooro}) was studied
in section \ref{sotwo} as a basic example of an oscillatory
integral with  symmetry. There were two main results in that
discussion.

 First, in eqns. (\ref{bombo}) and (\ref{orkoff}), 
 we analyzed the integral by a simple scaling argument.  According to (\ref{bombo}),
 the integral is $n^{-1/2}$ times a transcendental function (essentially an error function) of $n^{1/2}\epsilon$.
 In the limit $n\to\infty$, $\epsilon\to 0$ with $n^{1/2}\epsilon$ fixed,
 the colored Jones polynomial will equal $n^{1/2}$ times this error function (times
 a constant described in  (\ref{tysmo})).  The factor of $n^{1/2}$ is most obvious in the special
 case $\epsilon=0$: 
 according to (\ref{orkoff}),
 compared to the behavior with
$\epsilon\not=0$, the  large $n$ behavior of the integral gains an extra factor of $n^{1/2}$
at $\epsilon=0$.  We can immediately apply this to the invariant
$J_n(k)$ at the values of $n/k$ (such as $1/6$ or $5/6$) that
correspond to $\epsilon=0$ or $\m^2+\m^{-2}=1$.   We expect
$J_n(k)$ to grow with a factor of $n^{1/2}$ at $\epsilon=0$,
relative to its behavior for generic real $\epsilon$:
\begin{equation}\label{omsky} \frac{\left.J_n\right|_{\epsilon=0}}{\left. J_n\right|_{\epsilon\not=0}}\sim
n^{1/2},~n\to\infty.\end{equation} This is a known result (Theorem
1.2 of \cite{km}).  We have expressed the result in terms of a
ratio of values of $J_n$ because our normalization of $J_n$ (as a
path integral rather than a ratio of path integrals) is not the
most common one in the mathematical literature.  With the usual
normalization, the colored Jones polynomial is of order 1 for
$\epsilon<0$ and of order $n^{1/2}$ at $\epsilon=0$.

Second, in section \ref{sotwo} we made a direct analysis of the
Stokes phenomena in passing through $\epsilon=0$ (or more
generally in crossing Stokes lines that pass through
$\epsilon=0$).   These results only depend on  the singularity at
$\epsilon=0$, so  they carry over immediately to Chern-Simons
theory. Thus, we can understand some Stokes phenomena in $SU(2)$
Chern-Simons theory without having to solve four-dimensional flow
equations.  Details are discussed  in section \ref{asjones}.

Finally let us return to (\ref{polyx}), but now consider it as a
model of the behavior of $SL(2,\C)$ gauge fields near the
singularity at $\epsilon=0$. In $SL(2,\C)$, $x$ and $y$ are
complex and the symmetry group is $\C^*$.  Unlike the case of
$SU(2)$, where the critical orbit $x^2+y^2=\epsilon$ becomes
equivalent to $x=y=0$ if we set $\epsilon=0$, in $SL(2,\C)$ the
equation $x^2+y^2=0$ can be solved with $x,y\not=0$.  Modulo the
$\C^*$ symmetry (which acts by $x\pm i y\to \lambda^{\pm
2}(x+iy)$), the locus with $x^2+y^2=0$ but $(x,y)\not=(0,0)$
consists of two points, $(x,y)=(1,\pm i)$.  These two points
correspond to the triangular flat connections $\A_\nab$,
$\A'_\nab$.  It is also possible to specialize the analysis to
$SL(2,\R)$ gauge fields -- the same singularity will also play a
role in $SL(2,\R)$ Chern-Simons theory --  but we leave this for
the reader.

\subsubsection{The Chern-Simons Function}\label{ormox}

 To understand possible Stokes phenomena and to
 determine the asymptotic behavior of the Chern-Simons path integral, we must evaluate
 the Chern-Simons function $W(\A)$ for the flat connection $\A_\nab$.
 For this, we use (\ref{csvalue}), which along with (\ref{orff}) gives $\delta
W(\A_\nab)=-(6/\pi)\delta\m/\m\cdot\ln \m=-(3/\pi)\delta\ln^2\m$.
So $W(\A_\nab)=-3\ln^2\m/\pi+ {\mathrm{ constant}}$.  To determine
the constant, we observe that in general $W(\A)$ vanishes mod
$2\pi$ for a representation with triangular structure group, which
we have in the present problem at the values of $\ln\m$ given in
(\ref{bordox}). Thus
\begin{equation}\label{polj}W(\A_\nab)=-\frac{3}{\pi}\ln^2\m-\frac{\pi}{12}-2\pi s,~~s\in\Z.\end{equation}
As we are planning to analytically continue in $k$, we have made explicit the possibility of
adding to $W(\A)$ an integer multiple of $2\pi$.
We can now evaluate $\I=ikW(\A)+(n-1)\ln\l$, using $\l=-\m^{-6}$ and $\ln\m=i\pi(n-1)/k$ to
evaluate $\ln\l$.
In the semiclassical limit of large $n,k$, we ignore the difference between $n-1$ and $n$.
However, we do want to take into account the multivaluedness of $\ln\l$, as explained at the end of section
\ref{zoknot}.  We get
\begin{equation}\label{morkop}\I(\A_\nab)=ik\left(\frac{3\pi n^2}{k^2}-\frac{\pi}{12}-
2\pi s\right)-in\left(6\pi\frac{n}{k} +2\pi (r-1/2)\right),~~r,s\in\Z.
\end{equation}
Alternatively, in terms of $\gamma=n/k$, we can write
\begin{equation}\label{orkop}\I(\A_\nab)=-in\bigl(3\pi\gamma+(\pi/12+2\pi s)\gamma^{-1}+2\pi(r-1/2)\bigr).
\end{equation}

We can  use these formulas to justify (\ref{polyx}).
 The relation
between $\m$ and $\gamma$ is $\ln\m=i\pi\gamma$, so
$\epsilon=1-\m^2-\m^{-2}$ vanishes precisely if
\begin{equation}\label{zelk}\gamma=\pm (1/6+p),~
p\in\Z.\end{equation}
For any number $y$ that is congruent to $\pm 1/6$ mod $\Z$, the nonabelian flat connection $\A_\nab$ with $\gamma=y$ and
\begin{align}\label{zorx}s&=\frac{3y^2}{2}-\frac{1}{24}\cr
                                                        r& = \frac{1}{2}-3y\end{align}
is equivalent to an abelian flat connection $\A_\ab$ with its
canonical lift (zero Chern-Simons invariant and $\ln\l=0$).  The
values of $r$ and $s$ were chosen to ensure vanishing of both
contributions to $\I(\A_\nab)$, namely $ik W(\A_\nab)$ and
$(n-1)\,\ln\,\l$.   With these values of $r$ and $s$, we find for
any $\gamma$
\begin{equation}\label{telk}\I(\A_\nab)=-3\pi i n\gamma^{-1}\bigl(\gamma-y\bigr)^2.\end{equation}
So for $\gamma=y+\epsilon$, $\I$ is of order $n\epsilon^2$. The
fact that the coefficient of $\epsilon^2$ is nonzero is the
content of (\ref{polyx}).  Any $SO(2)$-invariant function
$F(x,y,\epsilon)$ that vanishes at $x=y=0$, has only one critical
orbit with $(x,y)\not=0$ that approaches $(x,y)=0$ for
$\epsilon\to 0$, and is of order $\epsilon^2$ at this orbit, can
be put in the form (\ref{polyx}) modulo a change of variables (an
$SO(2)$-invariant redefinition of $x$ and $y$ that is the identity
near $x=y=0$). Possible higher order terms in $F$ are irrelevant
in the sense of singularity theory and do not affect the estimate
(\ref{omsky}) of the critical behavior of the Chern-Simons path
integral.

\subsubsection{The Integration Cycle And The Asymptotic Behavior Of The Path Integral}\label{asjones}

An obvious consequence of (\ref{morkop})  is that for real $k$ and $n$, the nonabelian critical points have
$\I$ imaginary.  Since  $\re\,\I$ is strictly increasing or decreasing along a flow line, and vanishes
on the real integration cycle $\CC_\R$, this implies
that for real $n$ and $k$, there are no flows between $\CC_\R$ and
a nonabelian critical point.  Hence for $\gamma$ real and small, the integration cycle $\CC_\R$ is
simply the Lefschetz thimble $\J_\ab$ associated to the real critical point $\A_\ab$.

Extending this reasoning slightly, for real $n$ and $k$, all
critical points, abelian or nonabelian, have $\re\,\I=0$. So
 there are no non-trivial flows between them.

How then can anything nontrivial happen for real values of $n,k$?
Something nontrivial must happen because for suitable real $\gamma$, the nonabelian
flat connection $\A_\nab$ is $SU(2)$-valued and hence must contribute to the $SU(2)$
Chern-Simons path integral.

The answer to this question is that on the real $\gamma$ axis,
jumping of the Lefschetz thimbles $\J_\sigma$, or of the
coefficients in the usual expansion $\CC=\sum_\sigma\n_\sigma
\J_\sigma$, occurs only at values of $\gamma$ where two different
flat connections become coincident. At these points, which were identified in
eqn. (\ref{zelk}), Chern-Simons
theory has jumping phenomena that can be understood from a
finite-dimensional model of the appropriate singularity, as
described in section \ref{singmod}.

Here we will analyze the integration cycle of the Chern-Simons path integral,
and its asymptotic behavior, for $\gamma$ close to the real axis.  For a full analysis of what
happens throughout the complex $\gamma$ plane, one must analyze the Stokes curves;
we postpone this to section \ref{sc}.

We want to express the integration cycle of the Chern-Simons path
integral for the trefoil knot in terms of Lefschetz thimbles
associated to critical points. To this aim, we let $\J_\ab$ be the
Lefschetz cycle associated to the abelian critical point $\A_\ab$
with its canonical lift.  $\J_\ab$ is the appropriate integration
cycle for sufficiently small $\gamma=n/k$.  It will become clear
that cycles associated to other ways of lifting the Chern-Simons
invariant $W(\A_\ab)$ will not play any role.  (This is actually a
general property for any knot.)

On the other hand, we do have to consider distinct cycles
associated to different ways of lifting the Chern-Simons invariant
$W(\A_\nab)$ of the nonabelian critical point $\A_\nab$, in other words
different choices of the integer $s$ in (\ref{polj}).  For each
number $x$ that is congruent to $\pm 1/6$ mod $\Z$, we write $\J_x$ for
the Lefschetz thimble associated to $\A_\nab$, with its
Chern-Simons invariant determined by the choice $s=3x^2/2-1/24$
as in (\ref{zorx}).  These cycles differ only by the lifting of the Chern-Simons invariant,
so they are equivalent when the Chern-Simons level $k$ is an integer.

We fix the relative orientations of the cycles $\J_x$ by requiring
that when $k$ is an integer, the Chern-Simons path integral over
$\J_x$ is independent of $x$.  Momentarily, we will fix the
overall orientations of the $\J_x$ by making a choice for
$\J_{1/6}$.

Near $\gamma=0$, the appropriate integration cycle of Chern-Simons
theory is simply $\J_\ab$.  Now we will explore what happens when
$\gamma$ departs from zero, but only near the real $\gamma$ axis, so that we
only need to worry about jumping due to singularities.
  As we increase $\gamma$ along the
positive real axis, nothing happens until we reach the value
$\gamma=1/6$.  At this point, we meet a singularity of a type
analyzed in sections \ref{sotwo} and \ref{singmod}, associated
with the fact that $\A_\ab$ and $\A_\nab$ become stably equivalent\footnote{This means
the following.  $\A_\nab$ is triangular at $\gamma=1/6$ and its diagonal part coincides with $\A_\ab$,
so the gauge orbit containing $\A_\nab$ has $\A_\ab$ in its closure.  Hence, to
construct a sensible (Hausdorff) moduli space of flat connections, one has to consider
$\A_\nab$ and $\A_\ab$ to be equivalent.}
at $\gamma=1/6$.  The resulting Stokes phenomenon was analyzed in
eqn. (\ref{klp}): $\J_\ab$ jumps by $\pm \J_{1/6}$, depending on
orientation.  ($\J_\ab$ and $\J_{1/6}$ correspond to $\J_p$ and
$\J_\O$ in (\ref{klp}).) We pick the orientation so that for
$\gamma$ slightly greater than $1/6$, the integration cycle of
Chern-Simons theory is $\J_\ab+\J_{1/6}$.

There is a simple intuitive meaning to this result.  In general,
for integer values of $n,k$, the path integral of Chern-Simons
theory is given asymptotically for large $n$ and fixed $n/k$ by a
sum over contributions of all $SU(2)$-valued flat connections, all
with weight 1. For $1/6<n/k<5/6$, there are two such connections,
namely $\A_\ab$ and $\A_\nab$, and both must contribute with
weight 1 to the integration cycle.  As long as $k\in\Z$, the
different $\J_x$ are equivalent, and it does not matter which of
them we include.  But to describe the analytic continuation away
from integer $k$ in the region just to the right of the Stokes
curve that passes through $\gamma=1/6$, the right integration
cycle is $\J_\ab+\J_{1/6}$.

Increasing $\gamma$ further, nothing new happens until we reach
$\gamma=5/6$.  In crossing $\gamma=5/6$, nothing happens to
$\J_{1/6}$, but $\J_\ab$ will jump again, picking up $\pm
\J_{5/6}$.  We claim that (with the orientation convention chosen
above) a minus sign is the right choice here, and hence that the
appropriate integration cycle just beyond $\gamma=5/6$ is
$\J_\ab+\J_{1/6}-\J_{5/6}$.

To explain why the sign is negative, consider again the behavior
for integer $n,k$ with $5/6<n/k<7/6$.  In this region, the only
$SU(2)$-valued flat connection is the abelian one, and hence the
large $n$, fixed $n/k$ behavior of the path integral must be given
by the contribution of $\A_\ab$ alone.  The contribution of
$\A_\nab$ must vanish for integer $k$.  Since $\J_{1/6}$ and
$\J_{5/6}$, which arise from two different liftings of $\A_\nab$,
are equivalent for integer $k$, the minus sign in the
integration cycle $\J_\ab+\J_{1/6}-\J_{5/6}$ ensures this.

Increasing $\gamma$ further, nothing happens until we reach
$\gamma=7/6$.  At this point, $\J_\ab$ jumps by $\pm\J_{7/6}$. In
this part of the real $\gamma$ axis, there is again an
$SU(2)$-valued flat connection that must contribute with weight 1.
So the sign is $+$ and the integration cycle is
$\J_\ab+\J_{1/6}-\J_{5/6}+\J_{7/6}$.  The next jump is at
$\gamma=11/6$, where we must add $-\J_{11/6}$ to the integration
cycle.

In general, if $x>0$ is a number of the form $\pm
(1/6+p),~p\in\Z$, then just to the right of $\gamma=x$, the
appropriate integration cycle is
\begin{equation}\label{ozmurk}\J_\ab+\J_{1/6}-\J_{5/6}\pm\dots\pm
\J_x.\end{equation} Similarly, for negative $\gamma$, the
integration cycle is $\J_\ab+\J_{-1/6}-\J_{-7/6}\pm \dots\pm
\J_x$, for the most negative relevant value of $x$.

Having found the right integration cycle, it is straightforward to
determine the asymptotic behavior of the Chern-Simons path
integral for $n\to\infty$ with fixed $\gamma$.
In general,  a flat connection $\A_\sigma$ with $\n_\sigma\not=0$
makes a contribution to the path integral which in the semiclassical limit is
of order $ \exp(\I_\sigma)$, where $\I_\sigma$ is the value of the exponent of the path
integral at $\A=\A_\sigma$.     Unless $\I_\sigma=0$, this contribution has an exponential
dependence on $n$ for large $n$, since $\I_\sigma$ is of order $n$ in the semiclassical
limit.  In general, $\exp(\I_\sigma)$ multiplies an asymptotic series in (not necessarily
integral) powers of $n$.  In the context of an expansion around a complex critical point,
this series has been studied in detail in \cite{DGLZ}.
The dominant contribution to the path integral, for a given $\gamma$, is the
one that maximizes $\re\,\I_\sigma$, subject to the condition $\n_\sigma\not=0$.

In the case of the trefoil knot, the abelian flat connection
$\A_\ab$ has $\I_\ab=0$. Its contribution $Z_\ab$ to the path
integral is bounded (in fact, of order $1/\sqrt n$) for
$n\to\infty$ with fixed $\gamma$. On the other hand, a non-abelian
connection $\A_\nab$, lifted so that it meets the abelian one at
$\gamma=y$, has $\I_y=-3\pi i n\gamma^{-1}(\gamma-y)^2$, according
to (\ref{telk}).  So, assuming the relevant coefficient -- which we will call $\n_y$ -- is not zero,  the asymptotic behavior of its contribution
to the Chern-Simons path integral is
\begin{equation}\label{melky}Z_y\sim\exp(-3\pi i n\gamma^{-1}(\gamma-y)^2).\end{equation}
These contributions are all oscillatory as long as $\gamma$ is on
the real axis. (On the real axis, they actually dominate $Z_\ab$
in absolute value by a factor of $\sqrt n$ for large $n$.) As soon
as $\gamma$ departs from the real axis, the $Z_y$ grow or decay
exponentially for $n\to\infty$, depending on the sign of the real
part of the exponent.    If $\gamma=a+ib$ with $a,b$ real, the real part of the exponent is
\begin{equation}\label{defreal}\re\,\log\,Z_y=3\pi n b\left(1-\frac{y^2}{a^2+b^2}\right).\end{equation}
Near the real axis, and assuming  $|a|>|y|$ (as otherwise
$\n_y=0$), this is positive if and only if $b>0$.  So near the
real axis, the condition for exponential growth of $J_n(q)$  is
that $|a|>1/6$ (so that one of the $\n_y$ is nonzero) and
$\im\,\gamma>0$.  In this region, $\re\,\log\,Z_y$ is maximal for
$y=\pm 1/6$, so this contribution dominates $J_n(q)$.   In section
\ref{sc}, we will analyze the Stokes curves and extend this
analysis to the rest of the complex $\gamma$ plane.

What we have just obtained (except so far restricted to the region near the real $\gamma$ axis)
is the standard picture of the colored
Jones polynomial of the trefoil knot.  For example, see eqn. (2.2) of \cite{Mm} or
Proposition 3.2 of \cite{km}, where this invariant, and its
cousins for other torus knots, is written as the sum of
contributions that correspond to our $Z_\ab$ and $Z_y$ (with the
same values of $y$). The starting point was an integral
representation for the colored Jones polynomial of a torus knot
which originally was deduced using formulas that come from braid
group representations, quantum groups, and conformal field theory
\cite{JR,KT}.  This integral representation has recently been
obtained directly from the Chern-Simons path integral
\cite{Beasley}.

\subsubsection{Stokes Curves}\label{sc}

By a Stokes curve, we mean a curve $l$ in the complex $\gamma$ plane on which distinct
critical points have the same value of the quantity $\im\,\I$ that is conserved in the flow equation.
There are many such curves.
But we only really care about Stokes curves across which some jumping occurs.  It turns
out that there are far fewer of these.

Our basic strategy will be to show that every Stokes curve intersects the real $\gamma$ axis.
But we already know from section \ref{asjones} what sort of jumping occurs along the real
axis.  It takes the form $\J_\ab\to\J_\ab\pm \J_y$, where $\J_y$ is a cycle associated to one
of the lifts of $\A_\nab$.  This will enable us to get a general description of the integration cycle.  In fact, it turns out that the integration cycle always
takes the form of (\ref{ozmurk}), or its cousin with negative $x$, with some $x$ that depends on
$\gamma$ in a way that will emerge.

A preliminary observation is that there are no flows between
abelian critical points differing in the values of $r,s$, since
all such critical points have $\re\,\I=0$, but $\re\,\I$ is never constant along a non-trivial flow.  So in any flow between two critical points,
at least one of them is nonabelian.  

To show that all Stokes curves intersect the real $\gamma$ axis, we
use the explicit formula for $\I$ and the fact that $\rm{Im}\,\I$
is conserved by the flow equations.  Equivalently,  the conserved quantity is
$\rm{Im}\,\I/\pi n$.  For a lift of $\A_\nab$, this can be conveniently expressed in terms
of $\gamma=n/k$:
\begin{equation}\label{zexp}\rm{Im}\,\I/\pi n=-\rm{Re}\,\left(3\gamma+\left(\frac{1}{12}+2
s\right)\gamma^{-1}+2r-1\right).\end{equation} If $\gamma=a+ib$ with $a,b$ real, then
\begin{equation}\label{mexp}\rm{Im}\,\I/\pi n=-3 a-\left(\frac{1}{12}+2
s\right)\frac{a}{a^2+b^2}-2r+1.\end{equation}

Let us first consider Stokes curves at which there might be
flows between nonabelian critical points with different values of
$r,s$.  The change in $\rm{Im}\,\I/\pi n$ in such a process is
\begin{equation}\label{pexx}\Delta(\rm{Im}\,\I/\pi n)=-2\left(\frac{a\,\Delta s
}{a^2+b^2}+\Delta r\right),\end{equation} where $\Delta s$,
$\Delta r$ are the jumps in $s$ and $r$. Setting this to zero, we
get
\begin{equation}\label{yexp} a^2+b^2+\frac{a \,\Delta s }{\Delta
r}=0.\end{equation} 
The case $\Delta s=0$ is not of interest, for then vanishing of (\ref{pexx}) implies
$\Delta r=0$, in which case the initial and final critical points are the same and the flow
is trivial.  If $\Delta s$ and $\Delta r$ are both nonzero, the Stokes curve is a circle that intersects
the real $\gamma$ axis in two points; if $\Delta r=0$, it becomes the line $a=0$, intersecting
the real $\gamma$ axis at one point.  
Either way, at an intersection point of one of these Stokes curves with the real axis, two
nonabelian critical points with different values of $r,s$ remain
different, and there are no flows between them, since in general
there are no non-trivial flows for real $\gamma$.

In general, for two critical points $p_\sigma,p_\tau$,
along a Stokes curve $l$ characterized by $\im\,\I_\sigma=\im\,\I_\tau$,
the integer $\m_{\sigma\tau}$ that counts flows between  $p_\sigma$ and
$p_\tau$ is constant, except possibly at a point along $l$ at which a third critical point
$p_\nu$ has $\im\,\I_\nu=\im\,\I_\sigma=\im\,\I_\tau$.  Such a point is a triple intersection
of Stokes curve (equality of any two of $\im\,\I_\sigma, \im\,\I_\tau,$ and $\im\,\I_\nu$ defines
such a curve).  At a triple intersection point there can be jumping $\m_{\sigma\tau}\to
\m_{\sigma\tau}+\m_{\sigma\nu}\m_{\nu\tau}$.   Eqn. (\ref{yexp}) implies that there are no
triple intersection points involving different lifts of $\A_\nab$.  (Two of the curves defined
by eqn. (\ref{yexp}) can intersect away from the real $\gamma$ axis only if they coincide
identically, in which case the effects they produce can again be analyzed from what
happens on the real axis.) We need not worry about the
case that $p_\nu$ is a lift of $\A_\ab$, since in this case $\m_{\nu\tau}=0$, as we learned in
section \ref{moreg}.

Hence the net
number of flows between  nonabelian critical points differing in
the values of $r$ and $s$ is zero anywhere along one of these
Stokes curves.   So there are no Stokes phenomena involving flows
between different lifts of $\A_\nab$.

The same type of argument can be used to constrain Stokes
phenomena involving a flow between
 the abelian critical point and
a nonabelian one. If we pick the canonical lift of $\A_\ab$, it
has $\I=0$ and the change in $\rm{Im}\,\I/\pi n$ is given in
(\ref{mexp}). Setting the right hand side of that equation to
zero, we get the equation for a Stokes curve:
\begin{equation}\label{morely}b^2=-\frac{a(3
a^2+(2r-1)a+(2s+1/12))}{3a+2r-1}.\end{equation}   If we pick a
different lift of $\A_\ab$, we get the same formula with shifted values of $r,s$.  However, a byproduct of the following
calculation will be to show that different lifts of $\A_\ab$ do not enter the integration
cycle; indeed, all jumping phenomena take the form of (\ref{klp}) and do not shift the
``abelian'' part of the integration cycle.

Eqn. (\ref{morely}) describes many Stokes curves, labeled by the
choice of $r$ and $s$ and a component of the part of the $a$ axis
on which the right hand side of (\ref{morely}) is nonnegative.
Such a component is either a closed interval whose endpoints are
zeroes of the numerator $a(3 a^2+(2r-1)a+(2s+1/12))$ or a
half-open interval whose endpoints are such a zero and  the zero of the denominator. Each component intersects the $a$ axis, either
in two points (the first case) or one point (the second case).

We now make the familiar argument. Since every Stokes curve $l$
intersects the $a$ axis,  the jumping that occurs in crossing $l$
anywhere can be deduced from what happens in crossing $l$ where it
meets that axis.  If $l$ intersects the $a$ axis at a value of $a$
not of the form $\pm(1/6+p)$, with $p\in\Z$, then $l$ is irrelevant because no
jumping can occur at such a value of $a$. Even if $l$ intersects the $a$ axis only at a point of the form
$\pm(1/6+p)$, something non-trivial can happen only if $r$ and $s$ are of
the form given in (\ref{zorx}).  The reason for this last
statement is that as there are no non-trivial flows for real
$\gamma$, all that can happen in crossing $l$ at a point
$y=\pm(1/6+p)$ is that some lift of $\A_\nab$ may coincide
with $\A_\ab$ with its canonical lift. For this to happen, $r$ and
$s$ must be related to $y$ by (\ref{zorx}).

So the relevant Stokes curves are determined by their
intersection point $y=\pm(1/6+p)$ with the real axis; the curve is
described by  (\ref{morely}) with $r,s$ given in (\ref{zorx}).
In this special case, (\ref{morely}) simplifies to
\begin{equation}\label{torely}b^2=-\frac{a(a-y)^2}{a-2y}.\end{equation}

\begin{figure}
 \begin{center}
   \includegraphics[width=3in]{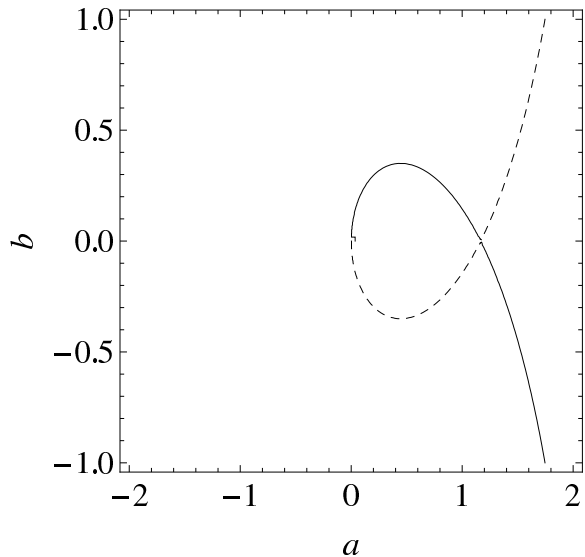}
 \end{center}
\caption{\small The Stokes curve (\ref{torely}) consists of both the solid and dashed curves, but just as
in fig. \ref{moresing}, jumping occurs only across the solid part of the curve. (Only the solid parts, therefore,
will be drawn in fig. \ref{lotsofcurves}.)}
 \label{toresing}
\end{figure}

We now meet a situation described at the end of section
\ref{sotwo}.  Eqn. (\ref{torely}) gives the condition that a flow
is possible between $\A_\ab$ (with its canonical lifting) and
$\A_\nab$ (with its chosen lifting).  The flow will go from
$\A_\ab$ to $\A_\nab$ if $\re\, \I(\A_\nab)$ is negative, and in
the other direction if $\re\,\I(\A_\nab)$ is positive.  However,
we also know from the analysis in section \ref{sotwo} of a
singularity of the type that occurs at $a=y$ (or from a more general argument in section
\ref{moreg}) that no jumping comes from flows from $\A_\nab$ to
$\A_\ab$.  So just as in fig.  \ref{moresing}, we can throw away
the part of the curve (\ref{morely}) with $\re\,\I(\A_\nab)>0$. As
sketched in fig. (\ref{toresing}), the part of the curve that we
keep is the part with $yb(y-a)>0$.

\begin{figure}
 \begin{center}
   \includegraphics[width=3in]{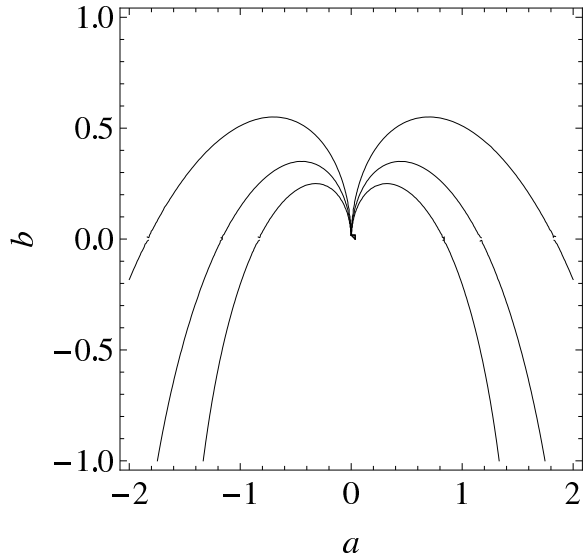}
 \end{center}
\caption{\small The Stokes curves across which jumping occurs for the trefoil knot
divide the complex $\gamma$ plane as shown here.  Sketched here are the curves
corresponding to $y=\pm 5/6,\pm 7/6$, and $\pm 11/6$. (All the other curves for larger or smaller values of $|y|$ are
obtained from these by suitable scalings of the $\gamma$-plane.)
Drawn are only the  parts of each curve that actually produce jumping; these portions meet only
at $\gamma=0$, where they all join.   }
 \label{lotsofcurves}
\end{figure}

Keeping only the ``good'' parts of these curves, they
 divide the complex $\gamma$ plane in a way that is
sketched in fig. \ref{lotsofcurves}. The important point here is
that any point $\gamma=\gamma_0$  that is not on one of the
relevant Stokes curves can be connected to the  $a$ axis in a
unique way (topologically) without crossing any of those curves.
The integration cycle $\CC$ of Chern-Simons theory analytically
continued to $\gamma=\gamma_0$ is given by the same formula as on
the  part of the $a$ axis to which $\gamma_0$ can be connected
without crossing a Stokes curve.    This result for the integration cycle is in full
accord with other analyses of the colored Jones polynomial of the trefoil knot.   What we call the Stokes curve (\ref{torely}) -- or more exactly the relevant portion of it
with $yb(y-a)>0$ -- is the curve across which an extra term appears in eqn. (2.2) of \cite{Mm}.

The analytically continued
function $J_n(q)$ grows exponentially in the semiclassical region
whenever $\re\,\I>0$ for a critical point that contributes to the
path integral.  This happens precisely if $\im\,\gamma>0$ and
$|\gamma|>1/6$.  (Those conditions are strong enough to ensure, using eqn. (\ref{torely}),
 that a critical point with
$y=1/6$ or $-1/6$ contributes to the path integral, and, using  eqn. (\ref{defreal}), that it
has $\re\,\I>0$.)

We are left with only one more issue, which is the behavior along
the positive imaginary $\gamma$ axis. As one approaches the
positive imaginary $\gamma$ axis from the right, one crosses all
Stokes curves of positive $y$.  As one approaches it from the
left, one crosses all Stokes curves of negative $y$.  A first
consequence is that, precisely on the positive imaginary axis, the
representation of the Chern-Simons integration cycle as a sum over
Lefschetz thimbles is an infinite sum.  A second consequence is
that, unless we can consider the Lefschetz thimble $\J_y$ to be
equivalent to $\J_{-y}$, this representation of the integration
cycle jumps in crossing the positive imaginary $\gamma$ axis.
Indeed,  in a complete circuit around the point $\gamma=0$, one
crosses exactly once each of the relevant Lefschetz thimbles of
fig. \ref{lotsofcurves}.  This will produce a monodromy of
infinite order, and hence an essential singularity in the
integral, unless $\J_y$ is equivalent to $\J_{-y}$.

Can we consider $\J_y$ and $\J_{-y}$ to be equivalent? Referring
back to (\ref{zorx}), we see that $y\to -y$ amounts to $s\to s$
with an integer change in $r$. In other words, it amounts to a
change in the choice of branch of $\ln\,\m$. The branch of
$\ln\,\m$ is irrelevant as long as $n\in\Z$ and we choose not to
replace the loop space of $G/T_\C$ by its universal cover. This
appears to be the right sense in which the integration cycle is
univalent near $\gamma=0$, but clearly, a study of more examples
is needed. For $n\not\in\Z$, the monodromy of the integration
cycle is unavoidable and an essential singularity of the integral
at $\gamma=0$ should be expected.

\subsection{The Figure-Eight Knot}\label{piecesofeight}

Here we will consider in a similar way the figure-eight knot (also known as the $4_1$ knot), which we denote
$K$.  The fundamental group of $S^3\backslash K$ again has two generators $\x$, $\y$,
either of them conjugate to the meridian $\hat\m$, with one relation.  The relation is
$\omega \x=\y\omega$, with $\omega=\x\y^{-1}\x^{-1}\y$.

For a generic eigenvalue $\m$ of the meridian, there are up to gauge transformation
three $SL(2,\C)$ flat connections on $S^3\backslash K$.  As usual, one is an abelian
flat connection $\A_\ab$, whose monodromies are up to conjugacy $\x=\y=\rm{diag}(\m,\m^{-1})$.
In addition, there are two nonabelian flat connections with monodromies
\begin{equation}\label{wimon}\x=\begin{pmatrix}\m&1\cr 0&\m^{-1}\end{pmatrix},~~
\y=\begin{pmatrix}\m&0\cr -d&\m^{-1}\end{pmatrix},\end{equation}
with
\begin{equation} \label{dis}d=\frac{1}{2}\left(\m^2+\m^{-2}-3\pm\sqrt{(\m^2+\m^{-2}+1)(\m^2+\m^{-2}-3)}\right).\end{equation}
The longitude is $\hat \l=\x\y^{-1}\x\y\x^{-2}\y\x\y^{-1}\x^{-1}$, leading to
\begin{equation}\label{wis}\l=\frac{1}{2}\left(\m^4-\m^2-2-\m^{-2}+\m^{-4}\right)\pm\frac{\m^2-\m^{-2}}{2}\sqrt{(\m^2+\m^{-2}+1)(\m^2+\m^{-2}-3)}.\end{equation}
We denote as $\A_\pm$ a flat connection with the monodromies just
indicated. These are the $SL(2,\C)$ flat connections that, if
$\m=\pm 1$, are related to the complete hyperbolic metric on
$S^3\backslash K$.

The longitude $\l$ is a zero of the $\AA$ polynomial, which is
\begin{equation}\label{gurz}\AA(\l,\m)=\l-\left(\m^4-\m^2-2-\m^{-2}+\m^{-4}\right)+\l^{-1}.\end{equation}
Apart from the standard symmetries $\m\to -\m$ and $\m\to\m^{-1}$,
this has an additional symmetry $\l\to\l^{-1}$, which exchanges
the two branches of the square root in (\ref{wis}). This symmetry
reflects the existence of a  diffeomorphism $\tau$ of $S^3$ that
maps $K$ to itself, reversing the orientation of $K$ but
preserving the orientation of its normal bundle.  ($\tau$ acts on
the above presentation of the fundamental group by
$\x\leftrightarrow \y$.) Such a symmetry maps $\m\to\m$,
$\l\to\l^{-1}$.  It  reverses the orientation of $S^3$, so it
changes the sign of the Chern-Simons function $W(\A)$ and is  not
a symmetry of Chern-Simons gauge theory. Nevertheless, it will
help us in understanding the behavior of the theory. A knot for
which such a symmetry exists is called amphichiral.  Actually, the
figure-eight knot is amphichiral in the strong sense that $S^3$
admits another diffeomorphism $\tilde\tau$ that maps $K$ to itself
(pointwise) while reversing the orientation of its normal bundle.
(It acts on the presentation of the fundamental group by $\x\to
\y^{-1}$, $\y\to \x^{-1}$.) Again, $\tilde\tau$ reverses the
orientation of $S^3$ and reverses the sign of the Chern-Simons
invariant.  It acts by $\m\to\m^{-1}$, $\l\to\l$.  (We leave it to
the reader to verify that the unknot is a simpler example of a
knot in $S^3$ that admits diffeomorphisms of both of these types.)
The product $\tau\tilde\tau$ preserves the Chern-Simons invariant
and acts by $\m\to\m^{-1}$, $\l\to\l^{-1}$.

Unlike the trefoil, where all of the interesting action occurs for $|\m|=1$, for the figure-eight
knot both the real $\m$ axis and the circle $|\m|=1$ are worthy of careful study.   The
eigenvalues of an element of $SL(2,\R)$ are either real or of modulus 1, so in general
flat connections with $SL(2,\R)$ structure group may have $|\m|=1$ (as  for the trefoil)
or real $\m$ (as happens for the figure-eight knot).  The eigenvalues of an element of $SU(2)$
are of modulus 1, so flat connections with $SU(2)$ structure group will arise only for
$|\m|=1$.

\subsubsection{Real $\m$}

Suppose first that $\m$ is real and that
\begin{equation}\label{gufon}\m^2+\m^{-2}>3.\end{equation}
Then by inspection, for both flat bundles $\A_\pm$, the matrices $\x$ and $\y$ are real, so the entire representation
of the fundamental group has its image in $SL(2,\R)$.   The diffeomorphism $\tau$ exchanges $\A_+$ with
$\A_-$ (since it maps $\l\to\l^{-1}$).  It leaves $\A_\ab$ invariant.

The region with $\m$ real and
\begin{equation}\label{zuton}\m^2+\m^{-2}<3 \end{equation}
is also easy to understand qualitatively.  The flat connections $\A_+$ and $\A_-$ are still
exchanged by $\tau$.  Now, however, they are no longer real (since the square roots in the
monodromy formulas are imaginary) and are exchanged by complex conjugation.  In fact,
the region $\m^2+\m^{-2}<3$  includes the points $\m=\pm 1$ where $\A_+$ and $\A_-$ are the
$SL(2,\C)$ flat connections that are related to a hyperbolic metric on $S^3\backslash K$.

As one might expect, among the most interesting points on the real $\m$ axis are the points with
\begin{equation}\label{uton}\m^2+\m^{-2}=3.\end{equation}
For such $\m$, since $d$ vanishes, the matrices $\x$ and $\y$ are both upper triangular.
The flat bundles $\A_+$ and $\A_-$ become equivalent to each other.  They are unstable
upper triangular deformations of the abelian flat bundle $\A_\ab$.
$\A_\ab$, just as in (\ref{ormok}), is a diagonal flat connection
\begin{equation}\label{poodle}\begin{pmatrix}* & 0 \cr 0 & *\end{pmatrix}.\end{equation}
It has automorphism group $\C^*$; this symmetry group reduces to $\R^*\cong SO(1,1)$ if we consider $\A_\ab$
(for real $\m$) as a flat bundle with structure group $SL(2,\R)$.
The symmetry extends to $O(1,1)$ if we include the outer automorphism of $SL(2,\R)$
corresponding to conjugation by $\rm{diag}(1,-1)$ or $\rm{diag}(i,-i)$, and we will do so.\footnote{In $SL(2,\R)$ gauge theory, one would consider two critical points that differ
by this outer automorphism to be inequivalent.  But in $SL(2,\C)$ gauge theory, or analytic
continuation of $SU(2)$ gauge theory, which are our focus in this paper, they are equivalent,
since $\rm{diag}(i,-i)$ is an element of $SU(2)$.}
 $\A_\ab$ has no deformations
as a flat bundle with specified  $\m$, but as in (\ref{rmok}), it
has a one-dimensional space of deformations of upper triangular
form
\begin{equation}\label{zmok} \begin{pmatrix} 0 & u \cr 0 & 0 \end{pmatrix}.
\end{equation}
Here $u$ is a real parameter if we consider deformations in
$SL(2,\R)$.  The value of $u$, assuming it is not zero, does not
matter modulo the action of $O(1,1)$, and the upper triangular
deformation gives us the triangular flat bundle $\A_+$ or $\A_-$.
Just as in (\ref{rmoko}), $\A_\ab$ has a similar lower triangular
deformation,
 schematically \begin{equation}\label{polgo}\begin{pmatrix}0 & 0 \cr -v & 0 \end{pmatrix}.\end{equation}
If we take $v\not=0$ but $u=0$, we get a lower triangular flat bundle, and again the value
of $v$ is inessential assuming that it is nonzero.

As in our discussion of the trefoil, there is no deformation with $u,v$ nonzero and
$\m^2+\m^{-2}=3$, since a flat $SL(2,\C)$ bundle at such a value of $\m$ is triangular
(or abelian).  If we do turn on both $u$ and $v$, the $O(1,1)$-invariant parameter is the product $uv$.  One might expect as in the discussion of the trefoil that we would have $uv\sim\epsilon$
where $\epsilon=\m^2+\m^{-2}-3$.  However, in (\ref{wimon}) the upper triangular parameter
is 1 and the lower triangular parameter is $-d$, so this case corresponds to $uv\sim -d$.  Moreover,
from (\ref{dis}), we have $d\sim\pm\sqrt\epsilon$ near $\epsilon=0$.  Critical points therefore
are at $uv=0$ or $\pm\sqrt\epsilon$.  Apart from a multiplicative constant, the function with
this critical behavior is
\begin{equation}\label{horsy}f(u,v)=(uv)^3/3-\epsilon uv.\end{equation}
The oscillatory integral associated to this function was studied in section \ref{onemore}.

Why have we arrived at this peculiar structure, with no $(uv)^2$
term in $f(u,v)$? The answer has to do with amphichirality.  As
$\tau$ leaves fixed $\A_\ab$ and exchanges $\A_\pm$, at a point at which one of these
becomes triangular and meets $\A_\ab$, the other must do so as
well.  This structure is lost if we add a $(uv)^2$ term to $f$.
More fundamentally, $\tau$ reverses the sign of the Chern-Simons
function and hence of $f$, which describes its critical behavior.
The fact that $\tau$ exchanges $\A_\pm$ implies that if we lift
the action of $\tau$ to the flat bundle with connection $\A_\ab$
in such a way that $\tau(u)=u$, then $\tau(v)=-v$.  (By composing
with the outer automorphism of $SL(2,\R)$, one can also choose a
lifting so that $\tau(u)=-u$, $\tau(v)=v$.)  Thus, $\tau$ reverses
the sign of $uv$.  Since $\tau$ reverses the sign of both $uv$ and
$f$, $f$ is an odd function of $uv$, and this accounts for the
structure that we have found.

As in our analysis of the trefoil, some of the Stokes phenomena of Chern-Simons theory of the figure-eight knot can be described in terms of critical behavior of the function $f$.  However, for brevity, the only
issue that we will discuss here will be the large $n$ behavior of the colored Jones polynomial
near $\epsilon=0$, that is, near $\m^2+\m^{-2}=3$.  First we simply set $\epsilon=0$.  The parameter $\lambda$ in eqn. (\ref{zorkoff})
corresponds in Chern-Simons theory to $n$ (or equivalently $k$, since the ratio $n/k$ is fixed when we specify $\epsilon$).  So according to (\ref{zorkoff}),
\begin{equation}\label{kilpo}\frac{\left. J_n(k)\right|_{\epsilon=0}}{\left. J_n(k)\right|_{\epsilon<0}}
\sim n^{2/3},~~~n\to\infty.\end{equation}
Accordingly, if we normalize the colored Jones polynomial of the
figure-eight knot so that it is of order 1 for $\epsilon<0$, then it is of order $n^{2/3}$ for $\epsilon=0$.
This is a known result
(\cite{km}, Theorem 1.1).  Instead of simply setting $\epsilon=0$, we can consider
the limit $n\to\infty$, $\epsilon\to 0$ with $n^{2/3}\epsilon$ fixed.  The singular contribution
to the colored Jones polynomial 
is this limit follows from (\ref{donko}) and is an Airy function (times $n^{2/3}$ in the usual
mathematical normalization), as one can see by introducing
polar coordinates in the $xy$ plane.

\subsubsection{$\m$ Of Modulus 1}\label{mmod}

Now let us analyze what happens for $|\m|=1$.  When are the flat connections $\A_\pm$
unitary?  For this to be the case, $\l$ must be of modulus 1.  From (\ref{wis}), the condition
for this (assuming that $|\m|=1$) is
\begin{equation}\label{rmod}\m^2+\m^{-2}+1\leq 0.\end{equation}
Conversely, if this condition is obeyed and $|\m|=1$, one can explicitly conjugate
the flat bundle with monodromies (\ref{wimon}) into $SU(2)$ by the same steps that
we used in discussing the trefoil knot.

For $|\m|=1$ and $\m^2+\m^{-2}+1>0$, the bundles $\A_\pm$ have
structure group $SL(2,\C)$. A reduction to $SL(2,\R)$ cannot occur
because $\hat\m$ has eigenvalues of unit modulus while $\hat\l$
has real eigenvalues; commuting (and noncentral) elements of
$SL(2,\R)$ do not have this property.  Also, the region $|\m|=1$,
$\m^2+\m^{-2}>1$ again contains the points $\m=\pm 1$ where the
connections $\A_\pm$ are $SL(2,\C)$ connections related to the
hyperbolic metric on $S^3\backslash K$.

As usual, we have to look more closely at what happens for
$\m^2+\m^{-2}+1=0$. It is convenient to set
$\epsilon=-\m^2-\m^{-2}-1$, so that $\epsilon$ is real for
$|\m|=1$ and the flat connections $\A_\pm$ have $SU(2)$ structure
group precisely when $\epsilon\geq 0$. Unlike our previous
examples, the nonabelian flat connections $\A_\pm$ remain
irreducible at those points and as a result the abelian flat
connection $\A_\ab$ will play no role in critical behavior near
$\epsilon=0$.  All that happens when $\epsilon=0$ is that the two
critical points $\A_\ab$ become coincident.  Moreover, near
$\epsilon=0$, the difference between the two critical points is of
order $\epsilon^{1/2}$; this is evident from the formulas
(\ref{wimon}) for the monodromies.  Finally,  the critical points
are real (in the natural real structure of $SU(2)$ gauge theory)
if $\epsilon>0$.  These are the properties of the familiar
function
\begin{equation}\label{polkox}f(x)=x^3/3-\epsilon x \end{equation}
associated to the Airy integral.
This function describes the critical behavior of the Chern-Simons function $W(\A)$ near
the points $\epsilon=0$, and hence in studying Stokes phenomena, we will be able to borrow
results from section \ref{twocrit}.

For now, we simply describe qualitatively the behavior of the
colored Jones polynomial $J_n(q)$ for real $\epsilon$ near
$\epsilon=0$.  First we consider the limit $n\to\infty$ with $\epsilon$ fixed:

(1) For $\epsilon<0$, the only real critical point is the abelian
one.  Its contribution to the path integral, for any knot, is of
order $1/\sqrt n$, so we have
\begin{equation}\label{tondo}J_n(q)\sim
\frac{C}{n^{1/2}},~~\epsilon<0  \end{equation} with a constant
$C$.  (Actually, in a more precise description, $C$ is the first
term in an asymptotic expansion $C+C'/n+\dots$ in integer powers
of $1/n$.)

(2) For $\epsilon>0$, the flat bundles $\A_\pm$ appear as real
critical points in the path integral.  They give oscillatory
contributions, but without the factor  $1/\sqrt n$ that appears in
the contribution of $\A_\ab$.  Including also the abelian
contribution, the integral behaves as
\begin{equation}\label{ondo}J_n(\epsilon)\sim
\frac{C}{n^{1/2}}+C_+\exp(in f_+)+C_-\exp(inf_-),~~\epsilon>0\end{equation}
with constants $C_\pm$ (which again are the leading terms of asymptotic series).  Here we
have written the action $\I(\A_\pm)$ as $in f_\pm(\epsilon)$,
where $f_\pm(\epsilon)$ is independent of $n$.  We have
$f_\pm(0)=0$, since $\A_\pm$ coincide with $\A_\ab$ at
$\epsilon=0$. Also, as explained in section \ref{csfun},
$f_+(\epsilon)=-f_-(\epsilon)$ for real $\epsilon$.

(3) What happens at $\epsilon=0$?   Typically the $n$-dependence
of the integral over any one mode of the path integral is that of
a Gaussian integral $\int \d x\exp(in x^2)\sim 1/n^{1/2}$.  For
the path integral around an irreducible flat connection with
vanishing cohomology (such as $\A_\pm$ at $\epsilon\not=0$), these
factors, after taking into account the regularization and the
volume of the gauge group, ultimately cancel in a complete
evaluation of the path integral. This is why the oscillatory terms
in (\ref{ondo}) do not have a power of $n$ as a prefactor. But at
$\epsilon=0$, the integral over $x$, rather than being Gaussian,
becomes $\int\d x\exp(in x^3)\sim 1/n^{1/3}=n^{1/6}/n^{1/2}$.  The
denominator participates in the usual cancellation, but the
numerator does not, so we expect
\begin{equation}\label{biondo} J_n(0)\sim n^{1/6}.\end{equation}
(In the usual mathematical normalization, one divides by the path
integral (\ref{zolly}) of the unknot, and then the colored Jones polynomial grows as
$n^{2/3}$.)

So far we have described what happens for $n\to\infty$ with fixed $\epsilon$.
If instead we consider the limit $n\to\infty$ with $n^{2/3}\epsilon$ fixed, then
in view of (\ref{urtz}),
the singular behavior of the colored Jones polynomial is an Airy function (times
$n^{2/3}$ in the usual mathematical normalization).

\subsubsection{The Real Part Of The Chern-Simons Function}\label{csfun}

For the figure-eight knot, we will study only the simplest
properties of the Chern-Simons function.

\def\CCC{{\cmmib C}}
First we focus on the region $|\m|=1$.  We denote as $\CCC$ the
operation of complex conjugation of a flat $SL(2,\C)$ bundle.
$\CCC$ maps $(\m,\l)$ to $(\bar \m,\bar \l)$. Let us compose this
operation with the symmetry $\tilde\tau$, which maps $(\m,\l)$ to
$(\m^{-1},\l)$. The composite acts by $(\m,\l)\to
(\bar\m\,^{-1},\bar \l)$.  If $|\m|=1$, then $\bar\m\,^{-1}=\m$
and the action is $(\m,\l)\to (\m,\bar\l)$.

There are two somewhat different cases:

(1) If $\m^2+\m^{-2}+1<0$, the two flat bundles are $SU(2)$
bundles with complex conjugate values of $\l$.  The composite
$\CCC\tilde\tau$ exchanges the two values of $\l$, so it exchanges
the two flat bundles.

(2) If $\m^2+\m^{-2}+1>0$, the two flat bundles are $SL(2,\C)$
bundles with real and unequal values of $\l$.  The composite
$\CCC\tilde\tau$ leaves fixed the two values of $\l$, so it also
leaves fixed the isomorphism class of each of these two flat
bundles.

Now let us analyze the Chern-Simons invariant $W(\A)$ in the two
cases.  $\CCC$ transforms $W$ to its complex conjugate, while
$\tilde\tau$ (because it reverses the orientation of $S^3$) maps
it to $-W$.  So the composite maps $W$ to $-\bar W$.

In case (1), $W$ is real, since in general the Chern-Simons
invariant of an $SU(2)$ connection is real.  So $W$ transforms
simply to $-W$ under the exchange of the two flat bundles. Thus
they have equal and opposite real values of $W$.

In case (2), $W$ is complex.  The fact that the two flat bundles
are invariant under $\CCC\tilde\tau$ tells us that their
Chern-Simons functions are invariant under $W\to -\bar W$.  This
tells us nothing about $\im\,W$, but it implies that
$\re\,W=-\re\,W$.  Before concluding that $\re\,W=0$, we should
recall that $\re\,W$ is only well-defined modulo $2\pi$.  So the
correct conclusion is only that $\re\,W$ vanishes mod $\pi$.

In case (2), we are left with two possible cases: $\re\,W$ may
vanish mod $2\pi$ or may equal $\pi$ mod $2\pi$.  The region
$\m^2+\m^{-2}+1>0$, $|\m|=1$, has two connected components, as
$\m$ may be near either 1 or $-1$.  It turns out that $\re\,W=0$
mod $2\pi$ if $\m$ is near 1, and $\re\,W=\pi$ mod $2\pi$ if $\m$
is near $-1$.

To show this, we observe first that if $\m^2+\m^{-2}+3=0$, then
the flat bundle is triangular, which guarantees that $W=0$ mod
$2\pi$.  Let $\m_0$ be a positive root of the  equation
$\m^2+\m^{-2}+3=0$.  Then $-\m_0$ is a negative root.  We can
calculate $W(\A)$ for $\m=1$ by integrating eqn. (\ref{csvalue})
along the real $\m$ axis from $\m_0$ to 1.  We get
\begin{equation}\label{nurz}W(\A;\m=1)=W(\A;\m=\m_0)+(1/\pi)\int_{\m_0}^1\ln\m\,\d \ln\l.\end{equation}
Along the integration path, $\ln\,\m$ is real.  On the other hand,
 according to eqn. (\ref{wis}),
$\l$ is of modulus 1 for $\m$ real,\footnote{Fundamentally, this
reflects the fact that for $\m$ real, the two flat bundles are
invariant under the composite $\CCC\tau$, which leaves real $\m$
fixed while mapping $\l$ to $\bar\l\,^{-1}$.}  so $\ln\,\l$ is
imaginary.  Hence, the integral is imaginary.  Since the
integration constant $W(\A;\m=\m_0)$ vanishes, we learn that
\begin{equation}\label{turx} \re\,W(\A;\m=1)=0~{\mathrm
{mod}}~2\pi.\end{equation}  Similarly we have
$W(\A;\m=-1)=W(\A;\m=-\m_0)+ (1/\pi)\int_{-\m_0}^{-1}\ln\m\,\d
\ln\l$, and again the integration constant vanishes.  But now we
get a different result because the imaginary part of $\ln\m$ along
the negative $\m$ axis is equal to $\pi$.  From (\ref{wis}), we
see that as $\m$ varies from $-\m_0$ to $-1$, $\l$ makes a
half-circuit around the unit circle from 1 to $-1$.  So $\ln\,\l$
changes by $i\pi$ and finally we get
\begin{equation}\label{zurx} \re\,W(\A;\m=-1) =\pi~{\mathrm
{mod}}~2\pi.\end{equation}

\def\Arg{{\mathrm {Arg}}}
Now let us discuss the behavior for $\m=\exp(i\pi\gamma)$ as a
function of real $\gamma$ with $|\gamma|\leq 1$. The condition
$\m^2+\m^{-2}+1<0$ amounts to $1/3<|\gamma|<2/3$.  So case (2) in
the above analysis corresponds to $|\gamma|<1/3$ or
$|\gamma|>2/3$.  In either of those regions, $\re\,W$ can only
take the values 0 or $\pi$, so it is constant by
continuity.\footnote{This constancy is also a direct consequence
of eqn. (\ref{csvalue}).  In the region in question, $\ln\m$ is
imaginary and $\ln \l$ is real, so the real part of
$\ln\m\,\d\ln\l$ vanishes.}  Hence $\re\,W=0$ if $|\gamma|\leq
1/3$, and $\re\,W=\pi$ if $|\gamma|\geq 2/3$.

For $1/3\leq |\gamma|\leq 2/3$, there are up to isomorphism two
flat $SU(2)$ connections, which we denote as $A_\pm$.    At
$|\gamma|=1/3$, $A_+$ and $A_-$ both coincide with the flat
$SL(2,\C)$ connection considered in the last paragraph. So in that
case, $\re\,W=0~{\rm {mod}}\,2\pi$.  It is convenient to lift the
Chern-Simons invariants so that $A_\pm$ have $\re\,W=0$ at
$|\gamma|=1/3$.

 Now let us follow the two $SU(2)$ flat connections $A_+$ and
 $A_-$ continuously from $|\gamma|=1/3$ to $|\gamma|=2/3$.  At
 $|\gamma|=2/3$, we can use (\ref{zurx}), along with the constancy
 of $\re\,W$ for $|\gamma|\geq 2/3$, to deduce that $W(A_+)$ and
 $W(A_-)$ are each congruent to $\pi$ mod $2\pi$.  Since we also
 showed above that $W(A_+)=-W(A_-)$, it follows that for some
 integer $w$, we have $W(A_\pm)=\pm(2w+1)\pi$ at $|\gamma|=2/3$.

\begin{figure}
 \begin{center}
   \includegraphics[width=1.5in]{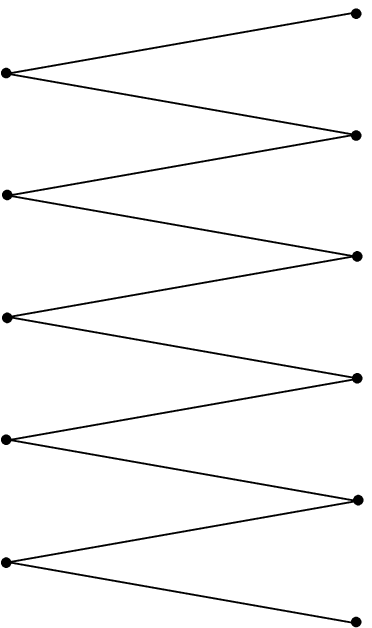}
 \end{center}
\caption{\small  Qualitative behavior of the conserved quantity $\im\,I/k$ for  real critical
points in the region $1/3\leq|\gamma| \leq 2/3$.  The structure is in perfect
parallel with fig. \ref{critflow} of section \ref{singrho}, except now $\gamma$ is plotted
on the horizontal axis.   At $|\gamma|=1/3$ (left), there are a pair of real critical
points for every value of $\im\,\I/k$ of the form $2\pi w$, $w\in\Z$.  Increasing $|\gamma|$, $\im\,\I/k$
increases for one of these critical points and reduces for the other.  At $|\gamma|=2/3$ (right), the critical
points recombine in pairs, but now with $\im\,\I/k$ of the form $(2w+1) \pi $, $w\in\Z$. } \label{lines}
\end{figure}

 It can be shown by integrating (\ref{csvalue}) that $w=1$.  (This
 is greatly facilitated by symmetrizing the integrand between
 $\gamma$ and $1-\gamma$ before attempting to integrate.)   Now let us discuss
 what happens when we include all possible
 lifts of $W(A_\pm)$.

For $|\gamma|<1/3$, critical points come in complex conjugate
pairs.
 In passing through $|\gamma|=1/3$, a pair of complex conjugate critical points meet
 and recombine into two  inequivalent real critical points
 $A_\pm$.  Actually, this occurs infinitely many times, once
 corresponding to each complex lift of $W$.

Increasing $\gamma$ past $|\gamma|=2/3$, the reverse process
occurs: a pair of inequivalent real critical points recombine into
a complex conjugate pair of critical points.  Again this actually
occurs infinitely many times.

The flow in $W$ between $|\gamma|=1/3$ and $|\gamma|=2/3$ leads to a crucial detail.
Actually, this detail also arose in analytic continuation of the Bessel function in section
\ref{anak}, and was illustrated in fig. \ref{critflow}.  That figure is repeated for convenience
in fig. \ref{lines}, with a caption adapted to the present problem.
The point is that a  pair of real critical points that
are ``born'' at $|\gamma|=1/3$ with the same value of $W$ no
longer have the same value of $W$ by the time they ``arrive'' at
$|\gamma|=2/3$.  Rather they have values of $W$ that differ by
$2\pi$.  The pairs of critical points that separate at $|\gamma|=1/3$ are not the same
as the pairs that rejoin at $|\gamma|=2/3$.

Precisely the same phenomenon led in section \ref{sotwo} to
exponential growth of the analytically continued Bessel function
for negative and non-integer values of $\k$.  It should therefore
come as no surprise that this mechanism will lead to exponential
growth of the function $J_n(q)$ associated to the figure-eight
knot in the region $|\gamma|>2/3$, as long as the Chern-Simons
level $k$ is not an integer.

\subsubsection{The Imaginary Part Of $W$}\label{impar}

We will now discuss much more briefly the simplest properties of
$\im\,W$ along the unit circle $|\m|=1$.

For $\m^2+\m^{-2}+1\leq 0$, the nonabelian flat connections are
$SU(2)$-valued flat connections $A_\pm$.  As we have already
noted, like any $SU(2)$-valued connections, these have $\im\,W=0$.

For $\m^2+\m^{-2}+1>1$, there are a pair of $SL(2,\C)$ flat
connections $\A_\pm$, exchanged by complex conjugation.  Since
$\im\,W$ is odd under complex conjugation, we have
$\im\,W(\A_+)=-\im\,W(\A_-)$.

The most important property of the invariant $\im\,W$ is its
relation to the hyperbolic volume of the knot complement
$S^3\backslash K$.  $S^3\backslash K$ admits a complete hyperbolic
metric, unique up to diffeomorphism.  Though this metric is
complete, it has the property that the volume is
convergent.\footnote{Near $K$, the metric can be modeled by $\d
s^2= z^2(\d x^2+\d y^2)+\d z^2/z^2$, where $x$ and $y$ parametrize
a two-torus and $K$ is at $z=0$. Thus the region near $K$ is the
``infrared'' region in the context of the AdS/CFT correspondence.
This description can  be used to compute the conjugacy class of
$\hat\m$ given in eqn. (\ref{tollfus}) below. Both $\hat\m$ and
$\hat\l$ arise in this description from shifts in $x$ and $y$;
such shifts correspond to unipotent elements of $PSL(2,\C)$.}  We
denote this volume as $V$.

Like a hyperbolic metric on any three-manifold, the complete
hyperbolic metric on $S^3\backslash K$ determines a flat
$PSL(2,\C)$ bundle on this manifold. Upon picking a spin structure
on $S^3\backslash K$, the flat $PSL(2,\C)$ bundle can be lifted to
a flat $SL(2,\C)$ bundle. The conjugacy class of the meridian for
this flat bundle is
\begin{equation}\label{tollfus}\hat\m=\pm \begin{pmatrix} 1 & 1 \cr 0
& 1\end{pmatrix},\end{equation} where the sign depends on the
choice of spin structure (in other words, the two flat bundles
with different signs of $\m$ differ by twisting by a flat line
bundle of order 2 -- the operation considered in section
\ref{tensoring}). The flow equations in the presence of such
unipotent monodromy were discussed in section \ref{further}; for
our present purposes, all we need to know is that the eigenvalue
of $\hat \m$ is $\m=\pm 1$.  On the other hand, from eqn.
(\ref{wis}), it follows that\footnote{From this we can actually
give a new demonstration of one result of section \ref{csfun},
which was that $\re\,W$ shifts by $\pi$ under $\m\to -\m$. In
section \ref{further}, we showed that the shift is $\Delta
W=-i\ln\l$, which here equals $\pi$ mod $2\pi$ as $\l=-1$.}
$\l=-1$, for either sign of $\m$.

The relation between the hyperbolic volume and the Chern-Simons
invariant is simply that, if we label $\A_\pm$ correctly,
\begin{equation}\label{zonkox}\re\left(iW(\A_\pm;\m=\pm 1)\right)=\pm\frac{V}{2\pi}.\end{equation}  This means that
$\im\,W(\A_\pm;\m=\pm 1)=\mp V/2\pi$.  The sign may look unnatural, but
it ensures that $\A_\pm$ make exponentially growing and decaying
contributions to the path integral, respectively; this seems like
the most useful mnemonic.

If we depart from $\m=\pm 1$, there is no longer a complete
hyperbolic metric on $S^3\backslash K$ with $\m$ as the eigenvalue
of the meridian.  But there is an incomplete hyperbolic metric
with this property (it has a conical singularity along $K$), and a
statement along the lines of (\ref{zonkox}) remains valid for this
incomplete metric.  We will not describe details here.

We have explained the qualitative properties of $\re\,W$ and
$\im\,W$ that will be of most concern for the limited application
that we consider next.  We should note, however, that (as
summarized in \cite{Mura}, for example) by integrating the usual
formula (\ref{csvalue}) and using (\ref{wis}), one can express
$W(\A)$ for flat connections on the complement of the figure-eight
knot in terms of a dilogarithm function.

\subsubsection{The Volume Conjecture For The Figure-Eight
Knot}\label{volfig}

The volume conjecture for knots emerged from the following
observation \cite{Kash,MM} about the analytically continued
invariant $J_n(q)$ of the figure-eight knot. For $\gamma$
approaching 1 from a direction away from the real axis, the large
$n$ behavior of $J_n(q)$ is, roughly speaking,
\begin{equation}J_n(q)\sim\exp(nV/2\pi),\end{equation}
where $V$ is the hyperbolic volume of the knot complement,
discussed in section \ref{impar}.  A more precise statement will
emerge below.

A partial explanation of this was given in \cite{Gu}.  It was
noted that $nV/2\pi$ is the value of the exponent $\I=ik W(\A)$ of
the analytically-continued Chern-Simons path integral at the
complex critical point $\A_+$.  So an expansion around this critical point would
give a contribution growing as $\exp(V/2\pi)$.   (In this paper, and subsequent
literature, generalizations of the statement to other values of
$\gamma$ were also made.  For example, see \cite{GL,KT,MM}.)

There has always been a puzzle about this partial explanation. For
physical values of $k$ and $n$ -- positive integers -- the
Chern-Simons path integral is dominated in the semiclassical limit
by real critical points.  Contributions, if any, of complex
critical points are exponentially small.  Therefore, at least for
positive rational values of $\gamma=n/k$, there are no
exponentially growing contributions to the path integral.  How
then can exponential growth be observed just because $\gamma$ is
infinitesimally displaced from 1?

The reader who has paid close attention to sections \ref{sotwo}
and \ref{impar} will hopefully anticipate the answer.  Once we
analytically continue away from integer values of $k$, we need to
work not on the space $\U_\C$ of complex-valued connections modulo
gauge transformations, but on a cover $\hat\U_\C$ on which the
Chern-Simons function $W(\A)$ is $\C$-valued (rather than valued
in $\C/2\pi\Z$). A given flat connection $\A$ corresponds to
infinitely many critical points that differ only by the choice of
a complex lift of $W(\A)$. When the integration cycle of the
analytically continued theory is expressed in terms of Lefschetz
thimbles associated to critical points, it is possible for contributions from
critical points that differ only in this lift to cancel for
integer $k$, but give exponential growth as soon as $k$ is not an
integer.

Let us first express in terms of Lefschetz thimbles the
integration cycle  $\CC$ of Chern-Simons theory for what proves to
be the simplest case: positive integer $k$ and $n$ with
$1/3<\gamma<2/3$, where as usual $\gamma=k/n$. This is easy to do
because in this range there are no complex critical points, only
real ones. Up to gauge transformation, there is an abelian flat
connection $A_\ab$, and  two nonabelian but $SU(2)$-valued flat
connections $A_\pm$. As they are all $SU(2)$-valued, they all
contribute to the integration cycle with a coefficient 1.  This
cycle is then in an obvious notation
\begin{equation}\label{orzk}\CC=\J_\ab+\J_++\J_-.\end{equation}

As long as $k$ and $n$ are integers, there is also no problem in
extending this answer outside the range $1/3<\gamma<2/3$.  At
either $\gamma=1/3$ or $\gamma=2/3$, there is an Airy-type
singularity.  The two real flat connections $A_\pm$ recombine to a
complex conjugate pair of $SL(2,\C)$-valued flat connections,
which we will call $\A'_\pm$ for $\gamma<1/3$ or $\A''_\pm$ for
$\gamma>2/3$.  We choose the subscript so that, as in
(\ref{zonkox}) (where $\m=\pm 1$ are treated together),
$\re\,(iW(\A'_+))>0$ for $\gamma<1/3$ and $\re\,(iW(\A''_+))>0$ for
$\gamma>2/3$.  This ensures that the integrals over the Lefschetz
thimbles $\J'_-$ and $\J''_-$ associated to $\A'_-$ and $\A''_-$
are exponentially decaying, while the integrals over the other
Lefschetz thimbles $\J'_+$, $\J''_+$ associated to $\A'_+$,
$\A''_+$ are exponentially growing.  When we cross the Airy-like
singularities at $\gamma=1/3$ or $2/3$, according to eqn.
(\ref{zelign}), the sum $\J_++\J_-$ recombines to a Lefschetz
thimble $\J'_-$ or $\J''_-$, whose contribution to the path
integral is exponentially decaying.  The integration cycle thus
becomes
\begin{equation}\label{intcy}\CC=\begin{cases}\J_\ab+\J'_-&\text{for~}\gamma<1/3\\
                                              \J_\ab+\J''_-&\text{for~}\gamma>2/3.\end{cases}\end{equation}

Now let us continue away from integer values of $k$ and $n$.
Starting in the region $1/3<\gamma<2/3$, to make an analytic
continuation, we must pick lifts of the Chern-Simons invariants
$W(A_\pm)$. When we do this, we write, for example, $\J_{+,w}$ for
the Lefschetz thimble associated to $A_+$ with a Chern-Simons
invariant $2\pi w$; we extend this notation to the other flat
connections in an obvious way.  Here for $\J_\pm$, the value of
$w$ mod $\Z$ depends smoothly on $\gamma$ (varying from $0$ mod
$\Z$ at $\gamma=1/3$ to $1/2$ mod $\Z$ at $\gamma=2/3$); for
$\J'_\pm$, $w$ takes values in $\Z$; and for $\J''_\pm$, $w$ takes
values in $1/2+\Z$.

If we do not want exponential growth for $n\to\infty$ and
$\gamma<1/3$, we must pick the lifts of $W(A_+)$ and $W(A_-)$ to
be equal at $\gamma=1/3$; otherwise, according to (\ref{zelign}),
in continuing past the Airy singularity, $\J'_+$ as well as
$\J'_-$ will appear in the integration cycle, giving exponential
growth.  Similarly, if we do not want exponential growth for
$\gamma>2/3$, we need $W(A_+)=W(A_-)$ at $\gamma=2/3$. But it is
impossible to obey both of these conditions, since as was
illustrated in fig. \ref{lines}, if $W(A_+)$ and $W(A_-)$ are
equal at $\gamma=1/3$, then they differ by $2\pi$ at $\gamma=2/3$,
and vice-versa.

Exponential growth is therefore unavoidable, once $k$ is not an
integer, either for $\gamma<1/3$ or for $\gamma>2/3$.  As
explained in section \ref{preliminaries}, we wish to pick an
analytic continuation that leads to no exponential growth for
$\gamma<1/3$. A further clue is that, to preserve $\tau$ and
$\tilde\tau$ symmetry, $J_n(q)$ should be real for real $k$.
(Complex conjugation of the path integral reverses the sign of
$k$, but as the figure-eight knot is amphichiral, one can
compensate for this with the symmetry $\tau$ or $\tilde\tau$.)
This suggests that we should take the lift of $\A'_-$ such that
$\re\,W(\A'_-)=0$.
 Then the integration cycle for $\gamma$ just less than 1/3  is
\begin{equation}\label{tomx} \CC=\J_\ab+\J'_{-,0}.\end{equation}

Continuing past $\gamma=1/3$, cycles $\J_{+,w}$ and $\J_{-,w}$
appear, initially with $w=0$. But by the time we reach
$\gamma=2/3$, the values are $w=\pm 1/2$, as we learned in section
\ref{impar}. So the integration cycle as $\gamma$ approaches $2/3$
from below is
\begin{equation}\label{vomx}
\CC=\J_\ab+\J_{+,1/2}+\J_{-,-1/2}.\end{equation} To continue past
$\gamma=2/3$, we use eqn. (\ref{helign}) to express $\J_{+,1/2}$
and $\J_{-,-1/2}$ in terms of corresponding cycles $\J''_{\pm,w}$.
Since $w$ does not change in this process, the result is
$\J_{+,1/2}\to -\J''_{+,1/2}+\J''_{-,1/2}$, and $\J_{-,-1/2}\to
\J''_{+,-1/2}$.  The integration cycle for $\gamma$ just greater than 1/3 is
therefore
\begin{equation}\label{omox}\CC=\J_\ab-\J''_{+,1/2}+\J''_{-,1/2}+\J''_{+,-1/2}.\end{equation}

Near or on the real $\gamma$ axis, with $\re\,\gamma>2/3$,
exponential growth comes from the terms
$\J''_{+,-1/2}-\J''_{+,1/2}$ in (\ref{omox}). Just as in the
analytic continuation of the Bessel function, the relative minus
sign is the reason that exponential growth is avoided when $k$ is
an integer.  For $\gamma$ near 1, using (\ref{zonkox}), the value
of the path integral is
\begin{equation}\label{tomox}J_n(q)\sim
\exp(nV/2\pi)\left(\exp(-i\pi k)-\exp(i\pi
k)\right).\end{equation} Here the exponential factors $\exp(\pm
i\pi k)$ reflect the values $w=\pm 1/2$ for the relevant Lefschetz
thimbles.  Also, near $\gamma=1$, we have written the
exponentially growing factor as $\exp(nV/2\pi)$ (with $n$ rather
than $k$ in the exponent) to put the formula in a more
recognizable form for some readers.

The expression $\log J_n(q)/n$ does not have a simple large $n$
limit if $\gamma$ is on the real axis, because of the oscillatory
factor $\left(\exp(-i\pi k)-\exp(i\pi k)\right)$.  As soon as
$\gamma$ is off the real axis, there is a simple limit, but it is
discontinuous on the real $\gamma$ axis.  The reason for that is
that the large $n$ limit of $\log \left(\exp(-i\pi k)-\exp(i\pi
k)\right)/n$ is $\mp i\pi k/n=\mp i\pi/\gamma$, depending on the
sign of $\im\,k$.

In the region $2/3<\gamma<4/3$, the flat connection $\A_+$, with
its various possible lifts, has the maximum value of $\re\,\I$,
among all flat connections. This ensures that, once the cycles
 $\J''_{+,\pm 1/2}$ appear in $\CC$ at $\gamma=2/3$, they cannot
 disappear at least until $\gamma=4/3$.  At
 $\gamma=4/3$, and more generally at every value of $\gamma$ of the form $\pm 1/3$ mod $\Z$,
 one meets another family of Airy-type
 singularities.  To analyze the consequences, it may be necessary
 to take into account Stokes phenomena beyond what follows
 from the singularities.  (For example, there appears to be a
 Stokes line crossing the real $\gamma$ axis between the values
 $2/3$ and 1; if there is a flow from $\A_\ab$ to $\A_-$ along
 this line, this might cause $\J''_{-,1/2}$ to disappear from
 $\CC$.)

The normalization of $J_n(q)$ that is most often used in the
mathematical literature differs from ours by dividing by a factor
(namely the right hand side of eqn.  (\ref{zolly})) that has no exponential growth in the
semiclassical limit but does have  a simple zero at $\gamma=1$.
This makes the proper statement about the semiclassical limit
simpler at $\gamma=1$, though not at other real values near
$\gamma=1$.  The function $J_n(q)$, normalized as we do, vanishes
 (assuming that $n$ is an integer)
 for $q^n=1$ or $n=k+2$.
One may ask what that vanishing means in the context of
integration over Lefschetz thimbles.  The answer is that it may be
hard to see the vanishing exactly, but we can certainly see it
approximately. The exponentially growing contributions cancel if
$n=k+2$, since after all this implies that $k=n-2$ is an integer.
The remaining contribution is then dominated by $\J_\ab$, but this
contribution vanishes at $n=k+2$ since at this value, the
dimension of the automorphism group of the abelian flat connection
$A_\ab$ increases.

\section{Four Or Five-Dimensional Interpretation}\label{fourfive}

In section \ref{flowchern}, we observed that the flow equations of three-dimensional Chern-Simons
theory have a natural four-dimensional symmetry.  This is a familiar fact for the flow equations
of Chern-Simons theory with a compact gauge group; what is novel and perhaps surprising
is that four-dimensional symmetry also emerges in flows of complex connections.   Here we will make only a few
simple remarks about how this fact may be interpreted.  The ultimate goal of
such considerations might be to give a gauge theory interpretation to Khovanov homology,
which is a four-dimensional extension of three-dimensional Chern-Simons theory.

One observation is that, in the case of a finite-dimensional Kahler manifold $X$ with a holomorphic superpotential $W$, the Morse theory flow equations can be interpreted as
equations for supersymmetric solitons \cite{CV}.   To interpret the Chern-Simons flow equations
in this way, we need a two-dimensional theory with $\N=2$ supersymmetry in which the chiral superfields are a gauge
field $\A$ on a three-manifold $M$ (with some gauge group $H$), the superpotential is the Chern-Simons function $W(\A)$, and the gauge group is the group of maps from $M$
to  $H$.     There is no problem in constructing such a theory.
We simply start with five-dimensional maximally supersymmetric Yang-Mills theory
(of course, it may be better to start with the usual ultraviolet completion of this theory in
six dimensions) and compactify on $M$ with a topological twist that preserves some supersymmetry.  This
gives a two-dimensional theory with $\N=2$ supersymmetry and all the stated properties.  In this two-dimensional
theory, the Chern-Simons flow equations that we studied in section \ref{flowchern} are equations for
supersymmetric solitons.  This may be a useful starting point for a new perspective on
three-dimensional Chern-Simons theory.

Another perhaps more direct, though mysterious, relation between
three-dimensional Chern-Simons theory and a four-dimensional
theory is as follows.   As we observed in section \ref{relation},
for a given three-manifold $M$, the possible integration cycles of
Chern-Simons theory, and therefore the possible Chern-Simons path
integrals, form a vector space $\V$.   For a given choice of a
direction in the complex plane for the Chern-Simons coupling
parameter $k$, $\V$ has a basis corresponding to Lefschetz
thimbles associated to critical points.  Because of Stokes
phenomena, this basis is not really natural, but $\V$ does have a
natural lattice structure, since the condition for an element of
$\V$ to be an integer linear combination of basis vectors is
invariant under all jumping processes. We can interpret $\V$ as
the physical ``Hilbert'' space of a certain topological field
theory -- the twisted version of $\N=4$ super Yang-Mills theory
that was studied in \cite{KW} and is related to geometric
Langlands.  From this point of view, one studies the flow equation
on a general four-manifold $X$, not necessarily the product of a
three-manifold and a one-manifold. If $X$ is a four-manifold of
boundary $M$, then by counting solutions of the flow equations on
$X$, one would construct a vector in the physical Hilbert space of
$M$ and thus an integration cycle for Chern-Simons theory on $M$.
Of course, for this to make sense depends on the fact that the
flow equations are specializations of equations with
four-dimensional symmetry.

\vskip 1 cm
\bigskip\noindent
{\it Acknowledgments}  This research was supported in part by NSF
Grant Phy-0503584.  The project was stimulated by lectures at the
conference {\it Chern-Simons Gauge Theory: 20 Years After} at the
Hausdorff Institute in Bonn (August, 2009).
 I would like to thank the organizers for their hospitality, and  to acknowledge discussions during the meeting
 and on other occasions with
 J. Anderson, C. Beasley, T. Dimofte, D. Gaiotto, S. Garoufalidis, A. Givental,
 M. Goresky, S. Gukov, R. MacPherson, G. Moore,  and H. Murakami.

\bibliographystyle{unsrt}

\end{document}